\documentclass[12pt]{article}

\usepackage[cp1251]{inputenc}
\usepackage[english]{babel}
\usepackage{amsfonts}
\usepackage{amsmath}
\usepackage{amssymb}
\usepackage{latexsym}
\usepackage{amscd}

\newtheorem{theorem}{Theorem}[section]
\newtheorem{lemma}[theorem]{Lemma}
\newtheorem{prop}[theorem]{Proposition}

\newtheorem{cor}[theorem]{Corollary}

\newcommand{\rk}{{\rm rk}}
\newcommand{\id}{{\rm id}}
\newcommand{\Aut}{{\rm Aut}}
\newcommand{\Tr}{{\rm Tr}}
\newcommand{\tpr}{{\rm tpr}}
\newcommand{\Ker}{{\rm Ker\,}}
\newcommand{\im}{{\rm Im\,}}
\newcommand{\diag}{{\rm diag}}
\newcommand{\ann}{{\rm ann}}
\newcommand{\Sym}{{\rm Sym}}

\newcommand{\eps}{\varepsilon}
\newcommand{\cA}{{\cal A}}
\newcommand{\cB}{{\cal B}}
\newcommand{\cC}{{\cal C}}

\newcommand{\cL}{{\cal L}}
\newcommand{\cH}{{\cal H}}
\newcommand{\cO}{{\cal O}}
\newcommand{\cP}{{\cal P}}
\newcommand{\cR}{{\cal R}}
\newcommand{\cS}{{\cal S}}
\newcommand{\cX}{{\cal X}}
\newcommand{\cY}{{\cal Y}}

\newcommand{\bbF}{{\mathbb F}}

\newcommand{\bbN}{{\mathbb N}}

\newcommand{\normaleq}{\trianglelefteq}

\newcommand{\sdir}{\leftthreetimes}
\newcommand{\fe}{\varphi}
\newcommand{\lra}{\longrightarrow}
\newcommand{\ov}{\overline}
\newcommand{\ot}{\otimes}

\newcommand{\wt}{\widetilde}
\newcommand{\sse}{\subseteq}
\newcommand{\lu}{{\langle}}
\newcommand{\pu}{{\rangle}}

\newcommand{\eoproof}{\hfill $\square$ \vspace{1.5ex}}

\begin{document}

\title{Symmetries of matrix multiplication algorithms. I.}
\author{Vladimir P. Burichenko}
\date{}
\maketitle

\begin{center}
Institute of mathematics of National Academy of Sciences of Belarus \\
Kirov Street 32a, Gomel 246000, Republic of Belarus \\
vpburich@gmail.com
\end{center}

\begin{abstract}
In this work the algorithms of fast multiplication of
matrices are considered. To any algorithm there
associated a certain group of automorphisms. These
automorphism groups are found for some well-known
algorithms, including algorithms of Hopcroft,
Laderman, and Pan. The automorphism group is
isomorphic to $S_3\times Z_2$ and $S_4$ for Hopcroft
anf Laderman algorithms, respectively. The studying
of symmetry of algorithms may be a fruitful idea
for finding fast algorithms, by an analogy with
well-known optimization problems for codes, lattices,
and graphs.

\end{abstract}

\section{Introduction}
	\label{sec:intro}
In the present work we study algorithms of fast
multiplication
of matrices. This work is a continuation  of the previous
work of the author~\cite{Bur2014} (but it can be
read idependently of \cite{Bur2014}. It is even preferable
to read the present work before \cite{Bur2014}, because
some basic concepts are exposed here better than
in~\cite{Bur2014}).

In 1969 V.Strassen \cite{Strassen1969} found an
algorithm for multiplication of two $N\times N$ matrices,
requiring $O(N^\tau)$ (or, more exactly, $\leq 4.7N^\tau$)
arithmetical operations; here $\tau=\log_27=2.808...$. (Recall
that the usual algorithm (``multiplying a row by a column'')
requires $2N^3-N^2$ operations). This algorithm is based
on the fact, discovered by Strassen, that two $2\times2$
matrices with non-commuting elements, i.e., matrices over
an arbitrary associative ring~$R$, can be multiplied using
only 7 multiplications in~$R$.

Later some algorithms, asymptotically faster than Strassen's,
were found. We give a very short survey of the related works
in the end of this section.

The subject of the present work is studying symmetry of
algorithms. The author thinks that using symmetry may be a
fruitful way to find good algorithms.

Very short and clear exposition of the Strassen algorithm
(appropriate for a student) may be found in some textbooks
on linear algebra or computer algorithms. See, for example,
\cite{KM}, \S I.4, Ex.12, or \cite{AHU}, \S 6.2. An introduction
to the whole area of fast matrix multiplication may be found
in book \cite{BCS} or survey~\cite{Landsberg}. The books
\cite{de_Groote_book}, \cite{Blaser_book}, and Section 4.6.4
of \cite{Knuth} also should be mentioned. Nevertheless, the
author tried to make the present work self-contained, so in
this Introduction and the next section all necessary concepts,
related to matrix multiplication algorithms, are recalled.
Also, in Subsection~\ref{subs:intro_rem} are contained
some directions concerning literature in algebra.

\subsection{Definition of an algorithm}
	\label{subs:alg_def}
An algorithm for the multiplication of matrices of given size
may be described either in computational (i.e., as a
sequence of computations), matrix, or tensor form.

To give an example of an algorithm in computational form,
we recall the description of the Strassen algorithm. Let
$R$ be arbitrary (associative) ring, and let
$$ X=\begin{pmatrix} x_{11} & x_{12} \\ x_{21} & x_{22}
\end{pmatrix},\qquad Y=\begin{pmatrix} y_{11} & y_{12}
\\ y_{21} & y_{22} \end{pmatrix} $$
be matrices over~$R$. Consider the following products:
$$ p_1=x_{11}(y_{12}+y_{22}),\quad p_2=(x_{11}-x_{12})
y_{22},\quad  p_3=(-x_{21}+x_{22})y_{11}, $$
$$p_4=x_{22}(y_{11}+y_{21}),\quad p_5=(x_{11}+x_{22})
(y_{11}+y_{22}),$$
$$ p_6=(x_{11}+x_{21})(y_{11}-y_{12}),\quad p_7=(x_{12}
+x_{22})(y_{21}-y_{22}).$$
Next, take linear combinations
$$ z_{11}=-p_2-p_4+p_5+p_7,\quad z_{12}=p_1-p_2,$$
$$ z_{21}=-p_3+p_4,\quad z_{22}=-p_1-p_3+p_5-p_6.$$
It is easy to check that these $z_{ij}$ are nothing else but
the elements of the matrix $Z=XY$. Thus, we have computed
the product of $X$ and $Y$, using only 7 multiplications
(but 18 additions/subtractions) in~$R$.

Further,  describe what is the matrix form of an algorithm.
Let $m$, $n$, $p$, and $r$ be natural numbers, $K$ be
a field. Take symbols $x_{ij}$ and $y_{jk}$, where
$1\leq i\leq m$,
$1\leq j\leq n$, $1\leq k\leq p$ ($mn+np$ symbols total),
and let
$$ A=K\lu x_{ij}, y_{jk}\mid i,j,k\pu$$
be the free associative algebra over $K$ generated by these
symbols. Next, suppose we are given field elements $a_{ijl}$,
$b_{jkl}$, $c_{ikl}\in K$, for $1\leq i\leq m$, $1\leq j\leq n$,
$1\leq k\leq p$, $1\leq l\leq r$; we may think of them as
elements of $3r$ matrices
$$ a_l=(a_{ijl})_{1\leq i\leq m,\ 1\leq j\leq n}, \qquad
b_l=(b_{jkl})_{1\leq j\leq n,\ 1\leq k\leq p}, $$
$$ c_l=(c_{ikl})_{1\leq i\leq m,\ 1\leq k\leq p}, \qquad
l=1,\ldots,r. $$
Suppose that the following $mp$ relations in $A$ are true:
\begin{equation}    \label{f:rels_A}
\sum_{l=1}^rc_{ikl}(\sum_{\substack{ 1\leq u\leq m \\
1\leq v\leq n} } a_{uvl}x_{uv}) (\sum_{\substack{1\leq v\leq n
\\ 1\leq w\leq p}} b_{vwl}y_{vw})=\sum_{j=1}^nx_{ij}y_{jk}\,,
\end{equation}
for all $1\leq i\leq m$, $1\leq k\leq p$. Then we say that
the set of $r$ triples of matrices
$$ \cA=\{(a_l, b_l,c_l)\mid l=1,\ldots,r\} $$
is a {\em bilinear} (or {\em noncommutative}) algorithm
over $K$ for multiplication of an $m\times n$ matrix by an
$n\times p$ matrix, {\em requiring $r$ multiplications} (also
called an algorithm {\em of length $r$} or {\em of bilinear
complexity $r$}).

 Indeed, let $R$ be an arbitrary (associative) algebra over
$K$, and let $Q=(q_{ij})$ and $S=(s_{jk})$ be $m\times n$
and $n\times p$ matrices, respectively, over~$R$. Then
one can compute their product $T=QS$
in the following way. First compute all linear combinations
$$ d_l=\sum_{\substack{ 1\leq u\leq m \\ 1\leq v\leq n}}
a_{uvl}q_{uv},\qquad  f_l=\sum_{\substack{1\leq v\leq n \\
1\leq w\leq p}}b_{vwl}s_{vw}\,,$$
for all $1\leq l\leq r$; then compute all products $p_l=d_lf_l$,
and finally compute all linear combinations
$$ t'_{ik}=\sum_{l=1}^rc_{ikl}p_l\,,$$
for all $1\leq i\leq m$, $1\leq k\leq p$. Then it follows from
the relations (\ref{f:rels_A}) that $t'_{ik}=t_{ik}$ are precisely
the elements of~$T$. Thus, we have computed $T$, using
$r$ ``nontrivial'' (also called {\em non-scalar}) multiplications
in~$R$ (here a {\em scalar} multiplication means a 
multiplication by an element of~$K$). 
(It is well known that for algorithms of matrix
multiplication the number of multiplications is most
important. In particular, if there exists a non-commutative
algorithm $\cA$ for multiplication of an $m\times n$ matrix
by an $n\times p$ matrix, requiring $r$ multiplications, then
there is an algorithm for multiplication of two $N\times N$
matrices of complexity $O(N^\tau)$, where
$\tau=3\log_{mnp}r$. On the other hand, the number of
additions/subtractions and scalar (i.e., by an elements of $K$)
multiplications in $\cA$ affects only the constant factor in
$O(N^\tau)$. The details may be found in the literature).

{\bf Example.} It is easy to see that  the Strassen algorithm
may be written in matrix form as the following set of seven
triples of matrices:
$$ \left( \begin{pmatrix} 1  & 0  \\ 0   & 0  \end{pmatrix},
\begin{pmatrix} 0  & 1  \\  0  & 1 \end{pmatrix},
 \begin{pmatrix} 0  & 1  \\ 0  & -1  \end{pmatrix}\right),\quad
\left( \begin{pmatrix} 1  & -1  \\ 0   & 0   \end{pmatrix},
\begin{pmatrix} 0  & 0  \\ 0  & 1  \end{pmatrix},
 \begin{pmatrix}  -1 & -1  \\ 0  & 0  \end{pmatrix} \right),$$
$$\left( \begin{pmatrix} 0  & 0  \\ -1   & 1   \end{pmatrix},
\begin{pmatrix} 1  & 0  \\ 0  & 0 \end{pmatrix},
 \begin{pmatrix} 0  & 0  \\ -1  & -1 \end{pmatrix} \right),
\quad  \left( \begin{pmatrix} 0  & 0  \\ 0  & 1  \end{pmatrix},
\begin{pmatrix} 1  & 0  \\ 1   & 0  \end{pmatrix},
\begin{pmatrix} -1  & 0  \\ 1  & 0  \end{pmatrix} \right), $$
 $$\left( \begin{pmatrix} 1  & 0  \\ 0  & 1  \end{pmatrix},
\begin{pmatrix} 1  & 0  \\ 0  & 1 \end{pmatrix},
\begin{pmatrix} 1  & 0 \\ 0  & 1  \end{pmatrix} \right), \quad
\left( \begin{pmatrix} 1 & 0 \\ 1  & 0  \end{pmatrix},
\begin{pmatrix} 1 & -1 \\  0 & 0 \end{pmatrix},
\begin{pmatrix} 0  & 0  \\ 0  & -1  \end{pmatrix} \right), $$
$$\left( \begin{pmatrix} 0  & 1 \\ 0  & 1 \end{pmatrix},
\begin{pmatrix} 0 & 0 \\ 1 & -1 \end{pmatrix},
 \begin{pmatrix} 1 & 0 \\ 0  & 0 \end{pmatrix} \right). $$

We will denote the Strassen algorithm by~$\cS$.

A description of what is an algorithm in tensor form will be
given in Section~\ref{sec:alg_tens}.

\subsection{Motivation, the aim of the work, and the results}
	\label{subs:aims}
By $r_K(m,n,p)$ we denote the minimal number of
multiplications in a bilinear algorithm over a field $K$ for
multiplication of an $m\times n$ matrix by an $n\times p$
matrix. In principle, $r_K(m,n,p)$ may depend on $K$, but
the author does not know any particular example of $m$,
$n$, $p$, $K_1$ and $K_2$ such that $r_{K_1}(m,n,p)\ne
r_{K_2}(m,n,p)$ (but, in the author's opinion, such examples
certainly must exist).

It is widely recognized that finding of $r_K(m,n,p)$ for small
$m$, $n$, $p$ is an important problem, both from
theoretical and practical viewpoint. The greatest interest
at the moment is attracted by $r(3,3,3)$.

For small $m$, $n$, $p$ the following estimates are known
(for any $K$).
\begin{itemize}
\item $r(2,2,2)=7$.  The inequality $r(2,2,2)\leq7$ follows
from the existence of the Strassen algorithm. The opposite
inequality $r(2,2,2)\geq7$ was first proved in~\cite{Win71},
and later several other proofs were found.
\item $r(2,2,3)=11$, $r(2,2,4)=14$, $17\leq r(2,2,5)\leq 18$.
Here the upper estimates easily follow from the Strassen
algorithm, and the lower ones were proved by V.B.Alekseev
in works \cite{Al1}, \cite{Al-Sm}, \cite{Al3}, respectively.
\item $14\leq r(2,3,3)\leq15$, $19\leq r(3,3,3)\leq23$.
Here the upper estimates follow from the algorithms
contained in the works \cite{Hop-Kerr} and \cite{Laderman}
respectively (we recall these algorithms in Sections
\ref{sec:hop} and \ref{sec:lad}). The lower estimates
were proved by Bl\"aser in works
\cite{Blaser1} and \cite{Blaser2}, respectively.
\end{itemize}
It is also well known that $r(m,n,p)$ is symmetric in $m$, $n$,
and $p$, and that $r(m,n,1)=mn$.

It should be noted that in the case when $K=GF(2)=\{0,1\}$,
or if the coefficients of algorithms are supposed to be
integers, there are some further results (see \cite{Hop-Kerr}
and~\cite{Hop-Mus}).

It may be a good idea in the search for economical
algorithms that such algorithms may have many symmetries,
that is, a large automorphism group. (It will be explained
later in the article what we mean by an automorphism of
an algorithm). Note that one faces the similar situation
when studying codes, lattices, or graphs. Good (that is,
dense) lattices and codes often have large group of
automorphisms (see \cite{CS} for numerous examples of
this phenomenon). Similarly, the graphs satisfying certain
regularity conditions (distance regular graphs, especially
those with ``extremal'' set of parameters) often have large
automorphism group; see~\cite{BCN}.

In \cite{Bur2014} the author has proved that the Strassen
algorithm $\cS$ has the automorphism group $\Aut(\cS)
\cong S_3\times S_3$, or even $S_3\times D_6$, if we
consider automorphisms in some ``extended'' sense.
Here $S_3$ is the symmetric group on 3 letters, and $D_6$
is the dihedral group of order~12 (i.e., the group of all
symmetries of a regular hexagon).

It should be noted that $\cS$ is, in a sense, unique: any
other algorithm for multiplication of two $2\times2$
matrices requiring 7 multiplications is conjugate to $\cS$
under certain transformation group. See \cite{deGroote1},
\cite{deGroote2}.

Before trying to find good algorithms with large
automorphism groups in unknown cases (say, for
multiplication of $3\times3$ matrices), it is a reasonable
first step to study automorphisms of some good algorithms
known so far. This is the aim of the present work.

J.E.Hopcroft found an algorithm for multiplying of
$3\times2$ matrix by a $2\times3$ matrix, requiring 15
multiplications. This algorithm is described in
\cite{Hop-Kerr}, and more accurately in~\cite{Hop-Mus}.
We denote the Hopcroft algorithm by~$\cH$.

J.Laderman~\cite{Laderman} found an algorithm for
multiplication of two $3\times3$ matrices requiring 23
multiplications. We denote this algorithm by~$\cL$.

V.Ya.Pan (see, for example, \cite{Pan_howto})
described several algorithms
for multiplication of matrices of arbitrary size, known as the
{\em trilinear aggregation algorithms}. The most known of
them is an algorithm for multiplication of two $n\times n$
matrices, where $n=2m$ is even, requiring $(n^3-4n)/3+
6n^2$ multiplications. We denote this algorithm
by~$\cP_{2m}$.

One of the main results of the present work is the following
theorem.
\begin{theorem} \label{th:main}
Let $\cH$, $\cL$, and $\cP_{2m}$ be the algorithms of
Hopcroft, Laderman and Pan, mentioned above. Then
$$ \Aut(\cH)\cong S_3\times Z_2\,,$$
$$ \Aut(\cL)\cong S_4\,, $$
and
$$ \Aut(\cP_{2m})\cong S_m\times Z_2\times S_3\,.$$
\end{theorem}
(Of course, we will give a description of automorphism
groups, mentioned in this theorem, not only up to
isomorphism, but in an explicit form).

\subsection{Some further remarks}
	\label{subs:intro_rem}
{\bf Remark 1.} Since Strassen's work, other estimates for
asymptotic complexity of matrix multiplication (better than
$O(N^{2.81})$) were found. The authors who contributed to
these investigations are (approximately in chronological
order) Pan, Bini/Capovani/Lotti/Romani, Sch\"onhage,
Strassen, and Coppersmith/Winograd. The most significant
progress was made due to the so-called ``laser method'' of
Strassen. The details and references may be found in the
literature, see for example Chapter 15 of \cite{BCS} and the
introduction to~\cite{Williams2}. The most recent works
belong to Stothers \cite{Stothers}, Vassilevska-Williams
\cite{Williams} (see also \cite{Williams2}), and 
Zhdanovich~\cite{Zhdan}.
In \cite{Williams} and \cite{Zhdan}, independently, the
estimation $O(N^\omega)$ with $\omega<2.373$ was
proved.

It should be said that in all these estimations the constant
factor in $O(N^\omega)$ is very large, so that the
corresponding algorithms are only of theoretical interest
and are useless in practice. For practical purposes only the
following algorithms
may be used: the usual algorithm, the Strassen algorithm,
the Pan trilinear aggregation method, and the ``compound''
algorithms (which will be mentioned in the next remark).

{\bf Remark 2.} There are other types of algorithms for
matrix multiplication, different from bilinear algorithms
as described above. Namely, there are
\begin{itemize}
\item  {\em commutative} (or {\em quadratic}) algorithms,
which may be used if we suppose that elements of matrices
belong to a commutative ring; see \cite{Win2} or \cite{Mak1}
for the examples of such algorithms, and \S 14.1 of \cite{BCS}
for general definition;
\item {\em approximate} algorithms, like in \cite{BCLR}
(see \cite{BCS}, \S 15.2 for further explanations);
\item {\em ``compound''} bilinear algorithms, that is, the
algorithms assembled, in an appropriate way, from several
algorithms of smaller formats. The work \cite{DIS} contains
a survey of such algorithms.
\end{itemize}
The algorithms of all these three types are {\em not}
considered in the present work.

{\bf Remark 3.} The present text is written in a manner a bit
different from the usual journal article. The author means
that he gives more details than it is usual in a journal. So
the reader may find some places trivial. The reason is that
the author wishes that the text could be readable both by
specialists in algebra
and by computer scientists; but these specialists may have
modest background in computer science or algebra,
respectively.

The author would like to give some references / reading
suggestions for readers who may be not very
experienced in algebra (say, computer scientists).

The reader can use textbooks \cite{Kostr}, \cite{Artin},
and \cite{KM} as a basic course in general and linear
algebra (including the basics of the group
representation theory). The book \cite{Yok} contains
a very lucid exposition of multilinear algebra
(i.e., the theory of tensors). The last chapter of
\cite{KM} is also devoted to multilinear algebra.
The book \cite{CR} is a classical (but not elementary)
source for group representation theory); chapters 1
and 2 are especially recommended to the reader.
There is also an elementary and application-oriented
textbook~\cite{JL}. Finally, we should list some
graduate-level algebra courses, namely \cite{Lang},
\cite{Hungerford}, and~\cite{DF}.

{\bf Structure of the work.} The work is organized
as follows. In Section~\ref{sec:alg_tens} we recall
the relations between matrix multiplication
algorithms and decompositions of tensors.
Section~\ref{sec:group_tens} contains general
considerations on symmetry of tensors and algorithms.
In a long Section~\ref{sec:mnp} we find the isotropy
group of the structure tensor of matrix multiplication
map (which is a necessary preliminary step for
studying automorphisms of any particular algorithm).
In Sections~\ref{sec:lad} and \ref{sec:hop} we find
automorphism groups of Laderman and Hopcroft algorithms,
respectively.

In Part II of the work Pan's trilinear aggregation
algorithm, and some other topics, will be considered.

{\bf Acknowledgement.} The author thanks A.S.Kleshchev
for useful literature directions.

\section{Tensor form of an algorithm}
	\label{sec:alg_tens}

Let $V_1,\ldots, V_l$ be vector spaces over a field $K$,
$\wt V=V_1\otimes \ldots\otimes V_l$ be their tensor
product. A tensor $t\in\wt V$ is {\em decomposable} if
$t=v_1\otimes\ldots\otimes v_l$, for some $v_i\in V_i$,
$i=1,\ldots,l$. We will consider representations of a tensor
$t\in\wt V$ in the form $t=t_1+\ldots+t_s$, where
$t_1\,,\ldots,t_s$ are decomposable tensors. The least
possible length $s$
of such a representation is called the {\em rank} of $t$, and
is denoted by $\rk(t)$. Obviously, $\rk(t)=1$ if and only if
$t$ is decomposable.

In the situation described the following terminology is also
used: the set $\{t_1\,,\ldots,t_s\}$ is called an {\em algorithm}
(of length $s$), {\em computing}~$t$.

By $M_{mn}(K)$ (or just $M_{mn}$) we denote the space of
all $m\times n$ matrices over~$K$. The basis of $M_{mn}(K)$
is $(e_{ij}\mid 1\leq i\leq m,\ 1\leq j\leq n)$, where $e_{ij}$
are usual matrix units.

In the sequel an important role is played by the tensor
$$ \lu m,n,p\pu= \sum_{1\leq i\leq m,\ 1\leq j\leq n,\
1\leq k\leq p} e_{ij}\otimes e_{jk}\otimes e_{ki}
\ \in M_{mn}\ot M_{np}\ot M_{pm}\,. $$
(Here two remarks concerning notation are in order: (1) Note
that we have abused notation a bit, by using the similar
symbols $e_{ij}$ and $e_{jk}$ for elements of different spaces;
(2) in \cite{Bur2014} the tensor $\lu m,n,p\pu$ was denoted
by $S(m,n,p)$. The notation $\lu m,n,p\pu$, which is now
classical, is due to Sch\"onhage.)

It is a classical fact (first established by Strassen) that there
is a bijection between the set of all algorithms for
multiplication of an $m\times n$ matrix by an $n\times p$
matrix requiring $r$ multiplications, and the set of all
algorithms of length~$r$
computing $\lu m,n,p\pu$. The following proposition gives
an explicit description of this bijection. Note that in
condition~(a) of this proposition we think of matrices as
tables whose elements are elements of $K$, so that
$$ a=(a_{ij})_{1\leq i\leq m,\ 1\leq j\leq n}\,,$$
whereas in condition (b) we think of matrices as elements
of linear spaces, so that
$$ a=\sum_{1\leq i\leq m,\ 1\leq j\leq n} a_{ij}e_{ij}\,.$$

As usually, by $x^t$ and $\delta_{ab}$ we denote the
transposed matrix and the Kronecker delta symbol.

\begin{prop}    \label{pr:mat_tens_brent}
Let $m,n,p,r\in\bbN$, and let $K$ be a field. Let
$$ a_l=(a_{ijl})_{1\leq i\leq m,\ 1\leq j\leq n}\ ,\quad  b_l=
(b_{jkl})_{1\leq j\leq n, \ 1\leq k\leq p}\ , \quad \text{and } c_l=
(c_{ikl})_{1\leq i\leq m,\ 1\leq k\leq p}\ ,$$
where $l=1,\ldots, r$, be matrices over $K$, of sizes $m\times n$,
$n\times p$, and $m\times p$, respectively.

Then the following three conditions are equivalent:

(a) $\{(a_l,b_l,c_l)\mid l=1,\ldots,r\}$ is a bilinear algorithm
over $K$ for multiplication of an $m\times n$ matrix by
an $n\times p$ matrix;

(b) $\{a_l\ot b_l \ot (c_l) ^t \mid l=1,\ldots,r\}$ is an algorithm
computing tensor $ \lu m,n,p\pu$;

(c) the following $(mnp)^2$ equations, for all
$1\leq i,i_1\leq m$, $1\leq j,j_1\leq n$, and $1\leq k,k_1
\leq p$, are satisfied:
\begin{equation}    \label{f:Brent}
\sum_{l=1}^r a_{ijl}b_{j_1kl}c_{i_1k_1l}=\delta_{ii_1}
\delta_{jj_1}\delta_{kk_1}\,.
\end{equation}
\end{prop}
{\em Proof.} First we prove that conditions (a) and (c)
are equivalent. Consider relation (\ref{f:rels_A}) of the
Introduction,
$$  \sum_{l=1}^rc_{ikl}(\sum_{\substack{ 1\leq u\leq m
\\ 1\leq v\leq n} } a_{uvl}x_{uv}) (\sum_{\substack{1
\leq v\leq n \\ 1\leq w\leq p}} b_{vwl}y_{vw})=
\sum_{j=1}^nx_{ij}y_{jk}, $$
for a given pair $(i,k)$. Clearly, this relation is true if and
only if the coefficients at $x_{ef}y_{gh}$ on both sides
coincide, for every quadruple $(e,f,g,h)$ such that
$1\leq e\leq m$, $1\leq f,g\leq n$, $1\leq h\leq p$. It is easy
to see that the coefficient on the left is $\sum_{l=1}^r c_{ikl}
a_{efl}b_{ghl}$, and the coefficient on the right is $\delta_{ie}
\delta_{fg}\delta_{kh}$. Thus we obtain the condition
$$ \sum_{l=1}^r c_{ikl}a_{efl}b_{ghl}= \delta_{ie}\delta_{fg}
\delta_{kh}\,, $$
for all $i$, $k$, $e$, $f$, $g$, $h$ such that $1\leq i,e\leq m$,
$1\leq f,g\leq n$, $1\leq h,k\leq p$. But the latter condition
coincides with the equality in condition~(c), up to names
of indices (namely, $e$, $f$, $g$, $h$, $i$, $k$ should be
changed to $i$, $j$, $j_1$, $k$, $i_1$, $k_1$, respectively).

In a similar way one can prove that conditions (b) and (c)
are equivalent too. Indeed, condition (b) means that
$$ \sum_{l=1}^r a_l\ot b_l\ot (c_l)^t=\lu m,n,p\pu. $$
Now it is sufficient to observe that the tensors of the form
$e_{ij}\ot e_{j_1k}\ot e_{k_1i_1}$, where $1\leq i,i_1\leq m$,
$1\leq j,j_1\leq n$, and $1\leq k,k_1\leq p$, form the basis of
$M_{mn}\ot M_{np}\ot M_{pm}$, and to calculate the
coefficients at basis elements in both sides of the latter
relation.

Finally, the conditions (a) and (b) are equivalent, because
each of them is equivalent to~(c). \eoproof
\vspace{2ex}

Let $\cA=\{(a_l,b_l,c_l)\mid l=1,\ldots,r\}$ be an algorithm
for multiplication of an $m\times n$ matrix by an
$n\times p$ matrix, and let $ \cA'=\{a_l\ot b_l\ot c_l^t
\mid l=1,\ldots,r\}$ be the corresponding algorithm
computing $\lu m,n,p\pu$. Then we say that $\cA'$ is the
{\em tensor form} of~$\cA$.

{\bf Example.} In the Introduction we have recalled
Strassen algorithm and have written it in matrix form.
It is readily seen that in the tensor form this algorithm
is
\begin{eqnarray*}
\cS=\{ && e_{11}\ot(e_{12}+e_{22})\ot(e_{21}-e_{22}),\
(e_{11}-e_{12})\ot e_{22}\ot(-e_{11}-e_{21}),\\
&&  (-e_{21}+e_{22})\ot e_{11}\ot(-e_{12}-e_{22}),\
e_{22}\ot(e_{11}+e_{21})\ot(-e_{11}+e_{12}),\\
&& (e_{11}+e_{22})\ot(e_{11}+e_{22})\ot(e_{11}+e_{22}),\
(e_{11}+e_{21})\ot(e_{11}-e_{12})\ot(-e_{22}),\\
&&  (e_{12}+e_{22})\ot(e_{21}-e_{22})\ot e_{11}\ \}.
\end{eqnarray*}

{\bf Remark.} The reader can check directly that the sum of
the tensors of the latter set is $\lu 2,2,2\pu$. Such a checking
can be considered as an evidence that we had made no
mistakes when finding the matrix form of the Strassen
algorithm from its computational form, and then the tensor
form from the matrix form.
\medskip

The equations (\ref{f:Brent}) first appeared in \cite{Brent}, so
they are called {\em Brent equations} (but it is possible that
similar equations appeared earlier in studying
decompositions of general tensors, see~\cite{KB}).

One of approaches to finding algorithms for matrix multiplication
is to solve Brent equations, usually by computer calculations. To
do this, one usually reduces solving the Brent equations to
finding minima of certain real-valued function of many (several
hundreds) variables, and then solves this optimization problem by
numerical methods. See \cite{Brent}, \cite{JM} for more details.
Other works in this direction are \cite{3korejca} and~
\cite{Smirnov}. In works \cite{Bard_ea} and \cite{Laderman} the
Brent equations are also used, but in a different way (without
numerical optimization).

\section{Group actions on tensors and algorithms}
	\label{sec:group_tens}
Let $U_1,\ldots, U_m$ and $V_1,\ldots,V_n$ be spaces over
a field $K$, and let $\wt U=U_1\ot\ldots\ot U_m$  and
$\wt V=V_1\ot\ldots\ot V_n$ be their tensor products. By a
{\em decomposable isomorphism} we mean an isomorphism
of vector spaces $\fe:\wt U\to\wt V$ such that there are a
bijection $\tau:\{1,\ldots,m\}\to\{1,\ldots,n\}$ (whence
$m=n$) and isomorphisms $\fe_i:U_i\to V_{\tau(i)}$
(whence $\dim V_{\tau(i)}=\dim U_i$ for all $i=1,\ldots,m$)
such that
$$ \fe(u_1\ot\ldots\ot u_m)=\fe_{\tau^{-1}(1)} (u_{\tau^{-1}
(1)})\ot\ldots\ot \fe_{\tau^{-1}(m)} (u_{\tau^{-1}(m)})$$
for all $u_i\in U_i$, $i=1,\ldots,m$.

For example, the usual permutation of factors $\pi: X\ot Y
\lra Y\ot X$, $x\ot y\mapsto y\ot x$, is a decomposable
isomorphism.

It is clear that the composition of two decomposable
isomorphisms is a decomposable isomorphism also, and the
isomorphism inverse to a decomposable isomorphism is
decomposable too.
It follows that a decomposable isomorphism $\fe:\wt U\lra
\wt V$ maps the set of all decomposable tensors in $\wt U$
bijectively onto the set of all decomposable tensors
in~$\wt V$.

If $\fe$ is a decomposable isomorphism defined by data
$(\tau; \fe_1\,,\ldots,\fe_m)$, as in the first paragraph of this
subsection, then we say that $\tau$ is a permutation,
corresponding to $\fe$ (and $\fe$ is an isomorphism,
corresponding to $\tau$). For $\fe_i$ the similar
terminology is used.

It is often convenient to think of the permutation $\tau$,
corresponding to $\fe$, as a (bijective) map from set of
factors $\{U_1\, \ldots, U_m\}$ of $\wt U$ to the similar set
$\{V_1\,,\ldots,V_m\}$ of~$\wt V$. In particular, if $\wt U
=\wt V$, then we may think of $\tau$ as a permutation of
the set $\{U_1\, \ldots, U_m\}$.

It is clear that if $\fe$ and $\psi$ are decomposable
isomorphisms, and $\sigma$ and $\tau$ are permutations
corresponding to $\fe$ and $\psi$, respectively, then $\rho=
\sigma\tau$ is a permutation corresponding to the
isomorphism $\theta=\fe\psi$. Moreover, $\sigma^{-1}$
is a permutation corresponding to~$\fe^{-1}$. (Note that
we multiply permutations ``from right to left'', for example,
$(134)(2)\cdot(14)(23)=(1)(243)$).

In general, the permutation (as well as the isomorphisms
$\fe_i$), corresponding to a given decomposable
isomorphism $\fe$, is defined by $\fe$ not uniquely.
Consider two examples.

{\bf Example 1.} Let $m=2$, $U_1=\lu e_1\pu$,
$U_2=\lu e_2\pu$, $V_1=\lu f_1\pu$, and $V_2=\lu f_2\pu$
be one-dimensional spaces. Their tensor products
$\wt U=U_1\ot U_2=\lu e_1 \ot e_2\pu$ and  $\wt V=V_1
\ot V_2=\lu f_1\ot f_2\pu$ are one-dimensional also. Let
$\fe:\wt U\lra\wt V$ be the isomorphism taking $e_1\ot e_2$
to $f_1\ot f_2$. Then $\fe$ is the decomposable isomorphism
corresponding to the permutation
$\tau=e=\{1\mapsto 1,\ 2\mapsto2\}$ and isomorphisms
$\fe_1:e_1\mapsto f_1$, $\fe_2:e_2 \mapsto f_2$. On the
other hand, $\fe$ may be considered as the decomposable
isomorphism, corresponding to permutation $\tau'=(1,2)=
\{1\mapsto2,\ 2\mapsto1\}$ and isomorphisms $\fe'_1:e_1
\mapsto f_2$, $\fe'_2:e_2\mapsto f_1$.

{\bf Example 2.} Let $\fe_1:U_1\lra V_1$ and $\fe_2:U_2\lra
V_2$ be isomorphisms of spaces. Then the decomposable
isomorphism
$$ \fe=\fe_1\ot\fe_2\::\: \wt U=U_1\ot U_2\lra \wt V=
V_1\ot V_2 $$ can also be written as $\fe=\fe'_1\ot\fe'_2$, where
$\fe'_1 =\lambda\fe_1$ and $\fe'_2=\lambda^{-1}\fe_2$, for any
$\lambda\in K^\ast$.

It turns out that the possible ambiguity of data $(\tau;
\fe_1\,,\ldots,\fe_m)$, corresponding to a given
decomposable isomorphism $\fe$, may be only of these two
kinds. To prove this, we need two simple (and well-known)
statements.

\begin{lemma}   \label{l:scalar}
1) Suppose that $\fe\in GL(V)$ is an automorphism of a
space $V$ such that $\fe(\lu v\pu)=\lu v\pu$ for each
one-dimensional subspace $\lu v\pu\sse V$. Then $\fe$
is a multiplication by a scalar $\lambda\in K^\ast$.

2) Let $\fe,\fe':U\lra V$ be isomorphisms such that
$\fe(\lu u\pu) =\fe'(\lu u\pu)$ for each one-dimensional
subspace $\lu u\pu\sse V$. Then there exists  $\lambda
\in K^\ast$ such that $\fe'=\lambda\fe$.
\end{lemma}
{\em Proof.} 1) Let $v_1,\ldots,v_n$ be a basis of~$V$. As
$\fe$ preserves all $\lu v_i\pu$, we have $\fe(v_i)=\lambda_i
v_i$, for some elements $\lambda_i\in K^\ast$. Next, take
any $i\ne j$. The vector $\fe(v_i+v_j)=\lambda_iv_i+
\lambda_jv_j$ must be proportional to $v_i+v_j$, whence
$\lambda_i=\lambda_j$. Consequently, all $\lambda_1=
\lambda_2=\ldots=\lambda_n$ coincide, whence $\fe$
is the multiplication by a scalar $\lambda=\lambda_1$.

2) It is obvious that $\theta=\fe^{-1}\fe'$ is an
automorphism of $U$, and for any $u\in U$ we have
$\theta(\lu u\pu)=(\fe^{-1}\fe')(\lu u\pu)=\fe^{-1}(\fe'(\lu u
\pu))=\fe^{-1}(\fe(\lu u\pu))=\lu u\pu$. Now we have
$\theta=\lambda$ (=$\lambda\cdot\id_U$) by 1), whence 
$\fe'=\lambda\fe$.
\eoproof

\begin{lemma}   \label{l:prod_subsp}
Let $\wt U=U_1\ot\ldots\ot U_m$ be the tensor product of
several spaces, and for each $i=1,\ldots,m$ let $T_i,T'_i
\sse U_i$ be two nonzero subspaces. Then the subspaces
$$ \wt T=T_1\ot \ldots\ot T_m \quad \ \text{and}\ 
\quad \wt T'=T'_1\ot \ldots\ot T'_m\sse\wt U$$
coincide if and only if $T_i=T'_i$ for all~$i$.  In particular,
if $0\ne t=u_1\ot\ldots\ot u_m=u'_1\ot\ldots\ot u'_m$ are
two decompositions of a nonzero decomposable tensor,
then $\lu u_i\pu=\lu u'_i\pu$ for all $i=1,\ldots,m$.
\end{lemma}

(The proof is left to the reader.)

\begin{prop}    \label{pr:dec_is_eq}
Let $U_1\,\ldots,U_m$ and $V_1,\ldots,V_m$ be two sets of
$m$ spaces each, and let $\wt U= U_1\ot\ldots\ot U_m$ and
$\wt V= V_1\ot\ldots\ot V_m$ be their tensor products.
Let $\tau,\tau'\in S_m$ be permutations such that
$\dim V_{\tau(i)}=\dim V_{\tau'(i)}=\dim U_i$, let
$\fe_i:U_i\lra V_{\tau(i)}$ and $\fe'_i:U_i\lra V_{\tau'(i)}$
be isomorphisms, and let $\fe,\fe':\wt U\lra\wt V$ be the
decomposable isomorphisms corresponding to the data
$(\tau; \fe_1\,,\ldots,\fe_m)$ and $(\tau'; \fe'_1\,,\ldots,\fe'_m)$,
respectively. Suppose that $\fe=\fe'$. Then the following two
statements hold.

1) $\tau'=\tau\sigma$, where $\sigma$ is a permutation of
$\{1,\ldots,m\}$ such that $\sigma(i)=i$ for all $i$ such that
$\dim U_i>1$. In particular, $\tau'=\tau$, if at most one of
the spaces $U_i$ is one-dimensional.

2) Suppose that $\tau=\tau'$. Then there exist elements
$\lambda_1\,,\ldots,\lambda_m\in K^\ast$ such that
$\fe'_i=\lambda_i\fe_i$. For these $\lambda_i$ we have
$\lambda_1\ldots\lambda_m=1$.
\end{prop}
{\em Proof.} 1) Define $\sigma=\tau^{-1}\tau'$. Then $\tau'
=\tau\sigma$. It follows from the previous discussion that
$\sigma$ is a permutation corresponding to the
decomposable automorphism $\theta=\fe^{-1}\fe'$
of~$\wt U$. But, clearly, $\theta=1=\id_{\wt U}$. So it is
sufficient to prove the following: if the decomposable
automorphism $\theta$ of $\wt U$, defined by the data
$(\sigma; \psi_1\,,\ldots,\psi_m)$ (where $\psi_i:U_i\lra
U_{\sigma(i)}$ are some isomorphisms) coincides with
$\id_{\wt U}$, then $\sigma(i)=i$ for all $i$ such that
$\dim U_i>1$.

We may assume without loss of generality that $i=1$. For
each $j=2,\ldots,m$ take a one-dimensional subspace
$l_j\sse U_j$ and consider the subspace
$$W=U_1\ot l_2\ot\ldots \ot l_m\sse\wt U.$$
Then $\theta(W)=X_1\ot\ldots\ot X_m$,
where $X_{\sigma(1)}=\psi_1(U_1)=U_{\sigma(1)}$,
and $X_r=\psi_{\sigma^{-1}(r)}(l_{\sigma^{-1}(r)})$ for
$r\ne\sigma(1)$. In particular, $\dim X_{\sigma(1)}>1$
and $\dim X_j=1$ if $j\ne\sigma(1)$.

Since $\theta(W)=W$, it follows from Lemma~\ref{l:prod_subsp}
that $X_1=U_1$ and $X_j=l_j$ for $j\geq2$. In particular,
$\dim X_1>1$ and $\dim X_j=1$ for $j\geq2$. Consequently,
$\sigma(1)=1$.

2) Consider $\theta=\fe^{-1}\fe'=\id_{\wt U}$ again. It is
rather clear that $\theta$ is the decomposable
automorphism  of $\wt U$, corresponding to $(e; \psi_1\,,
\ldots, \psi_m)$, where $e$ is the identity permutation of
$\{1,\ldots,m\}$ and $\psi_i=\fe_i^{-1}\fe'_i$. That is, $\theta
=\psi_1\ot\ldots\ot\psi_m$. So it
suffices to show that if $\psi_i\in GL(U_i)$ are some
automorphisms such that $\theta=\psi_1\ot\ldots\ot\psi_m
=\id_{\wt U}$, then there exist $\lambda_1\,,\ldots,
\lambda_m\in K^\ast$ such that $\psi_i=\lambda_i
\id_{U_i}$, and that these $\lambda_i$ satisfy the relation
$\lambda_1\ldots\lambda_m=1$.

Take arbitrary nonzero elements $u_i\in U_i$, $u_i\ne0$.
Then $w=u_1\ot\ldots\ot u_m\ne0$, and $w=\theta(w)=
\psi_1(u_1)\ot\ldots\ot\psi_m(u_m)$. It follows from Lemma
\ref{l:prod_subsp} that $\lu \psi_i(u_i)\pu=\lu u_i\pu$ for all~$u_i$.
Since $u_i$ were taken arbitrarily, it follows that $\psi_i(x)$
is proportional to $x$ for all $x\in U_i$. By Lemma~\ref{l:scalar},
there exists $\lambda_i\in K^\ast$ such that $\psi_i=
\lambda_i\id_{U_i}$. Finally, $w=\psi_1(u_1)\ot\ldots\ot
\psi_m(u_m)=\lambda_1u_1\ot\ldots\ot\lambda_mu_m
=\lambda_1\ldots\lambda_mw$, whence $\lambda_1\ldots
\lambda_m=1$.
\eoproof
\vspace{2ex}

In particular, we see that if at most one of the spaces $U_i$
is one-dimensional, then for any decomposable
automorphism $\fe$ of the space $\wt U=U_1\ot\ldots\ot
U_m$ the corresponding permutation of $\{U_1,\ldots,
U_m\}$ is determined uniquely.

By $S(U_1\,,\ldots,U_m)$ we denote the group of all
decomposable automorphisms of $\wt U=U_1\ot\ldots
\ot U_m$. Next, by $S^0(U_1\,,\ldots,U_m)$ we denote the
subgroup of $S(U_1\,,\ldots,U_m)$ consisting of all
automorphisms that preserve each factor~$U_i$
(that is, corresponding to the trivial permutation of $\{U_1\,,
\ldots,U_m\}$). In other words, $S^0(U_1\,,\ldots,U_m)$
is the image of the homomorphism $GL(U_1)\times\ldots
\times GL(U_m)\to GL(\wt U)$ defined by
$$ (g_1,\ldots,g_m)\mapsto g_1\ot\ldots\ot g_m\,.$$
Clearly, $S^0(U_1\,,\ldots,U_m)$ is normal in $S(U_1\,,\ldots,
U_m)$. The corresponding quotient group $T=S(U_1\,,\ldots,
U_m)/S^0(U_1\,,\ldots,U_m)$ may be described as follows.
Let $\Omega=\{U_i\mid \dim U_i>1\}$ be the set of all
factors $U_i$ of dimension $>1$, and let
$$ T'=\{g\in\Sym(\Omega)\mid\dim g(X)=\dim X\,\ \ \forall\
X\in\Omega\} $$
be the group of all permutations of these factors, preserving
dimensions. Then it is easy to deduce from
Proposition~\ref{pr:dec_is_eq} that $T$ can be
identified with $T'$ (the details are left to the reader).

Let $t\in\wt U$ be an arbitrary tensor. We call the set of all
decomposable automorphisms of $\wt U$ that preserve
$t$ the {\em (full) isotropy group} of $t$, and denote it by
$\Gamma(t)$:
$$ \Gamma(t)=\{g\in S(U_1\,,\ldots,U_m)\mid g(t)=t\}.$$
We also consider the {\em small} isotropy group
$$ \Gamma^0(t)=\Gamma(t)\cap S^0(U_1\,,\ldots,U_m).$$
Clearly, $\Gamma^0(t)\normaleq \Gamma(t)$ and
$\Gamma(t)/\Gamma^0(t)$ may be identified with a
subgroup of the above-mentioned~$T$ (that is, with a
certain group of permutations of factors of dimension $>1$,
preserving dimensions).

Finally, let $\cA=\{t_1\,,\ldots,t_r\}$ be an algorithm
computing~$t$. Then
$$ \Aut(\cA)=\{g\in S(U_1\,,\ldots,U_m)\mid g(\cA)=\cA\}$$
will be called the {\em automorphism group} of~$\cA$.
Obviously, $\Aut(\cA)$ preserves $t_1+\ldots+t_r=t$, whence
$$ \Aut(\cA)\leq \Gamma(t). $$

If $u\in\wt U$, $v\in\wt V$, and $\fe:\wt U\to\wt V$
is a decomposable isomorphism such that $\fe(u)=v$, then
$\rk(u)=\rk(v)$. Moreover, if $\cA$ is an algorithm of length
$l$, computing $u$, then $\cB=\fe(\cA)$ is an algorithm
of length $l$, computing~$v$ (and conversely, if $\cB$ is
an algorithm of length $l$ for $v$, then $\cA=\fe^{-1}(\cB)$
is an algorithm of length $l$ for~$\cA$).
Therefore, $\fe$ bijectively maps the set
of all optimal algorithms computing $u$ to the set of all
optimal algorithms computing~$v$.

In particular, we see that $\Gamma(t)$ acts on the set of all
optimal algorithms computing~$t$. Obviously, the stabilizer
of a point (i.e., of an algorithm) with respect to this action
is the automorphism group of a given algorithm.

{\bf Example.} Let $U_1=U_2=U_3=M_{22}(K)$ and $t=\lu2,2,2\pu\in
U_1\ot U_2\ot U_3$. In this case it was shown by de Groote
\cite{deGroote2} that $\Gamma(t)$, and even $\Gamma^0(t)$,
acts on the set of all optimal algorithms transitively, so this
set is an orbit. The stabilizer (in the full $\Gamma(t)$)
of a point in this orbit is nothing else but $\Aut(\cS)$, the
automorphism group of the Strassen algorithm, which is
isomorphic to $S_3\times S_3$ by~\cite{Bur2014}.

{\bf Remark.} It is natural to consider more general
situation,
when studying decompositions of tensors, as it is done
in~\cite{deGroote1}. Let $\cR(\wt U)$ be the set of all
decomposable tensors in $\wt U=U_1\ot\ldots\ot U_m$. A
linear map $\fe:\wt U\to\wt V$ is called a {\em Segre
homomorphism}, if $\fe(\cR(\wt U))\sse\cR(\wt V)$. Next, let
$t\in\wt U$, let $\fe:\wt U\to\wt U$ be a Segre
endomorphism
such that $\fe(t)=t$, and let $\cA$ be an optimal algorithm
computing~$t$. Then $\fe(\cA)$ is also an optimal algorithm
computing~$t$. So the semigroup of all Segre
endomorphisms, preserving $t$, acts on the set of all
optimal algorithms computing~$t$. Thus one may think that
studying of general Segre endomorphisms may be useful in
algorithm analysis. However, it was, in fact, shown in \cite{deGroote1}
that such studying completely reduces to consideration of
decomposable automorphisms.

\section{The isotropy group of $\lu m,n,p\pu$}
	\label{sec:mnp}
Let $\cA=\{t_1,\,\ldots,t_r\}$ be an algorithm computing the
tensor $t=\lu m,n,p\pu$. It was observed above that its
automorphism group $\Aut(\cA)$ is contained in the (full)
isotropy group $\Gamma(t)$:
$$ \Aut(\cA)\leq\Gamma(t).$$
So, before studying group $\Aut(\cA)$ for any particular
algorithm $\cA$, it is natural to find the group~$\Gamma(t)$.
This is the aim of the present section.

In the case $m=n=p$ the group $\Gamma(t)$ was already
found, independently by Brockett-Dobkin, Strassen, and
de~Groote. Actually, Strassen and de Groote found
$\Gamma(t)$ in the case where $t$ is the structure tensor of
a finite-dimensional simple $K$-algebra, that is, a matrix
algebra over a skew field. See the comments after
Theorem~3.3 of~\cite{deGroote1}.

\subsection{Some known facts}
	\label{subs:mnp_prel}
We begin with several standard statements.

Let $V=K^l$ be the space of columns of height $l$ ($l\in\bbN$)
with elements in $K$, let $V'$ be the space of rows of the same
length $l$, and $(e_1\,,\ldots,e_l)$ and $(e^1\,,\ldots,e^l)$ be
the usual bases of $V$ and $V'$ (i.e., $e_i$ is the column whose
$i$-th element is $1$, the others are equal to~$0$). Note that the
rule $(v,v')\mapsto v'v$, where $v\in V$ and $v'\in V'$, defines
nondegenerate bilinear map $V\times V'\lra K$ (pairing),
and the bases $(e_i)$ and $(e^i)$ are dual with respect to
this pairing (note that $v'v$ is an $1\times1$ matrix, i.e., an
element of~$K$). Thus, we may identify $V'$ with $V^\ast$,
the dual space of~$V$.

Denote $G=GL_l(K)$. Then $G$ acts on $V$ on the left as
usually: $(g,v)\mapsto gv$, where $gv$ is the usual product
of a matrix and a column. Also, there is a left action of $G$
on $V'$ by the rule
$$ (g,v')\mapsto g\circ v':=v'g^{-1}.$$
(This is a left action indeed, that is, $(gh)\circ v'=g\circ
(h\circ v')$ for all $g,h\in G$ and $v'\in V'$. Indeed,
$g\circ(h\circ v')=g\circ(v'h^{-1})=(v'h^{-1})g^{-1}=v'h^{-1}
g^{-1}=v'(gh)^{-1}=(gh)\circ v'$.)  So there is a left action of
$G$ on $V\ot V'=V\ot V^\ast$ such that
$$ g(v\ot v')=gv\ot v'g^{-1}\qquad \forall g\in G,\ v\in V,
v'\in V'. $$

Consider the tensor
$$ \delta=\sum_{i=1}^le_i\ot e^i =\sum_{1\leq i,j\leq n}
\delta_{ij}e_i\ot e^j\in V\ot V^\ast $$
(so-called identity tensor, i.e., the tensor, associated with the
identity linear map of~$V$. Observe that the coefficients
of the tensor $\delta$ are precisely $\delta_{ij}$, where
$\delta_{ij}$ is the Kronecker symbol. This justifies using the
same symbol $\delta$ both for identity tensor and the
Kronecker symbol).

The following lemma is well-known (even trivial; cf. Remark~3 
in Subsection~\ref{subs:intro_rem}).

\begin{lemma}   \label{l:delta}
We have $g\delta=\delta$, for all $g\in G$.
\end{lemma}
{\em Proof.} Let $a_{ij}$ and $b_{ij}$ be the coefficients of the
matrices $g$ and $g^{-1}$, i.e.,
$$ g=\sum_{1\leq i,j\leq l} a_{ij}e_{ij}\quad {\rm and}\quad
g^{-1}=\sum_{1\leq i,j\leq l} b_{ij}e_{ij}.$$
Then $ge_i=\sum_{j=1}^l a_{ji}e_j$ and $e^ig^{-1}=
\sum_{j=1}^l b_{ij}e^j$. Hence
\begin{eqnarray*}
g\delta &=& g(\sum_{i=1}^le_i\ot e^i)= \sum_{i=1}^l
ge_i\ot e^ig^{-1}=\sum_{i=1}^l(\sum_{j=1}^l a_{ji}e_j)
\ot(\sum_{k=1}^lb_{ik}e^k) \\
&=& \sum_{1\leq j,k\leq l}(\sum_{i=1}^l a_{ji}b_{ik})e_j
\ot e^k=\sum_{1\leq j,k\leq l}(\delta_{jk})e_j\ot e^k=
\sum_{j=1}^le_j\ot e^j=\delta,
\end{eqnarray*}
as $\sum_{i=1}^l a_{ji}b_{ik}=\delta_{jk}$ for all $1\leq j,k
\leq l$ (because matrices $g$ and $g^{-1}$ are inverse).
\eoproof

\paragraph{Tensor products of group representations.}
Recall the notion of tensor product of representations
of a group (see \cite{Kostr}, \S VIII.7, or \cite{CR},
\S~12). Let $F$ be a field, $G$ be a group, and let
$U$ and $V$ be $FG$-modules, that is, $F$-spaces
endowed with $F$-linear action of~$G$. Let
$$ T:G\lra GL(U)\quad \text{and}\quad R:G\lra GL(V)$$
be the corresponding representations of $G$ on $U$
and~$V$. Put $W=U\ot_FV$. For an element $g\in G$
define linear map $S(g):W\lra W$ by $S(g)=T(g)\ot R(g)$.
Since both $T(g)$ and $R(g)$ are invertible, $S(g)$ is
invertible also. Moreover,
\begin{eqnarray*} S(g_1g_2) &=& T(g_1g_2)\ot R(g_1g_2)
= T(g_1)T(g_2)\ot R(g_1)R(g_2) \\
&=& (T(g_1)\ot R(g_1))(T(g_2)\ot R(g_2))=S(g_1)S(g_2). 
\end{eqnarray*}
That is,
$$S:g\mapsto S(g)\,,\qquad G\lra GL(W)$$
is a representation of~$G$. It is called the {\em tensor
product} of representations $T$ and~$R$ (and $W$ a
tensor product of $FG$-modules $U$ and~$V$).

If $U_1$, $U_2$, $V_1$, and $V_2$ are $FG$-modules,
and $\alpha:U_1\lra U_2$ and $\beta:V_1\lra V_2$
are $FG$-homomorphisms, then (as is easy to check)
the map $\gamma=\alpha\ot\beta:U_1\ot V_1\lra
U_2\ot V_2$ is a $FG$-homomorphism also.

\subsection{A subgroup of $\Gamma^0(t)$}
	\label{subs:Gamma0}
Let $D=K^m$, $E=K^n$, and $F=K^p$ be the spaces of
columns over $K$ of height $m$, $n$, and $p$, respectively,
and $D'$, $E'$, and $F'$ be the spaces of rows of the same
length. By $d_i$, $e_j$, $f_k$, $d^i$, $e^j$, and $f^k$ we
denote the elements of the usual bases of $D,\ldots,F'$.
As observed earlier, we may identify $D'$, $E'$ and $F'$
with $D^\ast$, $E^\ast$, and $F^\ast$, respectively.

Note that for any $d\in D$ and $e'\in E'$ their product $de'$
is an $m\times n$ matrix. In particular, $d_ie^j=e_{ij}$ are
matrix units, for all $1\leq i\leq m$, $1\leq j\leq n$.
Similarly for $ef'$ and $fd'$.

Denote $M_{mn}=M_{mn}(K)$, $M_{np}$ and $M_{pm}$ by
$L_1$, $L_2$, and $L_3$, respectively, and let
$$ L=L_1\ot L_2\ot L_3 =M_{mn}\ot M_{np}\ot M_{pm}\,. $$
Also denote
$$ N=D\ot D'\ot E\ot E'\ot F\ot F'\,. $$
Next, define the linear map $\tau:N\to L$ by the rule
$$ \tau: d\ot d'\ot e\ot e'\ot f\ot f'\mapsto de'\ot ef'\ot fd'. $$
This map is well-defined indeed, because $de'$, $ef'$, and
$fd'$ are in $L_1$, $L_2$, and $L_3$, respectively, and,
moreover, the expression $de'\ot ef'\ot fd'$ is linear in each
of the arguments $d,d',\ldots,f'$.

It is easy to see that $\tau$ is an isomorphism of vector
spaces.

Let $\delta_D=\sum_{i=1}^m d_i\ot d^i$, $\delta_E$ and
$\delta_F$ be the identity tensors of the spaces $D$, $E$,
and $F$. Consider the tensors $\delta_D\ot\delta_E\ot
\delta_F\in N$ and $\tau(\delta_D\ot\delta_E\ot\delta_F)
\in L$.
\begin{lemma}   \label{l:tau3delta}
The equality $\tau(\delta_D\ot\delta_E\ot\delta_F)
=\lu m,n,p\pu$ holds.
\end{lemma}
{\em Proof.} We have
\begin{eqnarray*}
\tau(\delta_D\ot\delta_E\ot\delta_F) &=& \tau\left(
(\sum_{i=1}^m d_i\ot d^i)\ot(\sum_{j=1}^n e_j\ot e^j)
\ot (\sum_{k=1}^p f_k\ot f^k) \right) \\
&=& \tau(\sum_{\substack{1\leq i\leq m \\ 1\leq j
\leq n \\ 1\leq k\leq p}} d_i\ot d^i\ot e_j\ot e^j\ot f_k\ot f^k)
=\sum_{\substack{1\leq i\leq m \\ 1\leq j\leq n \\ 1\leq k
\leq p}} d_ie^j\ot e_jf^k\ot f_kd^i \\
&=& \sum_{1\leq i\leq m, \  1\leq j\leq n, \ 1\leq k\leq p}
e_{ij}\ot e_{jk}\ot e_{ki} =\lu m,n,p\pu.
\end{eqnarray*}
\hfill $\square$ \vspace{3ex}

Next we consider some group actions. Put $G_D=GL_m(K)$,
$G_E=GL_n(K)$, $G_F=GL_p(K)$. Then $G_D$ acts on $D$
and $D'$, $G_E$ --- on $E$ and $E'$, and $G_F$ acts on $F$
and~$F'$.

Form the direct product $G=G_D\times G_E\times
G_F$ and define actions of $G$ on $D,\ldots,F'$. For
example, if $g=(g_1,g_2,g_3)\in G_D\times G_E\times G_F$,
$d\in D$ and $d'\in D'$, then we define $g(d)=g_1d$ and
$g(d')=d'g_1^{-1}$ (here $g(d)$ and $g(d')$ mean the
result of group action, and $g_1d$ and $d'g_1^{-1}$ mean the
products of matrices). That is, the factor $G_D$ of $G$
acts on $D$ and $D'$ as usually, whereas $G_E$ and $G_F$
act trivially. The actions on $E$, $E'$, $F$, $F'$ are defined
similarly.

 Now we can consider $N=D\ot\ldots\ot F'$ as a
$G$-module.

Further, it is easy to see that the rules
$$ (g_1,g_2,g_3)x=g_1xg_2^{-1}, \ g_2xg_3^{-1}\,,
\quad \text{or}\quad  g_3xg_1^{-1} $$
define actions of $G$ on $L_1$, $L_2$, and $L_3$,
respectively. Hence we can define action of $G$ on $L=L_1
\ot L_2\ot L_3$.

In the proof of the following lemma (and later) we use
the following simple \\
{\bf Observation.} Let $K$ be a field, $G$ a group,
$X$ and $Y$ be $KG$-modules, and let $\fe:X\lra Y$
be a $K$-linear map. If $\fe$ is a $KG$-module homomorphism
and a $K$-spaces isomorphism, then $\fe$ is a
$KG$-module isomorphism (that is, the inverse map
$\fe^{-1}:Y\lra X$ is a homomorphism of $KG$-modules).
\begin{lemma}   \label{l:tau_Ghom}
The map $\tau:N\to L$, defined above, is a $G$-module
isomorphism.
\end{lemma}
 {\em Proof. } As $\tau$ is an isomorphism of vector spaces,
it remains to check that $\tau$ is a $G$-module
homomorphism, that is, $\tau(g(x))=g(\tau(x))$ for all
$g\in G$ and $x\in N$.

It is sufficient to consider  $x=d\ot d'\ot e\ot e'\ot f\ot f'$.
Let $g=(g_1,g_2,g_3)$. Then
\begin{eqnarray*}
\tau(g(x)) &=& \tau((g_1,g_2,g_3) (d\ot d'\ot e\ot e'\ot f
\ot f')) \\
&=& \tau(g_1d\ot d'g_1^{-1} \ot g_2e\ot e'g_2^{-1}
\ot g_3f\ot f'g_3^{-1}) \\
&=& (g_1d)(e'g_2^{-1})\ot(g_2e)(f'g_3^{-1})\ot(g_3f)
(d'g_1^{-1}) \\
&=& g_1de'g_2^{-1}\ot g_2ef'g_3^{-1}\ot g_3fd'g_1^{-1}
=(g_1,g_2,g_3)(de'\ot ef'\ot fd') \\
&=& g(\tau(x)).
\end{eqnarray*}

(Note that there is an alternative way to prove that $\tau$
is a $G$-homomorphism. Namely, observe that the formula
$x\ot y\mapsto xy$ defines $G$-homomorphisms
$\alpha:D\ot E'\lra L_1$, $\beta:E\ot F'\lra L_2$ and $\gamma:
F\ot D'\lra L_3$. So their product
$$ \alpha\ot\beta\ot\gamma: D\ot E'\ot E\ot F'\ot F\ot D' \lra
L_1\ot L_2\ot L_3 $$
is a $G$-homomorphism also. Also, the ``permutation map''
$$\zeta: D\ot D'\ot E\ot E'\ot F\ot F' \lra  D\ot E'\ot E\ot F'
\ot F\ot D' $$
is obviously a $G$-homomorphism. Now it remains to
observe that $\tau$ coincides with $(\alpha
\ot\beta\ot\gamma)\circ\zeta$.
\eoproof

The proof of the following simple lemma is left to
the reader.
\begin{lemma}   \label{l:one-sided}
Let $a\in GL_m(K)$, $b\in GL_n(K)$, and $x\in M_{mn}(K)$.
If either $ax=0$, or $xb=0$, or $axb=0$, then $x=0$.
So the map $y\mapsto ayb$ on $M_{mn}(K)$ is invertible.
\end{lemma}

\begin{prop}    \label{pr:H_constr}
Let $L=L_1\ot L_2\ot L_3$ be as above. For $a\in GL_m(K)$,
$b\in GL_n(K)$, and $c\in GL_p(K)$ let $T(a,b,c):L
\lra L$ be the linear map defined by
$$ T(a,b,c)(x\ot y\ot z)= axb^{-1}\ot byc^{-1}\ot cza^{-1}. $$
Then
$$ H=\{ T(a,b,c)\mid (a,b,c)\in GL_m(K)\times GL_n(K)\times
GL_p(K) \} $$
is a subgroup of $\Gamma^0(t)$.
\end{prop}
{\em Proof.} It follows from Lemma~\ref{l:one-sided}
that $T(a,b,c)$ is an automorphism of $L$. It is also easy
to see that $T(a,b,c)T(a_1,b_1,c_1)=T(aa_1,bb_1,cc_1)$
for any $a,\ldots,c_1$. Therefore $H$ is a subgroup
of $S^0(L_1,L_2,L_3)$.

Let $D$, $E$, $F$, $\ldots$ be as above. The group $G=G_D
\times G_E\times G_F=GL_m(K)\times GL_n(K)\times
GL_p(K)$ preserves $\delta_D\in D\ot D'$, when
acting on $D\ot D'$, according to Lemma~\ref{l:delta}.
Similarly
$G$ preserves $\delta_E$ and $\delta_F$, and so preserves
$\delta_D\ot\delta_E\ot\delta_F$. As $\tau:N\lra L$ is a
$G$-homomorphism and $t=\tau(\delta_D\ot\delta_E\ot
\delta_F)$, we see that $G$ preserves~$t$.

It remains to note that the image of $(a,b,c)\in G_D\times
G_E\times G_F$ in $GL(L)$ coincides with $T(a,b,c)$. So
$T(a,b,c)$ preserves $t$ for any $a$, $b$, and $c$, whence
$H\leq\Gamma^0(t)$.
\eoproof

\subsection{Structure of the full $\Gamma(t)$}
	\label{subs:full_Gamma}
The full isotropy group $\Gamma(t)$, where $t=\lu m,n,p\pu$,
may be larger than $\Gamma^0(t)$. However, the relations
between $\Gamma(t)$ and $\Gamma^0(t)$ can be easily
described. This is the aim of the present subsection.

{\em In the rest of this section we assume that at least one
of the numbers $m$, $n$, and $p$ is different from~1.}
If this is the case, then at most one of the three spaces
$L_1$, $L_2$, and $L_3$ is one-dimensional. So for any
decomposable automorphism $\fe\in S(L_1,L_2,L_3)$ the
permutation of $\{L_1,L_2,L_3\}$, corresponding to $\fe$,
is uniquely determined, by Proposition~\ref{pr:dec_is_eq}.

First of all, we construct some elements of $\Gamma(t)$, not
belonging to $\Gamma^0(t)$. For a permutation $g$ of
$\{L_1,L_2\,, L_3\}$ we define certain decomposable
automorphism $\rho_g:L\lra L$ (however, $\rho_g$ will be
defined not for any triple $(m,n,p)$).

Suppose that $m=n$. Define $\rho_{(23)}:L\lra L$ by the formula
$\rho_{(23)}(x\ot y\ot z)=x^t\ot z^t\ot y^t$.
(Note that we use the same symbol $t$ for the tensor
$t=\lu m,n,p\pu$ and the transpose map, but we hope
this will not lead to a confusion). Note that
$\rho_{(23)}$ is well-defined indeed, because the formula
$x\mapsto x^t$ defines an isomorphism of the space $L_2=
M_{np}=M_{mp}$ onto $L_3=M_{pm}$, and also this formula
defines an isomorphism
of $L_3$ onto $L_2$, and an automorphism of~$L_1$.
Observe next that $\rho_{(23)}^2=1$($=\id_L$), as
$\rho^2_{(23)}
(x\ot y\ot z)=\rho_{(23)}(\rho_{(23)}(x\ot y\ot z))=\rho_{(23)}
(x^t\ot z^t\ot y^t)=(x^t)^t\ot (y^t)^t\ot (z^t)^t
=x\ot y\ot z$. Finally, we have $\rho_{(23)}\in\Gamma(t)$,
because
$$\rho_{(23)}(t)=\rho_{(23)}(\sum_{\substack{ 1\leq i,j
\leq m \\ 1\leq k\leq p}} e_{ij}\ot e_{jk}\ot e_{ki}) =
\sum_{\substack{ 1\leq i,j\leq m \\ 1\leq k\leq p}} e_{ji}
\ot e_{ik}\ot e_{kj}=t$$
(note that if we change names of indices in the latter sum by
$i\lra j$, $j\lra i$, $k\lra k$, then we obtain the sum for~$t$).

Similarly, if $m=p$ or $n=p$, then we may define
$\rho_{(12)}$ and $\rho_{(13)}$ by formulae 
\begin{eqnarray*}
\rho_{(12)}(x\ot y\ot z) &=& y^t\ot x^t\ot z^t\,,\qquad 
\text{and} \\ 
\rho_{(13)}(x\ot y\ot z) &=& z^t\ot y^t\ot x^t\,, 
\end{eqnarray*}
respectively.

Next suppose that $m=n=p$. Then we define $\rho_{(123)}$
and $\rho_{(132)}$ by the formulae 
$$\rho_{(123)}(x\ot y\ot z)=z\ot x\ot y,$$ 
resp.  
$$\rho_{(132)}(x\ot y\ot z)=y\ot z\ot x.$$
Clearly,  $\rho_{(123)}^2=\rho_{(132)}$ and
$\rho_{(123)}^3=1$. Also, it is easy to see that $\rho_{(123)}
\in\Gamma(t)$.

Finally, for any triple $m$, $n$, $p$ define $\rho_e=\id_L$.

Observe that the permutation of the factors $L_1$, $L_2$,
and $L_3$, corresponding to $\rho_g$, is precisely~$g$.
Hence $\rho_g\ne1$ if $g\ne1$, and also $\rho_g\ne
\rho_h$, if $g\ne h$.

Let $Q$ be the set of all $\rho_g$, that are defined, for given
$m$, $n$, and~$p$. Thus,
$$ Q=\begin{cases}
\{\rho_e=1\}, & \text{if $m\ne n\ne p\ne m$,} \\
\{1,\rho_{(23)}\}, &\text{if $m=n\ne p$,} \\
\{1,\rho_{(12)}\}, &\text{if $m=p\ne n$,} \\
\{1,\rho_{(13)}\}, &\text{if $n=p\ne m$,} \\
\{1,\rho_{(12)}, \rho_{(13)}, \rho_{(23)}, \rho_{(123)},
\rho_{(132)} \}, &\text{if $m=n=p$.}
\end{cases} $$

\begin{lemma}   \label{l:Q}
For any $m$, $n$, and $p$ the set $Q$ is a subgroup
of~$\Gamma(t)$. Let $R\leq S_3$ be the group of all
permutations of $\{L_1,L_2,L_3\}$, preserving dimensions.
Then the rule $g\leftrightarrow \rho_g$ defines isomorphism
$R\leftrightarrow Q$. Thus, $Q\cong S_3$, $Z_2$, or $1$,
when $|\{m,n,p\}|=1$, $2$, or $3$, respectively.
\end{lemma}
{\em Proof.} First suppose that $m$, $n$, and $p$ are
pairwise distinct. Then the numbers $\dim L_1=mn$,
$\dim L_2=np$, and $\dim L_3=pm$ are pairwise distinct
also, whence $R=1$. Thus, in this case both $Q$ and $R$
are trivial groups, and the statement is trivial too.

Further suppose that $|\{m,n,p\}|=2$. We consider, as an
example, the case $m=n\ne p$ only. In this case
$Q=\{ \rho_e=1, \rho_{(23)}\}$. Moreover, $\dim L_1\ne
\dim L_2=\dim L_3$, whence $R=\{e, (23)\}$. As $\rho_{(23)
}^2=1$ and $\rho_{(23)}\ne1$, we see that $Q$ is a group
isomorphic to $Z_2$, and that the bijection $e\leftrightarrow
\rho_e=1$, $(23)\leftrightarrow\rho_{(23)}$ is an isomorphism
between $R$ and~$Q$.

Finally consider the case $m=n=p$. In this case $mn=np
=pm$, whence $R\cong S_3$ consists of all permutations
of $\{L_1\,, L_2\,,L_3\}$. Moreover, all $\rho_g$ are pairwise
distinct, and any $\rho_g$ is an automorphism
of~$L$. So it is sufficient to prove that $\rho_g\rho_h=
\rho_{gh}$ for each pair of $g,h\in S_3$.

It is not hard to check the latter equality in all cases directly.
If $g=e$ or $h=e$, then this equality is trivial, as $\rho_e=1$.
It remains to check this equality for 25 pairs $(g,h)$ with
$g,h\ne e$, which is not too many.

There is, however, a shorter argument. For $g\in S_3$ let $\pi_g$
be the usual permutation of factors, for example $\pi_{(13)}(x\ot
y\ot z)=z\ot y\ot x$. Then $\pi_g\pi_h= \pi_{gh}$ for any $g$
and~$h$. Next, let $\tau:L\lra L$ be the componentwise transpose
map, i.e., $\tau(x\ot y\ot z) =x^t\ot y^t\ot z^t$. Note that
$\tau$ commutes with all $\pi_g$, and also that $\tau^2=1$. Next,
note that $\rho_g=\pi_g$ if $g$ is even, and $\rho_g=\tau\pi_g$ if
$g$ is odd. In other words, $\rho_g=\pi_g\tau^{\eps(g)}$, where
$\eps:S_3\lra Z_2=\{0,1\}$ is the parity homomorphism. Now for any
$g$ and $h$ we have
$\rho_g\rho_h=\pi_g\tau^{\eps(g)}\pi_h\tau^{\eps(h)}=
\pi_g\pi_h\tau^{\eps(g)}\tau^{\eps(h)}=\pi_{gh}
\tau^{\eps(g)+\eps(h)}=\pi_{gh}\tau^{\eps(gh)}=\rho_{gh}$, as
required. \eoproof
\vspace{2ex}

To state the next proposition it is convenient to use the
notion of semidirect product.

Recall that a group $G$ is the {\em product} of its subgroups $A$
and $B$, which is denoted by $G=AB$, if for each $g\in G$ there
exist $a\in A$ and $b\in B$ such that $g=ab$. If in addition
$A\cap B=1$, then it is easy to see that the representation of $g$
in the form $g=ab$ is unique.

A group $G$ is said to be a {\em semidirect product} of
$A$ by $B$, which is denoted by $G=A\sdir B$, if $G=AB$,
$A$ is normal in $G$, and $A\cap B=1$.
\begin{prop}    \label{pr:Gamma_sdir_Q}
Let $t=\lu m,n,p\pu$, and let $Q\leq\Gamma(t)$ be the
subgroup described above. Then $\Gamma(t)=\Gamma^0(t)
\sdir Q$.
\end{prop}
{\em Proof.} We know that $\Gamma^0(t)\normaleq
\Gamma(t)$. Further, $Q\cap\Gamma^0(t)=1$, because a
nontrivial element of $Q$ corresponds to a nontrivial
permutation of $L_1$, $L_2$, $L_3$. It remains to show
that $\Gamma(t)=\Gamma^0(t)Q$. Let $x\in\Gamma(t)$,
and let $g$ be the permutation of $L_1$, $L_2$, $L_3$,
corresponding to~$x$. Then $g$ preserves the dimensions of
factors, and therefore $\rho_g$ is well-defined and
$\rho_g\in Q$. Since the permutation of factors,
corresponding to $\rho_g$, is $g$, it follows that the
permutation of factors, corresponding to the element
$x'=x\rho_g^{-1}$, is trivial, that is, $x'\in\Gamma^0(t)$.
Thus, we have $x=x'\rho_g$, where $x'\in\Gamma^0(t)$
and $\rho_g\in Q$. Hence $\Gamma(t)=\Gamma^0(t)Q$.
\eoproof
\begin{prop}    \label{pr:Gamma0_H}
Let $T(a,b,c)$ and $H$ be the transformations and the
group introduced in Proposition~\ref{pr:H_constr}.
Then any element $g\in\Gamma^0(t)$ has the form
$g=T(a,b,c)$, for some
$a$, $b$, and $c$. Therefore, $\Gamma^0(t)=H$.
\end{prop}

This proposition will be proved later in this section.

Next we describe the group $\Gamma^0(t)$ as an abstract
group.

Recall that the {\em projective general linear group}
$PGL_n(K)$ is the quotient group
$$ PGL_n(K)=GL_n(K)/Z_n(K),$$
where $Z_n(K)=\{\lambda E_n\mid\lambda\in K^\ast\}$
is the subgroup of all nonzero scalar matrices.

(For reader's information we recall the following
standard facts on linear groups; they can be found
in many textbooks, see for instance \cite{Dieud}
(\S\S I.1, I.2, II.1, II.2), \cite{KargMer} (\S 13),
and \cite{Suzuki}(\S I.9).

The group $Z_n(K)$ is the center of $GL_n(K)$.
The group $PGL_n(K)$ contains the projective special linear
group $PSL_n(K)=SL_n(K)/(Z_n(K)\cap SL_n(K))$ as a normal
subgroup. The latter group is simple, except for the two
cases $(n,K)=(2,\bbF_2)$, $(2,\bbF_3)$. The quotient
$PGL_n(K)/PSL_n(K)$ is trivial, if $K$ is algebraically closed,
and is a finite cyclic group if $K$ is finite.)

Let $\fe:GL_m(K)\times GL_n(K)\times GL_p(K)\lra
\Gamma^0(t)$ be the map defined by $\fe((a,b,c))=T(a,b,c)$.
It was observed in the proof of Proposition~\ref{pr:H_constr}
that $\fe$ is a group homomorphism.
Proposition~\ref{pr:Gamma0_H} shows that
$\fe$ is surjective. Therefore, in order to describe its image
$\im\fe=H=\Gamma^0(t)$ (as an abstract group, i.e., up
to isomorphism), it is sufficient to know $\Ker\fe$, its kernel.

We need a lemma.
\begin{lemma}   \label{l:scalar2}
Suppose $A\in GL_m(K)$ and $B\in GL_n(K)$ be matrices
such that $AxB$ is proportional to $x$ for all matrices
$x\in M_{mn}(K)$. Then $A$ and $B$ are scalar matrices.
\end{lemma}
{\em Proof.} Consider the map $\alpha:x\mapsto AxB$ of
$M_{mn}(K)$ to itself. It follows from
Lemma~\ref{l:one-sided} that $\alpha$ is an isomorphism.
So $\alpha$ is a scalar map by Lemma~\ref{l:scalar} :
$AxB=cx$, for
a fixed $c\in K^\ast$. Next, let $a_{ij}$ and $b_{ij}$ be the
coefficients of $A$ and $B$, respectively:
$$ A=\sum_{i,j=1}^m a_{ij}e_{ij}\,,\quad
B=\sum_{i,j=1}^n b_{ij}e_{ij}\,. $$
Then for all $p$ and $q$ such that $1\leq p\leq m$ and
$1\leq q\leq n$
we have
$$ ce_{pq}=Ae_{pq}B=\sum_{1\leq i\leq m,\ 1\leq j\leq n}
a_{ip}b_{qj}e_{ij}\,, $$
whence $a_{pp}b_{qq}=c$, and $a_{ip}b_{qj}=0$ if $i\ne p$
or $j\ne q$. The former of these relations implies that
$a_{pp}, b_{qq}\ne0$ for all $p$ and $q$. Taking $i=p$,
$j\ne q$ in the second relation, we obtain $a_{pp}b_{qj}=0$,
whence $b_{qj}=0$. Similarly, we obtain $a_{ip}=0$ for
$i\ne p$. Thus, $A$ and $B$ are diagonal matrices. Next,
for any $q$ we have $a_{11}b_{qq}=c$, whence $b_{qq}=
c/a_{11}$, so $B$ is a scalar matrix. Similarly, $A$ is
a scalar matrix also.
\eoproof

\begin{prop}    \label{pr:G0_abst}
The kernel $\Ker\fe$ coincides with $Z_m(K)\times Z_n(K)
\times Z_p(K)$, and therefore the group $\Gamma^0(t)=H$
is isomorphic to $PGL_m(K)\times PGL_n(K)\times PGL_p(K)$.
\end{prop}
{\em Proof.} Let $(a,b,c)=(\lambda E_m,\mu E_n,\nu E_p)$,
where $\lambda,\mu,\nu\in K^\ast$, be an element of
$N=Z_m(K)\times Z_n(K)\times Z_p(K)$. Then for any
$x\in L_1$, $y\in L_2$ and $z\in L_3$ we have
$$ T(a,b,c)(x\ot y\ot z)=\lambda x\mu^{-1}\ot \mu y
\nu^{-1} \ot \nu z\lambda^{-1}=x\ot y\ot z, $$
so $T(a,b,c)=1$. Hence $N\leq\Ker\fe$.

Conversely, suppose that $(a,b,c)\in\Ker\fe$. Then $T(a,b,c)
(x\ot y\ot z)=x\ot y\ot z$ for all $x$, $y$, $z$, that is,
$axb^{-1}\ot byc^{-1}\ot cza^{-1}=x\ot y\ot z$. Hence
$axb^{-1}$ is proportional to $x$ by Lemma~\ref{l:prod_subsp}.
So both $a$ and $b$ are scalar matrices by
Lemma~\ref{l:scalar2}. The
matrix $c$ is scalar also by a similar argument, whence
$(a,b,c)\in N$. Therefore $\Ker\fe\leq N$.
\eoproof
\vspace{2ex}

It may be useful to have explicit formulae for conjugation
of an element of $H$ by an element of $Q$ (however,
we will not use these formulae in the present work).

For a matrix $x\in GL_l(K)$ we denote by $x^\vee$ the matrix
$x^\vee=(x^t)^{-1}=(x^{-1})^t$ (which is usually called the matrix
{\em contragradient} to~$x$).
\begin{prop}    \label{pr:Q_act_H}
The following relations hold:
$$ \rho_eT(a,b,c)\rho_e^{-1}=T(a,b,c),$$
$$ \rho_{(12)}T(a,b,c)\rho_{(12)}^{-1} \ \text{($=\rho_{(12)}
T(a,b,c)\rho_{(12)}$) } =T(c^\vee, b^\vee,a^\vee), $$
$$ \rho_{(13)}T(a,b,c)\rho_{(13)}=T(a^\vee, c^\vee,b^\vee), $$
$$ \rho_{(23)}T(a,b,c)\rho_{(23)}=T(b^\vee, a^\vee,c^\vee), $$
$$ \rho_{(123)}T(a,b,c)\rho_{(123)}^{-1}=T(c,a,b), $$
$$ \rho_{(132)}T(a,b,c)\rho_{(132)}^{-1}=T(b,c,a). $$
\end{prop}
{\em Proof.} The first relation is trivial, because $\rho_e=1$.
Prove the next relation, as an example. Note that
$\rho_{(12)}^{-1}=\rho_{(12)}$, as $\rho_{(12)}^2=1$.

For $x\in L_1$, $y\in L_2$, and $z\in L_3$ we have
$\rho_{(12)}(x\ot y\ot z)=y^t\ot x^t\ot z^t$, whence
\begin{eqnarray*} 
x\ot y\ot z  && \stackrel{\rho_{(12)}}\mapsto
y^t\ot x^t\ot z^t \stackrel{T(a,b,c)}\mapsto ay^tb^{-1}
\ot bx^tc^{-1}\ot cz^ta^{-1} \\ 
&& \stackrel{\rho_{(12)}}\mapsto
(bx^tc^{-1})^t \ot (ay^tb^{-1})^t \ot (cz^ta^{-1})^t
= (c^{-1})^txb^t\ot(b^{-1})^tya^t\ot (a^{-1})^tzc^t \\
&& =c^\vee x(b^\vee)^{-1} \ot b^\vee y(a^\vee)^{-1} \ot
a^\vee z(c^\vee)^{-1} =T(c^\vee, b^\vee,a^\vee)
(x\ot y\ot z), 
\end{eqnarray*}
whence
$$\rho_{(12)}T(a,b,c)\rho_{(12)}=T(c^\vee, b^\vee,a^\vee).$$
The other relations can be proved similarly.
\eoproof
\vspace{2ex}

We summarize the results of this section in the following
theorem. For the convenience of the future usage, we state
this theorem ``in full''.
\begin{theorem} \label{th:Gamma_t_full}
Let $m,n,p\in\bbN$, $(m,n,p)\ne(1,1,1)$,
let $L_1=M_{mn}=M_{mn}(K)$,
$L_2=M_{np}$, $L_3=M_{pm}$, let $L=L_1\ot L_2\ot L_3$,
and let
$$ t=\lu m,n,p\pu= \sum_{1\leq i\leq m, \ 1\leq j\leq n, \
1\leq k\leq p} e_{ij}\ot e_{jk}\ot e_{ki} \in L.$$

For elements $a\in GL_m(K)$, $b\in GL_n(K)$, $c\in GL_p(K)$
define transformation $T(a,b,c):L\lra L$ by the formula
$$T(a,b,c)(x\ot y\ot z)= axb^{-1}\ot byc^{-1}\ot cza^{-1}.$$
Put
$$ H=\{ T(a,b,c)\mid (a,b,c)\in GL_m(K)\times GL_n(K)\times
GL_p(K) \}. $$
Then $\Gamma^0(t)=H$. The transformations $T(a,b,c)$
and $T(a_1,b_1,c_1)$ coincide if and only if
$a_1=\lambda a$, $b_1=\mu b$, and $c_1=\nu c$, for some
$\lambda,\mu,\nu\in K^\ast$. The group $H$ is isomorphic
to $PGL_m(K)\times PGL_n(K)\times PGL_p(K)$.

For any element $g\in S_3$ and some $m$, $n$, $p$ define
transformation $\rho_g:L\lra L$ as follows:
\\ $\rho_e=1=\id_L$, for any $m$, $n$, $p$;
\\ $\rho_{(23)}(x\ot y\ot z)=x^t\ot z^t\ot y^t$, when $m=n$;
\\ $\rho_{(13)}(x\ot y\ot z)=z^t\ot y^t\ot x^t$, when $n=p$;
\\ $\rho_{(12)}(x\ot y\ot z)=y^t\ot x^t\ot z^t$, when $m=p$;
\\ $\rho_{(123)}(x\ot y\ot z)=z\ot x\ot y$ and $\rho_{(132)}
(x\ot y\ot z)=y\ot z\ot x$, when $m=n=p$.

Put
$$ Q=\{\rho_g\mid g\in S_3\,, \ \rho_g \ \text{is well-defined}
\}. $$
In other words,
$$ Q=\begin{cases}
\{\rho_e=1\}, & \text{if $m\ne n\ne p\ne m$,} \\
\{1,\rho_{(23)}\}, &\text{if $m=n\ne p$,} \\
\{1,\rho_{(12)}\}, &\text{if $m=p\ne n$,} \\
\{1,\rho_{(13)}\}, &\text{if $n=p\ne m$,} \\
\{1,\rho_{(12)}, \rho_{(13)}, \rho_{(23)}, \rho_{(123)},
\rho_{(132)} \}, &\text{if $m=n=p$.}
\end{cases} $$
Then $Q$ is a subgroup of $\Gamma(t)$, isomorphic to
$S_3$, $Z_2$, or $1$, when $|\{m,n,p\}|=1$, $2$, or $3$,
respectively.

The group $\Gamma(t)$ is a semidirect product of
$\Gamma^0(t)$ and $Q$:
$$ \Gamma(t)=\Gamma^0(t)\sdir Q. $$
The following relations, describing action of $Q$ on $H$
by conjugation, hold (in all cases where $\rho_g$ is
well-defined):
$$ \rho_eT(a,b,c)\rho_e^{-1}=T(a,b,c),$$
$$ \rho_{(12)}T(a,b,c)\rho_{(12)}^{-1} \ \text{($=\rho_{(12)}
T(a,b,c)\rho_{(12)}$) } =T(c^\vee, b^\vee,a^\vee), $$
$$ \rho_{(13)}T(a,b,c)\rho_{(13)}=T(a^\vee, c^\vee,b^\vee), $$
$$ \rho_{(23)}T(a,b,c)\rho_{(23)}=T(b^\vee, a^\vee,c^\vee), $$
$$ \rho_{(123)}T(a,b,c)\rho_{(123)}^{-1}=T(c,a,b), $$
$$ \rho_{(132)}T(a,b,c)\rho_{(132)}^{-1}=T(b,c,a). $$
\end{theorem}

\vspace{2ex}

This theorem is an immediate consequence of Propositions
\ref{pr:H_constr}, \ref{pr:Gamma0_H}, \ref{pr:Gamma_sdir_Q},
\ref{pr:G0_abst}, and~\ref{pr:Q_act_H}.

\vspace{2ex}

{\em The aim of the rest of this section is to prove
Proposition~\ref{pr:Gamma0_H}.}

\subsection{Structure tensors and contragradient maps}
	\label{subs:contragrad}

In this subsection we recall some well-known notions.

Let $V$ be a space and $V^\ast$ be its dual. For two
elements $v\in V$ and $l\in V^\ast$ it will be convenient
to denote $l(v)$ either by $\lu l,v\pu$ or by
$\lu v,l\pu$. Thus, the element $\lu u_1,u_2\pu$ is
defined, if one of elements $u_1$ and $u_2$ is in $V$,
the other is in $V^\ast$; and we always have
$\lu u_1,u_2\pu=\lu u_2,u_1\pu$. The symbol
$\lu u_1,u_2\pu$ is called the {\em pairing} of
$u_1$ and~$u_2$.

Let $f:X\lra Y$ be a linear map. The map $f^\ast:Y^\ast\lra
X^\ast$, taking an element $l\in Y^\ast$ to the element
$m=f^\ast(l)\in X^\ast$, defined by $m(x)=l(f(x))$, is
linear and is called the map, {\em dual} to~$f$.
Thus, $f^\ast$ is the unique map such that
$$\lu l,f(x)\pu=\lu f^\ast(l), x\pu,\qquad
\forall\ x\in X\,,\ l\in Y^\ast. $$

For any two maps $f:X\lra Y$ and $g:Y\lra Z$ the equality
$(gf)^\ast=f^\ast g^\ast$ holds.

Suppose $f:X\lra Y$ is an isomorphism. It is easy to see that
$f^\ast:Y^\ast\lra X^\ast$ is an isomorphism also. The
inverse isomorphism $(f^\ast)^{-1}:X^\ast\lra Y^\ast$
is called an isomorphism, {\em contragradient} to $f$, and
is denoted by $\check f$ or $f^\dag$. It will be
convenient for us to denote it by~$f^\vee$. This
isomorphism can be described as the unique isomorphism
$f^\vee:X^\ast\lra Y^\ast$ such that
$$ \lu f^\vee(l), f(x)\pu=\lu l,x\pu \quad \forall
\ x\in X,\ l\in X^\ast. $$

It is easy to see that $(gf)^\vee=g^\vee f^\vee $ for any
two isomorphisms $f:X\lra Y$ and $g:Y\lra Z$. Also,
$(f^{-1})^\vee=(f^\vee)^{-1}$. In particular, suppose that
$\fe:G\lra GL(X)$ is a representation of a group $G$ on
a space~$X$. Then the map $\fe^\ast:G\lra GL(X^\ast)$,
defined by $\fe^\ast(g)=\fe(g)^\vee$, is a representation
of $G$ also. It is called a representation {\em contragradient}
(or more often {\em dual}) to~$\fe$. (Thus, the usage of the
word ``dual'' for linear maps and for group representations
is somewhat different).

Finally note that taking contragradient map is involutive,
that is, $(f^\vee)^\vee=f$ for any isomorphism $f:X\lra Y$.
(Strictly speaking, $(f^\vee)^\vee$ is a map from
$(X^\ast)^\ast$ to $(Y^\ast)^\ast$, but we can identify
$(V^\ast)^\ast$ with $V$, because we consider only
finite-dimensional spaces.)

Let $X$, $Y$, $Z$ be spaces. By $\cL(X,Y)$ we denote the
space of all linear maps from $X$ to $Y$, and by
$\cL_2(X,Y;Z)$ the space of all bilinear maps $f:X\times Y
\lra Z$. The spaces $\cL(X,Y)$ and $\cL_2(X,Y;Z)$ may be
identified, in a canonical way, with $X^\ast\ot Y$ and
$X^\ast\ot Y^\ast\ot Z$, respectively (see \cite{KM}, \S 4.2).
Recall the description of this identification. Let
$l\in X^\ast$ and $y\in Y$.
Consider the map $\fe_{l,y}:X\lra Y$, defined by
$$ \fe_{l,y}(x)=l(x)y. $$
Clearly, $\fe_{l,y}$ is a linear map. Furthermore, the
expression $l(x)y$ is linear in all three arguments $l$, $x$,
and $y$, and therefore the rule $(l,y)\mapsto\fe_{l,y}$
defines a bilinear map from $X^\ast\times Y$ to $\cL(X,Y)$.
By the universal property of tensor product there exists a
unique linear map $\fe:X^\ast\ot Y\lra\cL(X,Y)$ such that
$\fe(l\ot y)=\fe_{l,y}$ for all $l$ and~$y$.

Show that this $\fe$ is an isomorphism. Let $e_1\,,\ldots,e_m$
and $f_1\,,\ldots,f_n$ be bases of $X$ and $Y$, respectively,
and $e^1,\ldots,e^m$ be the basis of $X^\ast$ dual to
$(e_i)$. Then $\{e^i\ot f_j\mid 1\leq i\leq m,\ 1\leq j\leq n\}$
is a basis of $X^\ast\ot Y$. Put $h_{ij}=\fe(e^i\ot f_j)$. It is
easy to see that $h_{ij}$ is the linear map that takes $e_i$
to $f_j$ and takes $e_l$ to $0$ for all $l\ne i$. Clearly,
$\{h_{ij}\mid i,j\}$ is a basis of $\cL(X,Y)$. Thus, $\fe$
takes a basis of $X^\ast\ot Y$ to a basis of $\cL(X,Y)$, and is
therefore an isomorphism. The map $\fe$ is called
the {\em canonical isomorphism} between $X^\ast\ot Y$
and $\cL(X,Y)$.

We can define the isomorphism $\fe:X^\ast\ot Y^\ast\ot Z
\lra\cL_2(X,Y;Z)$ in a similar way. Namely, $\fe$ is the
unique linear map such that
$$ (\fe(l\ot m\ot z))(x,y)=l(x)m(y)z \quad \forall\ x\in X,\
y\in Y, z\in Z,\ l\in X^\ast,\ m\in Y^\ast $$
(the details are left to the reader).

Let $f\in\cL(X,Y)$ (resp., $f\in\cL_2(X,Y;Z)$), and let
$h\in X^\ast\ot Y$ (resp., $h\in X^\ast\ot Y^\ast\ot Z$) be
a tensor such that $\fe(h)=f$. This $h$ is called the {\em
structure tensor} of $f$, and will be denoted by~$\wt f$.

Consider the group $G=GL(X)\times GL(Y)$. It acts on the
spaces $X^\ast\ot Y$ and $\cL(X,Y)$ as usually. That is, an
element $g=(g_1,g_2)\in G$ acts on $X^\ast\ot Y$ as
$g_1^\vee\ot g_2$, and the action of $g$ on $\cL(X,Y)$
is defined
by $g(f)=g_2fg_1^{-1}$ (we leave to the reader to show that
this is indeed a left action). Similarly, the group $G=GL(X)
\times GL(Y)\times GL(Z)$ acts on $X^\ast\ot Y^\ast\ot Z$
and on $\cL_2(X,Y;Z)$. The element $g=(g_1,g_2,g_3)\in G$
acts on $X^\ast\ot Y^\ast\ot Z$ as $g_1^\vee\ot g_2^\vee
\ot g_3$, and the action on $\cL_2(X,Y;Z)$ is described by
the rule
$$ (g(f))(x,y)=g_3(f(g_1^{-1}(x), g_2^{-1}(y))) $$
(i.e., $g$ takes $f$ to the map $f_1$ defined by $f_1(x,y)
=g_3(f(g_1^{-1}(x),g_2^{-1}(y)))$; we may also write this as
$g(f)=g_3\circ f\circ(g_1^{-1}\times g_2^{-1})$).
\begin{prop}    \label{pr:equivar}
Let $G=GL(X)\times GL(Y)$ (resp. $G=GL(X)\times GL(Y)
\times GL(Z)$), and let $\fe:X^\ast\ot Y\lra\cL(X,Y)$ (resp.
$\fe:X^\ast\ot Y^\ast\ot Z\lra\cL_2(X,Y;Z)$) be the canonical
isomorphism. Then $\fe$ is an isomorphism of
$KG$-modules.
\end{prop}
{\em Proof.} It was observed above that $\fe$ is an
isomorphism of vector spaces. By the observation preceding
Lemma~\ref{l:tau_Ghom} it is sufficient to check
that $\fe$ is a homomorphism of $KG$-modules.
We prove this statement only for $\fe:X^\ast\ot
Y\lra\cL(X,Y)$, leaving the second statement
to the reader.

We have to check that $g(\fe(u))=\fe(g(u))$ for all $g=(g_1,
g_2)\in G$ and $u\in X^\ast\ot Y$. By linearity, we may
assume that $u=l\ot y$. The condition $g(\fe(u))=\fe(g(u))$
means that $(g(\fe(u)))(x)=(\fe(g(u)))(x)$ for all $x\in X$.
Thus, we have to show that
\begin{equation}    \label{f:equivar}
((g_1,g_2)(\fe(l\ot y)))(x)=(\fe((g_1,g_2)(l\ot y)))(x)
\end{equation}
for all $g_1\in GL(X)$, $g_2\in GL(Y)$, $l\in X^\ast$, $y\in Y$,
and $x\in X$.

We have $((g_1,g_2)(\fe(l\ot y)))(x)=((g_1,g_2)(\fe_{l,y}))(x)$
(by the definition of $\fe$) $=g_2(\fe_{l,y}(g_1^{-1}x))$
(by the definition of the action of $G$ on $\cL(X,Y)$)
$=g_2(l(g_1^{-1}x)y)$ (by the definition of $\fe_{l,y}$)
$=l(g_1^{-1}x)g_2(y)$ (because $g_2$ is linear, and
$l(g_1^{-1}x)$ is an element of $K$).

On the other hand, $(\fe((g_1,g_2)(l\ot y)))(x)=(\fe
(g_1^\vee l\ot g_2y))(x)$ (by the definition of the action of
$G$ on $X^\ast\ot Y$) $=\fe_{g_1^\vee l, g_2y}(x)$
(by the definition of $\fe$) $=(g_1^\vee l)(x)\cdot g_2y$
(by the definition of $\fe_{l,y}$). Further, note that
$(g_1^\vee l)(x)=\lu g_1^\vee l,x\pu = \lu g_1^\vee l,
g_1(g_1^{-1}x)\pu$ (as $x=g_1(g_1^{-1}x)$) $=\lu l,
g_1^{-1}x \pu$ (by the property of contragradient maps)
$=l(g_1^{-1}x)$. Hence $(g_1^\vee l)(x)\cdot g_2y=
l(g_1^{-1}x)\cdot g_2y$.

Thus, both the left-hand and right-hand sides
of~(\ref{f:equivar}) are equal to $l(g_1^{-1}x)\cdot g_2y$,
and therefore (\ref{f:equivar}) is true.
\eoproof

\subsection{The isotropy group of a bilinear map}
	\label{subs:isotr_bil}
Let $X$, $Y$, and $Z$ be vector spaces and let $f\in\cL_2
(X,Y;Z)$ be a bilinear map. The group $G=GL(X)\times
GL(Y)\times GL(Z)$ acts on $\cL_2(X,Y;Z)$ in the way
described in the previous subsection. The
stabilizer of $f$ in $G$ with respect to this action will be
called the {\em isotropy group} of $f$, and will be denoted
by $\Delta(f)$. The reader can easily check that this
definition is equivalent to the following: $\Delta(f)$ is the
set of all triples $(A,B,C)\in G$ such that $f(Ax,By)=Cf(x,y)$
for all $x\in X$ and $y\in Y$. In other words, the diagram
$$ \begin{CD}  X\times Y @>f>> Z \\ @V{A\times B}VV
@ VV{C}V \\ X\times Y @>f>> Z \end{CD} $$
must commute.

{\bf Example.} Let $X$, $Y$, and $Z$ be three spaces,
let $U=\cL(X,Y)$, $V=\cL(Y,Z)$, $W=\cL(X,Z)$, and let
$f: U\times V\to W$ be the usual composition of mappings,
i.e., $f(x,y)=yx$. Clearly $f$ is bilinear. For $g=(g_1,g_2,g_3)
\in GL(X)\times GL(Y)\times GL(Z)$ put $R(g)=(A,B,C)$,
where $A:U\to U$, $B:V\to V$, and $C:W\to W$ are defined
by the rules $Ax=g_2xg_1^{-1}$,  $Bx=g_3xg_2^{-1}$,
and $Cx=g_3xg_1^{-1}$, respectively. Then it is easy to
see that $R(g)\in\Delta(f)$ for all~$g$. Moreover,
$R:g\mapsto R(g)$ is a group homomorphism. Later we
will show that $R$ is an {\em epimorphism}.

The following proposition shows that the isotropy group
of a bilinear map is closely related to the (small)
isotropy group of the structure tensor of this map.
\begin{prop}    \label{pr:isotr_bil}
Let $f:X\times Y\to Z$ be a bilinear map and let
$\wt f\in X^\ast\ot Y^\ast\ot Z$ be its structure tensor.
Let $(A,B,C)\in GL(X)\times GL(Y)\times GL(Z)$. Then
$(A,B,C)\in\Delta(f)$ if and only if $A^\vee\ot B^\vee\ot C
\in\Gamma^0(\wt f)$.
\end{prop}
{\em Proof.} By Proposition~\ref{pr:equivar}, the
map $h\mapsto\wt h$ is a $G$-isomorphism from $\cL_2(X,Y;Z)$
to $X^\ast\ot Y^\ast\ot Z$, where $G=GL(X)\times
GL(Y)\times GL(Z)$.
So $g=(A,B,C)\in G$ is in $\Delta(f)$ if
and only if $g$ fixes~$\wt f$. But $g(\wt f)=(A^\vee\ot
B^\vee\ot C)\wt f$ by the definition of the action of $G$
on $X^\ast\ot Y^\ast\ot Z$.
\eoproof

\subsection{Type of a space of linear maps}
	\label{subs:type}
Let $X$ and $Y$ be vector spaces, $\dim X=m$ and
$\dim Y=n$. Let $\cL=\cL(X,Y)$, and $L\sse\cL$ be some
space of linear maps from $X$ to~$Y$. It is natural to call
spaces
$$ \Ker L=\bigcap_{f\in L}\Ker f=\{x\in X\mid f(x)=0
\ \ \forall\ f\in L\} $$
and
$$ \im L=\sum_{f\in L}\im f $$
the {\em kernel} and {\em image} of $L$, respectively.

Put $\beta_1(L)=m-\dim\Ker L$ and $\beta_2(L)=\dim\im L$.
The pair $\beta(L)=(\beta_1(L),\beta_2(L))$ will be called the
{\em type} of~$L$.

Note that for any linear map $f\in\cL(X,Y)$ we have $\dim
\Ker f+\dim\im f=m$, so for a one-dimensional subspace
$\lu f\pu\sse \cL(X,Y)$ we have $\beta_1(\lu f\pu)=\beta_2
(\lu f\pu)=\dim\im f=\rk(f)$. Thus, $\beta(L)$ is a
generalization of the rank of a linear map.

Let $f\in\cL(X,Y)$, and let $a\in GL(X)$ and $b\in GL(Y)$ be
automorphisms of $X$ and $Y$, respectively. Then $bfa\in
\cL(X,Y)$. It is easy to see that
$$ \Ker bfa=a^{-1}(\Ker f), \quad \text{and}\quad \im bfa
=b(\im f). $$
Hence, if $\tau_{a,b}:\cL\lra\cL$ is a linear transformation
defined by $\tau_{a,b}(f)=bfa$, then
$$ \Ker\tau_{a,b}(L)=a^{-1}(\Ker L) \quad\text{and}\quad
\im\tau_{a,b}(L)=b(\im L), $$
for any subspace $L\sse\cL$. It follows that $\tau_{a,b}$
preserves the type:
$$ \beta(\tau_{a,b}(L))=\beta(L)\quad \forall\ L\sse\cL. $$

The aim of this subsection is to prove the converse
statement.
\begin{prop}    \label{pr:pres_type}
Let $\cL=\cL(X,Y)$, and let $h:\cL\lra\cL$ be a linear
transformation such that $\beta(h(L))=\beta(L)$ for all
subspaces $L\sse\cL$. Then there exist $a\in GL(X)$ and
$b\in GL(Y)$ such that $h=\tau_{a,b}$.
\end{prop}
{\em Proof.} It is useful to observe that the set of all
transformations of the form $\tau_{a,b}$ is a group, because
$\tau_{a,b}\tau_{c,d}=\tau_{ac,bd}$, and $\tau_{\id_X,\id_Y}
=\id_{\cL(X,Y)}$, whence also $(\tau_{a,b})^{-1}=\tau_{a^{-1},
b^{-1}}$. It is obvious that the group
$$\{\tau_{a,b}\mid a\in GL(X), b\in GL(Y)\}$$
is the image of $GL(X)\times GL(Y)$ under the
representation of the latter group on $\cL(X,Y)$, described
in Subsection~\ref{subs:contragrad}.

Take bases $(e_1,\ldots,e_m)$ and $(d_1\,,\ldots,d_n)$
of $X$ and $Y$, respectively. Moreover, let $(e^1,\ldots,
e^m)$ be the basis of $X^\ast$ dual to $(e_i)$. Let $f_{ij}:
X\lra Y$ be the map defined by $f_{ij}(e_i)=d_j$, $f_{ij}(e_l)
=0$ when $l\ne i$. Then $\{f_{ij}\mid 1\leq i\leq m,1\leq j
\leq n\}$ is a basis of~$\cL$. Moreover, $f_{ij}=\fe_{e^i,d_j}$,
where for $l\in X^\ast$ and $y\in Y$
$$ \fe_{l,y}=\fe(l\ot y):x\mapsto l(x)y $$
is the linear map, described in Subsection~\ref{subs:contragrad}.

It was noted above that $\beta(\lu f\pu)=(\rk(f), \rk(f))$ for any
$f\in\cL(X,Y)$. So $h$ preserves the rank: $\rk(h(f))=\rk(f)$
for all $f\in\cL$.

Put $q_{ij}=h(f_{ij})$. As $f_{ij}$ are of rank~1,  and $\{f_{ij}\}$
is a basis of $\cL$, it follows that all $q_{ij}$ are also of rank
1, and
$$\{q_{ij}\mid 1\leq i\leq m,\ 1\leq j\leq n\}$$
is a basis of~$\cL$.

It is easy to see that the maps $f\in\cL$ of rank~1 are
precisely the maps of the form $\fe_{l,y}$. In particular,
there exist $l_{ij}\in X^\ast$ and $y_{ij}\in Y$ such that
$q_{ij}=\fe_{l_{ij},y_{ij}}$ for all $1\leq i\leq m$, $1\leq j
\leq n$.

Show that $l_{i,j_1}$ and $l_{i,j_2}\in X^\ast$ are
proportional, for any $i$, $j_1$, and~$j_2$;
in other words, the
line $\lu l_{ij}\pu\sse X^\ast$ depends only on~$i$.
Consider the space $L=\lu f_{i,j_1}\,,f_{i,j_2}\pu$. Obviously,
$\Ker L=\lu e_l\mid l\ne i\pu$ has codimension~1 in
$X^\ast$, so the kernel of $h(L)=\lu q_{i,j_1}\,,q_{i,j_2}\pu$
must be of codimension 1 also. Now observe that $\Ker
\fe_{l,y}=\Ker l$ and $\im\fe_{l,y}=\lu y\pu$, for any
$l\in X^\ast$ and $y\in Y$ such that $l,y\ne0$. In particular,
$\Ker q_{ij}=\Ker l_{ij}$ and $\im q_{ij}=\lu y_{ij}\pu$, for all
$i$ and~$j$. Hence $\Ker\lu q_{i,j_1}\,,q_{i,j_2}\pu=\Ker
l_{i,j_1}\cap\Ker l_{i,j_2}$. If $l_{i,j_1}$ and $l_{i,j_2}$ are not
proportional, then the latter intersection is a subspace of
codimension~2 in $X^\ast$, a contradiction. Thus, $l_{i,j_1}$
and $l_{i,j_2}$ are proportional. Therefore, there exist
$l_1,\ldots,l_m\in X^\ast$ such that $\lu l_{ij}\pu=\lu l_i\pu$,
for all $i$ and~$j$.

Similarly, one can show that $\lu y_{ij}\pu$ depends only
on~$j$. Indeed, for given $i_1\ne i_2$, $1\leq i_1,i_2\leq m$,
consider the space $L=\lu f_{i_1,j},f_{i_2,j}\pu$. Its image
is $\lu d_j\pu$. So the image of the space $\lu q_{i_1,j},
q_{i_2,j}\pu=h(L)$ is of dimension 1 also, whence
$\lu y_{i_1,j}\pu=\lu y_{i_2,j}\pu$, whence $\lu y_{ij}\pu
=\lu y_j\pu$, for some $y_1,\ldots,y_n\in Y$.

Thus, we see that $q_{ij}=\fe_{l_{ij},y_{ij}}=\lambda_{ij}
\fe_{l_i,y_j}$, for some $\lambda_{ij}\in K^\ast$.

Show that $(y_1,\ldots,y_n)$ is a basis of~$Y$. Otherwise,
$\lu y_1\,,\ldots,y_n\pu=Y'$ is a proper subspace of~$Y$.
But the image of any $q_{ij}$ lies in $Y'$, so the image
of the whole $\cL(X,Y)
=\lu q_{ij}\mid 1\leq i\leq m,\ 1\leq j\leq n\pu$ also lies
in $Y'$, which is impossible when $Y'\ne Y$.

Similarly, $(l_1,\ldots,l_m)$ is a basis of~$X^\ast$.
Indeed, if this is not the case, then there exists
an element $x\in X$, $x\ne0$, annihilated by
all $q_{ij}$, i.e., by the whole $\cL(X,Y)$.

Further, we ``normalize'' basis $\{q_{ij}\}$ in an appropriate
way. To do this, we need to know how $\tau_{a,b}$ acts on
$\fe_{l,y}$. Prove that
\begin{equation}    \label{f:act_tau_fe}
\tau_{a,b}(\fe_{l,y})=\fe_{a^\ast(l),b(y)}\ ,
\end{equation}
where $a^\ast\in GL(X^\ast)$ is the map dual to~$a$. Indeed,
for all $x\in X$ we have $(\tau_{a,b}(\fe_{l,y}))(x)=(b\fe_{l,y}a)
(x)=b(\fe_{l,y}(a(x)))=b(l(a(x))y)=l(a(x))b(y)$. Since $l(a(x))
=(a^\ast(l))(x)$ by the definition of the dual map, we
finally get
$$ (\tau_{a,b}(\fe_{l,y}))(x)=(a^\ast(l))(x)b(y)=\fe_{a^\ast(l),
b(y)}(x), $$
which proves formula~(\ref{f:act_tau_fe}).
Let $b_1\in GL(Y)$ be the element taking basis $(y_1,\ldots,
y_n)$ to $(d_1,\ldots,d_n)$, let $c\in GL(X^\ast)$ takes
$(l_1,\ldots,l_m)$ to $(e^1,\ldots,e^m)$, and $a_1=c^\ast
\in GL(X)$ be the map dual to~$c$. Then $\tau_{a_1,b_1}$
takes $\fe_{l_i,y_j}$ to $\fe_{a_1^\ast(l_i),b_1(y_j)}=
\fe_{e^i,d_j}=f_{ij}$. As $q_{ij}=\lambda_{ij}\fe_{l_i,y_j}$,
we see that the map $h'=\tau_{a_1,b_1}h$ takes $f_{ij}$
to $\lambda_{ij}f_{ij}$. Moreover, $h'$ preserves types of
subspaces. Thus, it remains to show that a map $h'$, taking
$f_{ij}$ to $\lambda_{ij}f_{ij}$, where $\lambda_{ij}\in K^\ast$,
and preserving the types, must have the form $\tau_{a,b}$.

Multiplying $h'$ by a scalar we may assume that $\lambda_{11}=1$.
Put $\mu_i=\lambda_{i1}$, $\nu_j= \lambda_{1j}$ (whence
$\mu_1=\nu_1=1$), and show that $\lambda_{ij}=\mu_i\nu_j$. This is
evident if $i=1$ or $j=1$, so suppose $i,j>1$. Consider
$f=f_{11}+f_{i1}+f_{1j}+f_{ij} \in\cL$, then
$h'(f)=f_{11}+\mu_if_{i1}+\nu_jf_{1j}+ \lambda_{ij}f_{ij}$. As
$\rk(f)=1$, we must have $\rk(h'(f))=1$, which is the case only if
$\lambda_{ij}=\mu_i\nu_j$.

Let $a$ and $b$ be the linear maps on $X$ and $Y$ such
that $a(e_i)=\mu_ie_i$ and $b(d_j)=\nu_jd_j$. Then it is easy
to see that $bf_{ij}a=\mu_i\nu_j f_{ij}=\lambda_{ij}f_{ij}=
h'(f_{ij})$, for all $i$ and $j$, and therefore $h'=\tau_{a,b}$.
\eoproof

\subsection{Further lemmas on linear maps}
	\label{subs:lin_lemmas}
\begin{lemma}   \label{l:cap_ker_1}
Let $U$ and $V$ be spaces, $f_1\,,\ldots,f_n:U\to V$ be
linear maps such that $\cap_{i=1}^n\Ker f_i=0$. Then for
any linear function $l:U\to K$ there exist linear functions
$l_i:V\to K$ such that $l=\sum_{i=1}^nl_i\circ f_i\,$.
\end{lemma}
{\em Proof.} Let $T\sse U^\ast$ be the set of all linear
functions of the form $\sum_{i=1}^nl_i\circ f_i$, for some
$l_i\in V^\ast$. Clearly, $T\sse U^\ast$ is a subspace.
To prove that $T=U^\ast$ it is sufficient to show
that there is no a non-zero vector $u\in U$ such that
$t(u)=0$ for all $t\in T$.

Let $u\in U$, $u\ne0$. As $\cap_{i=1}^n\Ker f_i=0$, there exist $i$
such that $f_i(u)\ne0$. Next, there exist a linear function
$m\in V^\ast$ such that $m(f_i(u))\ne0$. Then
$m'=m\circ f_i\in T$ and $m'(u)=(m\circ f_i)(u)=m(f_i(u))
\ne0$.
\eoproof
\begin{lemma}   \label{l:cap_ker_2}
Let $U$, $V$, and $W$ be three spaces, $f_1\,\ldots,f_n
\in\cL(U,V)$ and $f\in\cL(U,W)$. Then $f$ can be represented
in the form $f=g_1f_1+\ldots+g_nf_n$, for some $g_i\in
\cL(V,W)$, if and only if $\cap_{i=1}^n\Ker f_i\sse \Ker f$.
\end{lemma}
{\em Proof.} Denote $\cap_{i=1}^n\Ker f_i$ by~$S$. It is
clear that if $f$ can be represented in the form
$f=g_1f_1+\ldots+g_nf_n$, then $S\sse\Ker f$.
We need to prove the converse implication.

First assume that $S=0$ and $\dim W=1$. Then we may
identify $W$ with $K$, so that $f$ and $g_i$ became linear
functions on $U$ and $V$, respectively, and the desired
statement follows from Lemma~\ref{l:cap_ker_1}.

Next consider the case when $S=0$, but $W$ may be of
arbitrary dimension. Let $(w_1,\ldots,w_m)$ be a
basis of~$W$. Then $f$ decomposes as
$$ f=\sum_{j=1}^mf^{(j)},$$
where $f^{(j)}\in \cL(U,\lu w_j\pu)$. According to the case
$S=0$ and $\dim W=1$, there exist linear maps
$g_i^{(j)}:V\to
\lu w_j\pu$, where $1\leq i\leq n$, $1\leq j\leq m$, such
that $f^{(j)}=\sum_{i=1}^ng_i^{(j)}f_i$ for all $j=1,\ldots,m$.
Now it is sufficient to take $g_i=\sum_{j=1}^mg_i^{(j)}$.
Thus the case $S=0$ is settled.

Finally consider the general case, when $S$ may be
nontrivial. Put $\ov U=U/S$, and let $\fe:U\to\ov U$
be the canonical factorization mapping. As $f$ and
all $f_i$ vanish on $S$, there exist unique maps
$\ov f\in\cL(\ov U,W)$ and $\ov f_i\in\cL(\ov U,V)$ such that
$f=\ov f\fe$ and $f_i= \ov f_i\fe$.
It is clear that $\Ker\ov f=(\Ker f)/S$ and
$\Ker \ov f_i=(\Ker f_i)/S$.

As $\cap_{i=1}^n\Ker f_i=S$,
we have $\cap_{i=1}^n\Ker\ov f_i=0$. Applying the
previous case we
see that $\ov f=\sum_{i=1}^ng_i\ov f_i$ for some linear maps
$g_i:V\to W$. Hence
$$ f=\ov f\fe=(\sum_{i=1}^ng_i\ov f_i)\fe=\sum_{i=1}^n
g_i(\ov f_i\fe)=\sum_{i=1}^ng_if_i\,. $$
\eoproof

The next lemma is, in a sense, dual to the previous one.
\begin{lemma}   \label{l:sum_im}
Let $U$, $V$ and $W$ be three spaces, $f_1,\ldots,f_n
\in\cL(V,W)$, and $f\in\cL(U,W)$. Then $f$ can be
represented in the form $f=\sum_{i=1}^n f_ig_i$, for some
$g_i\in\cL(U,V)$, if and only if $\im f\sse\sum_{i=1}^n
\im f_i$.
\end{lemma}
{\em Proof.}  Denote $\sum_{i=1}^n\im f_i$ by~$S$.
It is clear that if $f$ can be represented as
$f=\sum_{i=1}^n f_ig_i$, then $\im f\sse S$.
Prove the converse implication. Suppose that $\im f\sse S$.
There exists a basis $w_1,\ldots,w_m$ of $S$ such that each
$w_j$ has the form $w_j=f_p(v_j)$, for some $p=p(j)$ and
$v_j\in V$. Next, let $l_j:U\to K$ be the (uniquely defined)
linear functions such that
$$ f(u)=\sum_{j=1}^ml_j(u)w_j\,,$$
for all $u\in U$. Now for $j=1,\ldots,m$ define linear maps
$h_j:U\to V$ by $h_j(u)=l_j(u)v_j$. Then
$$f(u)=\sum_{j=1}^ml_j(u)w_j=\sum_{j=1}^ml_j(u)
f_{p(j)}(v_j)=\sum_{j=1}^mf_{p(j)}(l_j(u)v_j)= \sum_{j=1}^m
f_{p(j)}(h_j(u)),$$
for all $u\in U$. That is, $f=\sum_{j=1}^mf_{p(j)}h_j$.
But
$$ \sum_{j=1}^m f_{p(j)}h_j=\sum_{i=1}^nf_ig_i\,, $$
where $g_i=\sum_{\{j\mid p(j)=i\}}h_j\,$.
\eoproof
\vspace{2ex}

It is convenient to rewrite results of Lemmas
\ref{l:cap_ker_2} and \ref{l:sum_im} in a slightly
different form, using the notions of
kernel and image of a space of linear maps.

For subspaces $A\sse\cL(U,V)$ and $B\sse\cL(V,W)$ define
subspace $BA\sse \cL(U,W)$ by
$$ BA=\lu ba\mid a\in A,\ b\in B\pu. $$
In particular, we can consider subspaces $\cL(V,W)A$ and
$B\cL(U,V)$ of $\cL(U,W)$.
\begin{prop}    \label{pr:ker_im_gen}
Let $U$, $V$, and $W$ be three spaces, $A\sse\cL(U,V)$ and
$B\sse\cL(V,W)$ be subspaces, and let $f\in\cL(U,W)$.
Then the following statements hold.

1) $f\in\cL(V,W)A$ if and only if $\Ker f\supseteq\Ker A$.

2) $f\in B\cL(U,V)$ if and only if $\im f\sse\im B$.
\end{prop}

We leave to the reader to deduce statements 1) and 2) from
Lemmas \ref{l:cap_ker_2} and \ref{l:sum_im}, respectively.

\subsection{The isotropy group of the map of composition
of linear maps}
	\label{subs:comp_isotr}
Let $U$, $V$ and $W$ be three vector spaces, and let
$$\fe:\cL(U,V)\times\cL(V,W)\to \cL(U,W) $$
be the usual composition of mappings, i.e., $\fe(x,y)=yx$.
The aim of the present subsection is to find the isotropy
group~$\Delta(\fe)$.

For subspaces
$A\sse\cL(U,V)$ and $B\sse\cL(V,W)$ we define their
{\em annihilators} by
$$ \ann(A)=\{h\in\cL(V,W)\mid hA=0\}$$
and
$$ \ann(B)=\{h\in\cL(U,V)\mid Bh=0\},$$
respectively.

For a subspace $L\sse\cL(U,V)$ or $L\sse\cL(V,W)$ let
$\beta(L)=(\beta_1(L),\beta_2(L))$ be its type, as described in
Subsection~\ref{subs:type}.

We need a lemma.
\begin{lemma}   \label{l:anns}
For subspaces $A\sse\cL(U,V)$ and $B\sse\cL(V,W)$ the
following equalities hold:
\\ (1) $\dim \cL(V,W)A=\beta_1(A)\dim W$;
\\ (2) $\dim\ann(A)=(\dim V-\beta_2(A))\dim W$;
\\ (3) $\dim B\cL(U,V)=\beta_2(B)\dim U$;
\\ (4) $\dim\ann (B)=(\dim V-\beta_1(B))\dim U$.
\end{lemma}
{\em Proof.} (1) Let $S=\Ker A$. By statement 1) of
Proposition~\ref{pr:ker_im_gen} $\cL(V,W)A$ consists
of all $h\in\cL(U,W)$ such that
$h(S)=0$. Hence $\dim \cL(V,W)A=(\dim U-\dim S)\dim W
=\beta_1(A)\dim W$.

(2) Let $T=\im A$. Consider
$f\in\cL(V,W)$. If $f(T)=0$, then $fA=0$, that is, $f\in\ann(A)$.
On the other hand, if $f(T)\ne0$, then there exist $u\in U$
and $h\in A$ such that $f(h(u))\ne0$, whence $(fh)(u)\ne0$,
and therefore $fh\ne0$ and $f\notin\ann(A)$. Thus,
$f\in\ann(A)$ if and only if $f(T)=0$. Hence $\dim\ann(A)=
(\dim V-\dim T)\dim W=(\dim V-\beta_2(A))\dim W$.

(3) Let $T=\im B$. By statement 2) of
Proposition~\ref{pr:ker_im_gen} an element
 $f\in\cL(U,W)$ is in $B\cL(U,V)$ if and
only if  $\im f\sse T$, whence $B\cL(U,V)=\cL(U,T)$, and
therefore $\dim B\cL(U,V)=(\dim U)(\dim T)=(\dim U)
\beta_2(B)$.

(4) It is easy  to see that $\ann(B)$ consists of all $f\in\cL
(U,V)$ such that $\im f\sse\Ker B$. Hence $\ann(B)=\cL(U,
\Ker B)$ and $\dim\ann(B)=(\dim U)(\dim\Ker B)=
(\dim V-\beta_1(B))\dim U$.
\eoproof
\vspace{2ex}

We also need the following lemma, whose proof is left to
the reader.
\begin{lemma}   \label{l:cancel}
Let $U$, $V$, and $W$ be three spaces, and let $a,b\in
\cL(V,V)$ be maps such that $yax=ybx$ for all
$x\in\cL(U,V)$ and $y\in\cL(V,W)$. Then $a=b$.
\end{lemma}

Recall that the center of the group $GL(V)$ consists
of scalar transformations, for any $V$, see the
remark after Proposition~\ref{pr:Gamma0_H}.
\begin{prop}    \label{pr:comp_isotr}
Let $X$, $Y$ and $Z$ be vector spaces, $U=\cL(X,Y)$,
$V=\cL(Y,Z)$ and $W=\cL(X,Z)$, and let $A\in GL(U)$,
$B\in GL(V)$ and $C\in GL(W)$ be transformations of $U$,
$V$ and $W$ such that
\begin{equation}    \label{f:ABC_comp}
(Bv)(Au)=C(vu) \quad \forall\ u\in U,\, v\in V.
\end{equation}
Then there exist $p\in GL(X)$, $q\in GL(Y)$, and $r\in GL(Z)$
such that $A$, $B$ and $C$ are given by rules
$Au=qup^{-1}$, $Bv=rvq^{-1}$, and $Cw=rwp^{-1}$,
for $u\in U$, $v\in V$ and $w\in W$, respectively.
\end{prop}
{\em Proof.} First prove that $A$ and $B$ preserve type of
subspaces of $U$ and $V$, respectively. Clearly, $(B(N))(A(M))
=C(NM)$ for any subspaces $M\sse U$ and $N\sse V$,
whence $\dim B(N)A(M)=\dim NM$, for all $M$ and~$N$.
In particular, $\dim VA(M)=\dim VM$ for all~$M$. Applying
statement 1) of Lemma~\ref{l:anns} we see that
$\beta_1(A(M))\dim Z=
\beta_1(M)\dim Z$, whence $\beta_1(A(M))=\beta_1(M)$
for any~$M$.

Also, $B(N)A(M)=0$ if and only if $NM=0$, whence one
easily sees that $\ann(A(M))=B(\ann(M))$, whence
$$(\dim Y-\beta_2(A(M)))\dim Z=(\dim Y-\beta_2(M))
\dim Z,$$
whence finally $\beta_2(A(M))=\beta_2(M)$. Thus, $A$
preserves type of subspaces in~$U$. Similarly one can
prove that $B$ preserves type of subspaces in~$V$. So by
Proposition~\ref{pr:pres_type} $A$ and $B$ have the form
$Au=qup^{-1}$ and $Bv=r_1vq_1^{-1}$ for some $p\in GL(X)$,
$q,q_1\in GL(Y)$ and $r_1\in GL(Z)$. Substituting these
explicit formulae into condition~(\ref{f:ABC_comp}),
we see that
$$(r_1vq_1^{-1})(qup^{-1})=C(vu) \qquad \forall
\ u\in U, v\in V.$$

For any $d\in GL(Y)$ we have $(vd)(d^{-1}u)=vu$, for all
$u\in U$ and $v\in V$, whence
$$(r_1vdq_1^{-1})(qd^{-1}up^{-1})=C(vd\cdot d^{-1}u)
=C(vu)=(r_1vq_1^{-1})(qup^{-1}). $$
Multiplying by $r_1^{-1}$ on the left and by $p$ on the right,
we see that $vdq_1^{-1}qd^{-1}u=vq_1^{-1}qu$, for all
$u\in U$ and $v\in V$. Applying Lemma~\ref{l:cancel}
we see that
$dq_1^{-1}qd^{-1}=q_1^{-1}q$. Thus, the element
$q_1^{-1}q\in GL(Y)$ commutes with all $d\in GL(Y)$
and so must be a scalar, whence $q_1=\lambda q$,  for
some $\lambda\in K^\ast$. So we can rewrite the
formula for $Bv$ as $Bv=r_1vq_1^{-1}=rvq^{-1}$, where
$r=\lambda^{-1}r_1$. Thus, we have $Au=qup^{-1}$ and
$Bv=rvq^{-1}$, for all $u$ and~$v$. Now
$$ (rvq^{-1})(qup^{-1})=C(vu), \quad \forall\ u\in U,
v\in V, $$
whence $C(vu)=rvup^{-1}$ for all $u$ and~$v$. Since
$\lu vu\mid u\in U,\,, v\in V\pu=W$, we see that $Cw=
rwp^{-1}$ for all $w\in W$.
\eoproof
\vspace{2ex}

 The following statement may be considered by the reader
as an evident consequence of Proposition~\ref{pr:comp_isotr}.
Nevertheless, we give a formal proof.
\begin{cor} \label{cor:mm_isotr}
Let $N_1=M_{nm}$, $N_2=M_{pn}$, $N_3=M_{pm}$, and
suppose that transformations $A_i\in GL(N_i)$, $i=1,2,3$,
satisfy relations $(A_2x_2)(A_1x_1)=A_3(x_2x_1)$ for all
$x_1\in N_1$ and $x_2\in N_2$. Then there exist elements
$s\in GL_m(K)$, $q\in GL_n(K)$, and $r\in GL_p(K)$ such that
$A_1x_1=qx_1s^{-1}$, $A_2x_2=rx_2q^{-1}$, and $A_3x_3
=rx_3s^{-1}$, for all $x_i\in N_i$, $i=1,2,3$.
\end{cor}
{\em Proof.} Let $X=K^m$, $Y=K^n$, and $Z=K^p$ be the
column spaces. First we identify, in a usual way, $N_1$,
$N_2$, and $N_3$ with $\cL(X,Y)$, $\cL(Y,Z)$, and $\cL(X,Z)$, respectively.

Namely, for $h\in N_1$ let $\fe(h)\in\cL(X,Y)$ be the
multiplication by~$h$, i.e., $(\fe(h))(x)=hx$, for all $x\in X$.
Equivalently, for a map $g\in\cL(X,Y)$ the corresponding
element $\fe^{-1}(g)\in N_1$ is just the matrix of $g$
with respect to standard bases in $X$ and~$Y$. It is clear
that $\fe:N_1\lra\cL(X,Y)$ is an isomorphism of linear
spaces.

In a similar way we identify $N_2$ with $\cL(Y,Z)$, and
$N_3$ with $\cL(X,Z)$. Moreover, we identify $GL(X)$,
$GL(Y)$ and $GL(Z)$ with $GL_m(K)$, $GL_n(K)$ and
$GL_p(K)$, respectively. It is convenient to use the same
symbol $\fe$ for all these identifications. Thus, $\fe$ is a
bijection from
$$I=N_1\sqcup N_2\sqcup N_3\sqcup GL_m(K) \sqcup
GL_n(K)\sqcup GL_p(K) $$
to
$$ J=\cL(X,Y) \sqcup \cL(Y,Z )\sqcup\cL(X,Z) \sqcup GL(X)
\sqcup GL(Y)\sqcup GL(Z). $$

Further, it is easy to see that $\fe$ preserves multiplication,
for example, $\fe(x_2)\fe(x_1)=\fe(x_2x_1)$ for any
$x_1\in N_1$ and $x_2\in N_2$. Generally, if $x,y\in I$,
and if at least one of two expressions $xy$ and $\fe(x)\fe(y)$
is defined, then the other one is defined also and $\fe(x)
\fe(y)=\fe(xy)$. Moreover, if $x,y\in J$ and at least one of
two expressions $xy$ and $\fe^{-1}(x)\fe^{-1}(y)$ is defined,
then the other one is defined also and $\fe^{-1}(x)\fe^{-1}(y)
=\fe^{-1}(xy)$.

Let $B_1$, $B_2$ and $B_3$ be the transformations of the
spaces $\cL(X,Y)$, $\cL(Y,Z)$, and $\cL(X,Z)$, corresponding
to $A_1$, $A_2$ and $A_3$ under $\fe$; that is, $B_i=\fe A_i
\fe^{-1}$. Then for any $y_1\in\cL(X,Y)$ and $y_2\in\cL(Y,Z)$
we have $(B_2y_2)(B_1y_1)=B_3(y_2y_1)$. Indeed,
\begin{eqnarray*}
(B_2y_2)(B_1y_1) &=& ((\fe A_2\fe^{-1})y_2)((\fe A_1\fe^{-1})
y_1)=(\fe(A_2(\fe^{-1}(y_2))))(\fe(A_1(\fe^{-1}(y_1)))) \\
&=& \fe((A_2(\fe^{-1}(y_2)))(A_1(\fe^{-1}(y_1))) )
=\fe(A_3((\fe^{-1}(y_2))(\fe^{-1}(y_1)))) \\
&=& \fe(A_3(\fe^{-1}(y_2y_1)))=(\fe A_3\fe^{-1})(y_2y_1)
=B_3(y_2y_1).
\end{eqnarray*}
Applying Proposition~\ref{pr:comp_isotr}, we see
that there exist
$s_1\in GL(X)$, $q_1\in GL(Y)$ and $r_1\in GL(Z)$ such that
$B_1x=q_1xs_1^{-1}$, $B_2x=r_1xq_1^{-1}$, and
$B_3x=r_1xs_1^{-1}$, where $x\in\cL(X,Y)$, $\cL(Y,Z)$,
or $\cL(X,Z)$, respectively. Therefore for any $x\in N_1$
we have
\begin{eqnarray*}
A_1x &=& (\fe^{-1}B_1\fe)(x)=\fe^{-1}(B_1(\fe(x)))
=\fe^{-1}(q_1\fe(x)s_1^{-1}) \\
&=& \fe^{-1}(q_1) \fe^{-1}(\fe(x))\fe^{-1}(s_1^{-1})=
qxs^{-1},
\end{eqnarray*}
where $s=\fe^{-1}(s_1)$ and $q=\fe^{-1}(q_1)$. In a
similar way one can prove formulae for $A_2x$ and $A_3x$
(with $r=\fe^{-1}(r_1)$).
\eoproof

\subsection{Proof of Proposition~\ref{pr:Gamma0_H}}
	\label{subs:proof_G0}
We start with the following observation. Let $x$ and $y$ be
$a\times b$ and $b\times a$ matrices, respectively. Then
$\Tr(xy)=\Tr(yx)$. Moreover,
$$ (x,y)\mapsto \lu x,y\pu=\Tr(xy)=\Tr(yx) $$
is a nondegenerate bilinear pairing between $M_{ab}$
and~$M_{ba}$. Therefore we may identify $M^\ast_{ab}$
with $M_{ba}$, and $M^\ast_{ba}$ with~$M_{ab}$.

Further, the group $G=GL_a(K)\times GL_b(K)$ acts on both
$M_{ab}$ and $M_{ba}$ in a usual way, that is, $g=(g_1,g_2)$
takes $x\in M_{ab}$ and $y\in M_{ba}$ to $g_1xg_2^{-1}$
and $g_2yg_1^{-1}$, respectively. The pairing is invariant
under this action. Indeed, if $x\in M_{ab}$, $y\in M_{ba}$,
and $g=(g_1,g_2)\in G$, then
$$\lu gx,gy\pu=\Tr((g_1xg_2^{-1})(g_2yg_1^{-1}))=
\Tr(g_1xyg_1^{-1})=\Tr(xy)=\lu x,y\pu.$$
Therefore the transformations, induced by $g$ on $M_{ab}$
and $M_{ba}$, are contragradient each to the other.

Let $L_1=M_{mn}$, $L_2=M_{np}$ and $L_3=M_{pm}$
be as in the hypothesis of the Proposition, and let $N_1
=M_{nm}$ and $N_2=M_{pn}$. Then $N_i$ is dual to $L_i$,
$i=1,2$. Let $\fe:N_1\times N_2\lra L_3$ be the usual product
map, that is, $\fe(x,y)=yx$. Its structure tensor $\wt\fe\in
N_1^\ast\ot N_2^\ast\ot L_3$ may be considered as an
element of $L_1\ot L_2\ot L_3$. We show that $\wt\fe=t=
\lu m,n,p\pu$.

Indeed, we have
$$ t=\sum_{1\leq i\leq m,\ 1\leq j\leq n,\ 1\leq k\leq p}
e_{ij}\ot e_{jk} \ot e_{ki}\,. $$
Let
$$\psi:L_1\ot L_2\ot L_3=N^\ast_1\ot N^\ast_2\ot L_3
\lra \cL_2(N_1,N_2;L_3) $$
be the canonical map, described in
Subsection~\ref{subs:contragrad} (denoted by $\fe$ there).
We must show that the bilinear
map $\rho=\psi(t)$ coincides with~$\fe$. The bases of
$N_1$ and $N_2$ are $\{e_{lq}\mid 1\leq l\leq n,\ 1\leq q
\leq m\}$ and $\{e_{rs}\mid 1\leq r\leq p,\ 1\leq s \leq n\}$,
respectively. It follows from the definition of $\psi$ that
the value of $\rho$ on the pair $(e_{lq},e_{rs})$ equals
$$\sum_{\substack{1\leq i\leq m \\ 1\leq j\leq n \\
1\leq k\leq p}} \Tr(e_{ij}e_{lq})\Tr(e_{jk}e_{rs})e_{ki}
=\sum_{\substack{1\leq i\leq m \\ 1\leq j\leq n \\
1\leq k\leq p}} \delta_{jl} \delta_{iq}\delta_{kr}\delta_{js}
e_{ki}=\sum_{1\leq j\leq n} \delta_{jl}\delta_{js}e_{rq}
=\delta_{ls} e_{rq}\,.$$
On the other hand, $\fe(e_{lq},e_{rs})=e_{rs}e_{lq}=
\delta_{sl}e_{rq}$. Thus, $\rho(e_{lq},e_{rs})=\fe(e_{lq},
e_{rs})$ for any $l$, $q$, $r$, and $s$, that is, $\fe=\rho$.
Thus, $t=\wt\fe$.

Return to the proof of the proposition, and assume that
$g\in\Gamma^0(t)$. We have $g=A_1\ot A_2\ot A_3$,
for some $A_i\in GL(L_i)$, $i=1,2,3$. For $i=1,2$ we put
$B_i=A_i^\vee\in GL(L_i^\ast)=GL(N_i)$. Then
$A_i=B_i^\vee$, $i=1,2$. So we have $B_1^\vee\ot
B_2^\vee\ot A_3\in\Gamma^0(\wt\fe)$. Now
Proposition~\ref{pr:isotr_bil} implies that
$(B_1,B_2,A_3)\in\Delta(\fe)$.
In other words, $(B_2y)(B_1x)=A_3(yx)$ for any $x\in
M_{nm}$ and $y\in M_{pn}$. By Corollary~\ref{cor:mm_isotr},
there exist
$a\in GL_m(K)$, $b\in GL_n(K)$ and $c\in GL_p(K)$ such
that $B_1$, $B_2$ and $A_3$ are defined by the rules
$B_1x=bxa^{-1}$, $B_2x=cxb^{-1}$ and $A_3x=cxa^{-1}$,
where $x\in N_1$, $N_2$, or $L_3$, respectively.

It follows from the discussion in the beginning of the proof
that the transformation on $L_1$, contragredient to
transformation $x\mapsto bxa^{-1}$ on $N_1$, may be
described by the formula $x\mapsto axb^{-1}$. Similarly,
$A_2$ acts by the rule $x\mapsto bxc^{-1}$. Therefore,
$g$ acts by
$$ A(x\ot y\ot z)=axb^{-1}\ot byc^{-1}\ot cza^{-1}. $$
That is, $g=T(a,b,c)$.
\eoproof

\section{Automorphisms of Laderman algorithm}
	\label{sec:lad}
In this section we find automorphisms of the Laderman
algorithm. The structure of the section is as follows. First
we recall Laderman algorithm in its computational form,
and rewrite it in the tensor form. Then we produce a certain
subgroup $G$ of $\Gamma(t)$, the isotropy group of
$t=\lu3,3,3\pu$. Then we check that $G$ preserves the
algorithm. Finally, we prove that $G$ is the full automorphism
group of the algorithm. In the end of the section we give
a less formal explanation on how $G$ was found.

\subsection{Laderman algorithm}
	\label{subs:lad_the_alg}
Recall the description of the Laderman algorithm in
computational form, according to~\cite{Laderman}. Let
$$ X=\begin{pmatrix} x_{11} & x_{12} & x_{13} \\ x_{21} &
x_{22} & x_{23} \\ x_{31} & x_{32} & x_{33} \end{pmatrix} \quad
\text{and}\quad Y=\begin{pmatrix} y_{11} & y_{12} & y_{13} \\
y_{21} & y_{22} & y_{23} \\ y_{31} & y_{32} & y_{33}
\end{pmatrix}. $$
Consider the products
$$ p_1=(x_{11}+x_{12}+x_{13}-x_{21}-x_{22}-x_{32}-x_{33})
y_{22}\,, \quad p_2=(x_{11}-x_{21})(-y_{12}+y_{22}), $$
$$ p_3=x_{22}(-y_{11}+y_{12}+y_{21}-y_{22}-y_{23}-y_{31}
+y_{33}), \quad p_4=(-x_{11}+x_{21}+x_{22})(y_{11}-y_{12}
+y_{22}), $$
$$ p_5=(x_{21}+x_{22})(-y_{11}+y_{12}), \quad p_6=x_{11}
y_{11}\,, \quad p_7=(-x_{11}+x_{31}+x_{32})(y_{11}-y_{13}
+y_{23}), $$
$$p_8=(-x_{11}+x_{31})(y_{13}-y_{23}),\quad p_9=(x_{31}+
x_{32})(-y_{11}+y_{13}),$$
$$ p_{10}=(x_{11}+x_{12}+x_{13}-x_{22}-x_{23}-x_{31}-x_{32})
y_{23}\,, \quad p_{11}=x_{32}(-y_{11}+y_{13}+y_{21}-y_{22}
-y_{23}-y_{31}+y_{32}),$$
$$p_{12}=(-x_{13}+x_{32}+x_{33})(y_{22}+y_{31}-y_{32}),
\quad  p_{13}=(x_{13}-x_{33})(y_{22}-y_{32}), \quad p_{14}=
x_{13}y_{31}\,, $$
$$ p_{15}=(x_{32}+x_{33})(-y_{31}+y_{32}), \quad
p_{16}=(-x_{13}+x_{22}+x_{23})(y_{23}+y_{31}-y_{33}), $$
$$p_{17}=(x_{13}-x_{23})(y_{23}-y_{33}),\quad p_{18}=(x_{22}
+x_{23})(-y_{31}+y_{33}), \quad p_{19}=x_{12}y_{21}\,,$$
$$ p_{20}=x_{23}y_{32}\,,\quad p_{21}=x_{21}y_{13}\,,
\quad p_{22}=x_{31}y_{12}\,,\quad p_{23}=x_{33}y_{33}\,. $$
Then one can check that the coefficients of the matrix
$$ Z=XY= \begin{pmatrix} z_{11} & z_{12} & z_{13} \\
z_{21} & z_{22} & z_{23} \\ z_{31} & z_{32} & z_{33}
\end{pmatrix} $$
can be computed according to the formulae
$$ z_{11}=p_6+p_{14}+p_{19}\,,$$
$$ z_{12}=p_1+p_4+p_5+p_6+p_{12}+p_{14}+p_{15}\,, $$
$$ z_{13}=p_6+p_7+p_9+p_{10}+p_{14}+p_{16}+p_{18}\,, $$
$$ z_{21}=p_2+p_3+p_4+p_6+p_{14}+p_{16}+p_{17}\,, $$
$$ z_{22}=p_2+p_4+p_5+p_6+p_{20}\,, $$
$$ z_{23}=p_{14}+p_{16}+p_{17}+p_{18}+p_{21}\,, $$
$$ z_{31}=p_6+p_7+p_8+p_{11}+p_{12}+p_{13}+p_{14}\,, $$
$$ z_{32}=p_{12}+p_{13}+p_{14}+p_{15}+p_{22}\,, $$
$$ z_{33}=p_6+p_7+p_8+p_9+p_{23}\,. $$

(In \cite{Laderman} there are only a few words on how this
algorithm was found. The author of \cite{Laderman} promised
to publish more detailed description of his approach, but
this was never done.)

Write Laderman algorithm in tensor form. Let $M=M_{33}(K)$
($=M_3(K)$, in traditional notation, since we consider square
matrices), put $L_1=L_2=L_3=M$, and put next $L=L_1\ot
L_2\ot L_3$. Consider the following elements of $L$:
$$t_1=(e_{11}+e_{12}+e_{13}-e_{21}-e_{22}-e_{32}-e_{33})
\ot e_{22}\ot e_{21}\,,$$
$$t_2=(e_{11}-e_{21})\ot (-e_{12}+e_{22})\ot (e_{12}+e_{22})
\,, $$
$$t_3=e_{22}\ot (-e_{11}+e_{12}+e_{21}-e_{22}-e_{23}-e_{31}
+e_{33}) \ot e_{12}\,, $$
$$t_4=(-e_{11}+e_{21}+e_{22})\ot (e_{11}-e_{12}+e_{22})
\ot (e_{21}+e_{12}+e_{22})\,,$$
$$t_5=(e_{21}+e_{22})\ot (-e_{11}+e_{12})\ot (e_{21}+e_{22})
\,,$$
$$ t_6=e_{11}\ot e_{11}\ot (e_{11}+e_{21}+e_{31}+e_{12}+e_{22}
+e_{13}+e_{33})\,,$$
$$t_7=(-e_{11}+e_{31}+e_{32})\ot (e_{11}-e_{13}+e_{23})\ot
(e_{31}+e_{13}+e_{33})\,, $$
$$t_8=(-e_{11}+e_{31})\ot (e_{13}-e_{23})\ot (e_{13}+e_{33})
\,, $$
$$t_9=(e_{31}+e_{32})\ot (-e_{11}+e_{13})\ot (e_{31}+e_{33})
\,, $$
$$t_{10}=(e_{11}+e_{12}+e_{13}-e_{22}-e_{23}-e_{31}-e_{32})
\ot e_{23}\ot e_{31}\,,$$
$$t_{11}=e_{32}\ot (-e_{11}+e_{13}+e_{21}-e_{22}-e_{23}
-e_{31}+e_{32}) \ot e_{13}\,, $$
$$t_{12}=(-e_{13}+e_{32}+e_{33})\ot (e_{22}+e_{31}-e_{32})
\ot (e_{21}+e_{13}+e_{23})\,,$$
$$t_{13}=(e_{13}-e_{33})\ot (e_{22}-e_{32})\ot (e_{13}+e_{23})
\,, $$
$$ t_{14}=e_{13}\ot e_{31}\ot (e_{11}+e_{21}+e_{31}+e_{12}
+e_{32}+e_{13}+e_{23})\,,$$
$$t_{15}=(e_{32}+e_{33})\ot (-e_{31}+e_{32})\ot (e_{21}+
e_{23})\,, $$
$$t_{16}=(-e_{13}+e_{22}+e_{23})\ot (e_{23}+e_{31}-e_{33})
\ot (e_{31}+e_{12}+e_{32})\,,$$
$$t_{17}=(e_{13}-e_{23})\ot (e_{23}-e_{33})\ot (e_{12}+e_{32})
\,, $$
$$t_{18}=(e_{22}+e_{23})\ot (-e_{31}+e_{33})\ot (e_{31}+
e_{32})\,, $$
$$t_{19}=e_{12}\ot e_{21}\ot e_{11}\,,\quad t_{20}=e_{23}
\ot e_{32}\ot e_{22}\,,$$
$$t_{21}=e_{21}\ot e_{13}\ot e_{32}\,,\quad t_{22}=e_{31}
\ot e_{12}\ot e_{23}\,,$$
$$t_{23}=e_{33}\ot e_{33}\ot e_{33}\,. $$
\begin{prop}    \label{pr:tens_lad}
The set $\cL=\{t_1\,,\ldots,t_{23}\}$ is the tensor form of the
Laderman algorithm.
\end{prop}
{\em Proof.} A direct computation following discussion in
Section~\ref{sec:alg_tens}. Alternatively, the reader can check that
the sum $t_1+\ldots+t_{23}$ coincides with $t=\lu3,3,3\pu$.
\eoproof

\subsection{A subgroup of $\Gamma(t)$}
	\label{subs:lad_subgr}
For $a,b,c\in GL_3(K)$ let $T(a,b,c):L\lra L$ be the
transformation, described in Subsection~\ref{subs:Gamma0},
defined by
$$ T(a,b,c): x\ot y\ot z\mapsto axb^{-1}\ot byc^{-1}\ot
cza^{-1}\,. $$

Introduce notation for several special elements of $GL_3(K)$.
Let
$$ \pi_{12}=\begin{pmatrix} 0 & 1& 0\\ 1& 0&0\\ 0&0&1
\end{pmatrix}=e_{12}+e_{21}+e_{33} $$
be the transformation, intrchanging the basis vectors $e_1$
and $e_2$, and define $\pi_{13}$ and $\pi_{23}$ similarly.
Also put $\eps_1=\diag(-1,1,1)$, $\eps_2=\diag(1,-1,1)$, and
$\eps_3=\diag(1,1,-1)$. Note that all matrices $\pi_{ij}$
and $\eps_i$ are symmetric, and $\eps_i$ and $\pi_{jk}$
commute, if $\{i,j,k\}=\{1,2,3\}$.

Further, introduce the following decomposable
automorphisms $\Phi_i$, $i=1,2,3,4$, of~$L$. Put
$$ \Phi_1=T(\pi_{23},\pi_{13},1): x\ot y\ot z\mapsto
\pi_{23}x\pi_{13}\ot \pi_{13}y\ot z\pi_{23}\,, $$
and
$$\Phi_2=T(\pi_{23},1,\pi_{23}).$$
Next, define $\Phi_3$ and $\Phi_4$ by
$$\Phi_3(x\ot y\ot z)=y^t\eps_2\ot\eps_2x^t\ot z^t $$
(where $x\mapsto x^t$ is the transpose map) and
$$ \Phi_4(x\ot y\ot z)=\eps_1z\pi_{12}\ot\pi_{12}x\pi_{12}
\eps_1\ot\eps_1\pi_{12}y\eps_1\,.$$
\begin{prop}    \label{pr:Phi_rels}
The following relations hold:
$$\Phi_1^2=\Phi_2^2=\Phi_3^2=\Phi_4^3=1,\quad
\Phi_1\Phi_2=\Phi_2\Phi_1\,,\quad \Phi_3\Phi_1\Phi_3
=\Phi_1\Phi_2\,, $$
$$ \Phi_3\Phi_2\Phi_3= \Phi_2\,,\quad  \Phi_4\Phi_1
\Phi_4^{-1}=\Phi_1\Phi_2\,,\quad  \Phi_4\Phi_2 \Phi_4^{-1}
= \Phi_1\,,$$
$$\Phi_3\Phi_4\Phi_3=\Phi_4^{-1}\,.$$
\end{prop}
{\em Proof.} The relations $\Phi_1^2=\Phi_2^2=1$ and
$\Phi_1\Phi_2=\Phi_2\Phi_1$ hold because $\pi_{13}$ and
$\pi_{23}$ are elements of order $2$ in $GL_3(K)$ and
$(a,b,c)\mapsto T(a,b,c)$ is a homomorphism from
$GL_3(K)^{\times3}$ to $\Gamma(t)$, as was observed
in the proof of Proposition~\ref{pr:H_constr}.
The remaining relations may be proved by a direct
calculation. Show, for example, that $\Phi_4^3=1$ and
$\Phi_3\Phi_1\Phi_3=\Phi_1\Phi_2$.

We have
\begin{eqnarray*}
x\ot y\ot z &\stackrel{\Phi_4}\mapsto&  \eps_1z\pi_{12}\ot
\pi_{12}x\pi_{12}\eps_1\ot\eps_1\pi_{12}y\eps_1 \\
&\stackrel{\Phi_4}\mapsto& \eps_1(\eps_1\pi_{12}y\eps_1)
\pi_{12}\ot \pi_{12}(\eps_1z\pi_{12})\pi_{12}\eps_1\ot
\eps_1\pi_{12}(\pi_{12}x\pi_{12}\eps_1)\eps_1.
\end{eqnarray*}

Simplify the latter expression. We have
$$ \eps_1(\eps_1\pi_{12}y\eps_1)\pi_{12}= \eps_1^2\pi_{12}
y\eps_1\pi_{12}= \pi_{12}y\eps_1\pi_{12}\,, $$
$$\pi_{12}(\eps_1z\pi_{12})\pi_{12}\eps_1=\pi_{12}\eps_1z
\pi_{12}^2\eps_1=\pi_{12}\eps_1z\eps_1\,, $$
and similarly
$$ \eps_1\pi_{12}(\pi_{12}x\pi_{12}\eps_1)\eps_1= \eps_1x
\pi_{12}\,,$$
whence
$$ \Phi_4^2(x\ot y\ot z)=\pi_{12}y\eps_1\pi_{12}\ot \pi_{12}
\eps_1z\eps_1\ot \eps_1x\pi_{12}\,.$$
Consequently,
$$\Phi_4^3(x\ot y\ot z)=\eps_1(\eps_1x\pi_{12})\pi_{12}
\ot \pi_{12}(\pi_{12}y\eps_1\pi_{12})\pi_{12}\eps_1\ot
\eps_1\pi_{12}(\pi_{12}\eps_1z\eps_1)\eps_1= x\ot y\ot z,$$
so the equality $\Phi_4^3=1$ is established.

Now we check that $\Phi_3\Phi_1\Phi_3=\Phi_1\Phi_2$. We
have
\begin{eqnarray*}
x\ot y\ot z &\stackrel{\Phi_3}\mapsto& y^t\eps_2\ot\eps_2
x^t\ot z^t \stackrel{\Phi_1}\mapsto \pi_{23}(y^t\eps_2)
\pi_{13}\ot \pi_{13}\eps_2x^t\ot z^t\pi_{23} \\
&\stackrel{\Phi_3}\mapsto& (\pi_{13}\eps_2x^t)^t\eps_2
\ot \eps_2(\pi_{23}(y^t\eps_2)\pi_{13})^t \ot
(z^t\pi_{23})^t.
\end{eqnarray*}
Simplify the latter expression. We have
$$ (\pi_{13}\eps_2x^t)^t\eps_2=(x^t)^t\eps_2^t\pi_{13}^t
\eps_2=x\eps_2\pi_{13}\eps_2=x\pi_{13}$$
(because $\pi_{13}$ and $\eps_2$ are symmetric
and commute). Similarly
$\eps_2(\pi_{23}(y^t\eps_2)\pi_{13})^t=\eps_2\pi_{13}\eps_2
y\pi_{23}=\pi_{13}y\pi_{23}$, and $(z^t\pi_{23})^t=
\pi_{23}z$. After all we obtain
$$(\Phi_3\Phi_1\Phi_3)(x\ot y\ot z)=x\pi_{13}\ot \pi_{13}y
\pi_{23}\ot \pi_{23}z=T(1,\pi_{13},\pi_{23})(x\ot y\ot z),$$
whence $\Phi_3\Phi_1\Phi_3=T(1,\pi_{13},\pi_{23})$. It
remains to observe that
$$T(1,\pi_{13},\pi_{23})=T(\pi_{23},\pi_{13},1)
T(\pi_{23},1,\pi_{23})=\Phi_1\Phi_2.$$
\eoproof
\begin{lemma}   \label{l:S4}
Let $G$ be a group containing four elements $a_1$, $a_2$,
$a_3$, $a_4$ such that

(1) $a_1$, $a_2$, $a_3$, $a_4$ generate $G$,

(2) $a_i$ satisfy relations
\begin{equation}    \label{f:S4_rels}
a_1^2=a_2^2=a_3^2=a_4^3=1,\quad a_1a_2=a_2a_1\,,
\quad a_3a_1a_3=a_1a_2\,,
\end{equation}
$$ a_3a_2a_3=a_2\,, \quad a_4a_1a_4^{-1}=a_1a_2\,,\quad
a_4a_2a_4^{-1}=a_1\,,\quad a_3a_4a_3=a_4^{-1}\,, $$
and

(3) $a_1\ne1$.

Then $G\cong S_4$.
\end{lemma}
{\em Proof.} First observe that $S_4$ contains elements
$b_1$, $b_2$, $b_3$, $b_4$, satisfying
relations~(\ref{f:S4_rels})
(with $a_i$ replaced by $b_i$), namely $b_1=(12)(34)$,
$b_2=(13)(24)$, $b_3=(13)$, and $b_4=(123)$ (checking
relations~(\ref{f:S4_rels}) is left to the reader). Also note that
the system of relations (\ref{f:S4_rels}) is equivalent
to the following system of relations:
\begin{equation}    \label{f:S4_lels_a}
a_1^2=a_2^2=a_3^2=a_4^3=1,\quad a_2a_1=a_1a_2\,,
\quad a_3a_1=a_1a_2a_3\,,
\end{equation}
$$ a_3a_2=a_2a_3\,, \quad a_4a_1=a_1a_2a_4\,,
\quad a_4a_2=a_1a_4\,,\quad a_3a_4=a_4^2a_3\,. $$
It follows that if $G$ is any group, satisfying conditions
(1) and (2) of the lemma (but not necessary (3)), then
any element of $G$ can be written in the form
\begin{equation}    \label{f:ai_can}
a_1^{l_1}a_2^{l_2}a_4^{l_4}a_3^{l_3}\,,
\end{equation}
where $0\leq l_1,l_2,l_3\leq1$ and $0\leq l_4\leq2$. Indeed,
any element of $G$ can be represented as a product
of elements $a_i$, for example
$$ a_2a_3^{-1}a_4^{-1}a_1=a_2a_3a_4a_4a_1\,.$$
Next, using relations $a_3^2=1$, $a_3a_1=a_1a_2a_3$,
$a_3a_2=a_2a_3$, and $a_3a_4=a_4^2a_3$, we can
transform such a product to the form $wa_3^{l_3}$, where
$w$ is a word involving only $a_1$, $a_2$, and $a_4$,
and $l_3\in\{0,1\}$. For example,
$$a_2\underline{a_3a_4}a_4a_1=a_2a_4a_4\underline
{a_3a_4}a_1=a_2a_4a_4a_4a_4\underline{a_3a_1}=a_2
\underline{a_4a_4a_4a_4}a_1a_2a_3=a_2a_4a_1a_2a_3$$
(in this chain of transformations we underline, in each step,
the piece of product being transformed in this step). Similarly,
using relations $a_4a_1=a_1a_2a_4$ and $a_4a_2=a_1a_4$,
we can drag all $a_4$ involved in $w$ to the right and
obtain an expression of the form $va_4^{l_4}$, where $v$
is a word involving only $a_1$ and $a_2$, and $0\leq l_4
\leq2$. Finally we can transform $v$ to the form $a_1^{l_1}
a_2^{l_2}$, where $0\leq l_1,l_2\leq1$, using relations
$a_1^2=a_2^2=1$ and $a_2a_1=a_1a_2$. After all these
transformations we arrive to the word of the
form~(\ref{f:ai_can}).

Therefore, any group $G$ that satisfies conditions (1) and
(2) is of order $\leq2\cdot2\cdot3\cdot2=24$; in particular,
$G$ is finite.

It is well known that the normal subgroups of the symmetric
group $S_4$ are the following: the trivial group $\{e\}$,
the Klein four-group
$$ V=\{e,\ (12)(34),\ (13)(24),\ (14)(23)\}, $$
the alternating group $A_4$, and the full~$S_4$. In
particular, any nontrivial normal subgroup of $S_4$
contains~$V$. Observe also that $\lu b_1,b_2\pu=V$,
$b_4\in A_4-V$, and $b_3\in S_4-A_4$. Hence it is easy
to see that $\lu b_1,b_2,b_4\pu=A_4$ and $\lu b_1,b_2,b_3,
b_4\pu=S_4$.

Consider the direct product $H=G\times S_4$, and let
$\pi_1:H\lra G$ and $\pi_2:H\lra S_4$ be the projections
onto factors. Consider next the elements $c_i=(a_i,b_i)\in H$,
$i=1,2,3,4$, and put $K=\lu c_1,c_2,c_3,c_4\pu$. Then
$$ \pi_1(K)=\lu\pi_1(c_1),\ldots,\pi_1(c_4)\pu= \lu a_1,a_2,
a_3,a_4\pu=G,$$
and similarly $\pi_2(K)=S_4$. Since the elements $a_i$,
as well as $b_i$, satisfy relations (\ref{f:S4_rels}),
the elements $c_i$ satisfy these relations also,
whence $|K|\leq24$. As $\pi_2|_K:K\lra S_4$ is
surjective, $|K|\leq 24$, and $|S_4|=24$,
we see that $\pi_2|_K$ must be an isomorphism.

Let $\rho=(\pi_2|_K)^{-1}:S_4\lra K$ be the isomorphism,
inverse to $\pi_2|_K$. Then $\sigma=\pi_1|_K\circ\rho:S_4
\lra G$ is an epimorphism. Now it is sufficient to show that
the kernel of $\sigma$ is trivial. It follows from the definitions
that $\pi_2(c_i)=b_i$, whence $\rho(b_i)=c_i$ and $\sigma
(b_i)=\pi_1(\rho(b_i))=\pi_1(c_i)=a_i$. If $\Ker\sigma\ne1$,
then $\Ker\sigma\supseteq V\ni b_1$, because $V$ is the
only minimal normal subgroup of~$S_4$. But $\sigma(b_1)
=a_1\ne1$, a contradiction. Hence $\Ker\sigma=1$.
\eoproof

\begin{prop}    \label{pr:G_is_S4}
The group $G=\lu\Phi_1,\Phi_2,\Phi_3,\Phi_4\pu$ is
isomorphic to~$S_4$.
\end{prop}
{\em Proof.} It follows from Proposition~\ref{pr:Phi_rels}
that the elements $\Phi_i=a_i$ satisfy all conditions of
Lemma~\ref{l:S4}.
\eoproof

\subsection{Invariance of $\cL$ under $G$}
	\label{subs:lad_inv}
In this subsection we prove that the Laderman algorithm
$\cL$ is invariant under the group $G=\lu\Phi_i\mid i=1,2,
3,4\pu$. To prove this, we need the following quite general
statement.
\begin{lemma}   \label{l:orbits}
Let $G$ be a group acting on a set $X$, and $N\normaleq G$
be a normal subgroup. Let $\cO$ be an orbit of $N$ on~$X$.
Then for any $g\in G$ the set $g\cO=\{gm\mid m\in\cO\}$
is also an $N$-orbit. In particular, if $\cO_1$ and $\cO_2$
are two $N$-orbits, $x_i\in\cO_i$, and $g\in G$ an element
such that $gx_1=x_2$, then $g$ bijectively maps $\cO_1$
onto~$\cO_2$.
\end{lemma}

This statement is well known, and we leave its proof to the
reader (for a proof it suffices to note that $g(nx)=(gng^{-1})
(gx)$ for any $x\in X$, $n\in N$, and $g\in G$; so, if $x$
and $y$ are in the same $N$-orbit, then $gx$ and $gy$ are
in the same $N$-orbit also).

We also need to know the images of some of the tensors
$t_1,\ldots,t_{23}$ under some of the
transformations~$\Phi_j$.
\begin{lemma}   \label{l:Phi_act}
The following relations hold:
\begin{eqnarray*}
\Phi_1:&& t_1\mapsto t_1,\ t_2\mapsto t_{13},\  t_3\mapsto
t_{11},\  t_4\mapsto t_{12},\  t_5\mapsto t_{15},\
t_6\mapsto t_{14},\\
&& t_7\mapsto t_{16},\  t_8\mapsto t_{17},\ t_9\mapsto t_{18},
\ t_{10}\mapsto t_{10},\ t_{19}\mapsto t_{19},\
t_{20}\mapsto t_{22}, \\
&& t_{21}\mapsto t_{23}\, ;
\end{eqnarray*}
\begin{eqnarray*}
\Phi_2: && t_1\mapsto t_{10},\  t_2\mapsto t_8,\  t_3\mapsto
t_{11},\  t_4\mapsto t_7,\  t_5\mapsto t_9,\
t_6\mapsto t_6,\\
&& t_{19}\mapsto t_{19},\  t_{20}\mapsto t_{23}\,;
\end{eqnarray*}
$$\Phi_3:\ t_6\mapsto t_6,\  t_2\mapsto t_5,\  t_4\mapsto
t_4,\  t_{19}\mapsto t_{19},\  t_{23}\mapsto t_{23};$$
$$\Phi_4:\ t_1\mapsto t_3\mapsto t_6,\  t_2\mapsto t_2, \
t_4\mapsto t_4,\  t_5\mapsto t_5,\  t_{19}\mapsto t_{19},
\ t_{23}\mapsto t_{23}\,.$$
\end{lemma}
{\em Proof.} A direct computation. Prove, for example, that
$\Phi_1(t_3)=t_{11}$, $\Phi_4(t_1)=t_3$, and $\Phi_3(t_2)=
t_5$.

We have $t_3=e_{22}\ot (-e_{11}+e_{12}+e_{21}-e_{22}
-e_{23}-e_{31}+e_{33}) \ot e_{12}$. The transformation
$\Phi_1$ acts by $\Phi_1(x\ot y\ot z)=\pi_{23}x\pi_{13}\ot
\pi_{13}y\ot z\pi_{23}$. Hence
$$ \Phi_1(t_3)=\pi_{23}e_{22}\pi_{13} \ot \pi_{13}(-e_{11}
+e_{12}+e_{21}-e_{22}-e_{23}-e_{31}+e_{33}) \ot
e_{12}\pi_{23}\,.$$
Simplify multiplicands in the latter product. We have
$e_{12}\pi_{23}=e_{12}(e_{11}+e_{23}+e_{32})=e_{12}e_{23}
=e_{13}$. Generally, for any $i=1,2,3$ we have $e_{i1}
\pi_{23}=e_{i1}$, $e_{i2}\pi_{23}=e_{i3}$, and $e_{i3}\pi_{23}
=e_{i2}$. That is, the multiplication of $e_{ij}$ by $\pi_{23}$
on the right affects only the index $j$, by the rule
$1\mapsto1$, $2\leftrightarrow3$. Similarly we can find any
product of the form $\pi e_{ij}\pi'$, where $\pi,\pi'\in\{
\pi_{12},\pi_{13},\pi_{23}\}$. In particular, $\pi_{23}e_{22}
\pi_{13}=e_{32}$.

Next, $\pi_{13}(-e_{11}+e_{12}+e_{21}-e_{22}-e_{23}-e_{31}
+e_{33})=-e_{31}+e_{32}+e_{21}-e_{22}-e_{23}-e_{11}+e_{13}$.
Thus we obtain
$$ \Phi_1(t_3)=e_{32}\ot (-e_{31}+e_{32}+e_{21}-e_{22}
-e_{23}-e_{11}+e_{13})\ot e_{13}\,,$$
which coincide with $t_{11}$.

Similarly we have $t_1=(e_{11}+e_{12}+e_{13}-e_{21}-e_{22}
-e_{32}-e_{33})\ot e_{22}\ot e_{21}$ and $\Phi_4(x\ot y\ot z)
=\eps_1z\pi_{12}\ot \pi_{12}x\pi_{12}\eps_1\ot \eps_1\pi_{12}
y\eps_1$, whence
$$ \Phi_4(t_1)=\eps_1e_{21}\pi_{12}\ot \pi_{12}(e_{11}+
e_{12}+e_{13}-e_{21}-e_{22}-e_{32}-e_{33})\pi_{12}\eps_1
\ot\eps_1\pi_{12}e_{22}\eps_1\,. $$
Simplify. For any $i$ and $j$ we have $\eps_1e_{1j}=-e_{1j}$
and $\eps_1e_{ij}=e_{ij}$ if $i=2$ or $3$; similarly $e_{i1}
\eps_1=-e_{i1}$ and $e_{ij}\eps_1=e_{ij}$ if $j=2,3$. Hence
$$\eps_1e_{21}\pi_{12}=e_{21}\pi_{12}=e_{22}\,,\quad
\eps_1\pi_{12}e_{22}\eps_1=\eps_1\pi_{12}e_{22}=
\eps_1e_{12}=-e_{12}\,,$$
and
\begin{eqnarray*}
\pi_{12}(e_{11}+e_{12}+e_{13}-e_{21}-e_{22}-e_{32}
-e_{33})\pi_{12}\eps_1 &=& (e_{22}+e_{21}+e_{23}-e_{12}-e_{11}
-e_{31}-e_{33})\eps_1 \\
&=&e_{22}-e_{21}+e_{23}-e_{12}+e_{11}+e_{31}-e_{33}\,.
\end{eqnarray*}

Thus we obtain
$$\Phi_4(t_1)=e_{22}\ot (e_{22}-e_{21}+e_{23}-e_{12}+e_{11}
+e_{31}-e_{33})\ot (-e_{12}), $$
which is equal to~$t_3$.

Finally, $t_2=(e_{11}-e_{21})\ot (-e_{12}+e_{22})\ot (e_{12}
+e_{22})$ and $\Phi_3(x\ot y\ot z)=y^t\eps_2\ot\eps_2x^t
\ot z^t$, whence
\begin{eqnarray*}
\Phi_3(t_2) &=& (-e_{12}+e_{22})^t\eps_2\ot \eps_2(e_{11}
-e_{21})^t\ot (e_{12}+e_{22})^t=(-e_{21}+e_{22})\eps_2\ot
\eps_2(e_{11}-e_{12})\ot (e_{21}+e_{22}) \\
&=& (-e_{21}-e_{22})\ot (e_{11}-e_{12})\ot (e_{21}+e_{22})=(e_{21}+e_{22})\ot
(-e_{11}+e_{12})\ot (e_{21}+e_{22})\\
&=&t_5\,.
\end{eqnarray*}
\eoproof

\begin{prop}    \label{pr:L_inv}
The set $\cL$ is invariant under the group $G=\lu\Phi_1,
\Phi_2, \Phi_3,\Phi_4\pu$.
\end{prop}
{\em Proof.} We will consider the action of $G$ on the set
of all decomposable tensors in $L=L_1\ot L_2\ot L_3$.

First show that $\cL$ is invariant under $\Phi_1$, and break
it into the orbits under the cyclic group $\lu\Phi_1\pu_2$.
By Lemma~\ref{l:Phi_act} we have $\Phi_1(t_2)=t_{13}$. As
$\Phi_1^2=1$, we have also $\Phi_1(t_{13})=\Phi_1(\Phi_1
(t_2))=t_2$, so $\{t_2,t_{13}\}$ is a $\lu\Phi_1\pu_2$-orbit.
We abbreviate $\{t_2,t_{13}\}$ to $\{2,13\}$. Similarly it
follows from Lemma~\ref{l:Phi_act} that
$$ \omega_1=\{1\},\quad \omega_2=\{2,13\},\quad
\omega_3=\{3,11\},\quad \omega_4=\{4,12\},$$
$$\omega_5=\{5,15\},\quad \omega_6=\{6,14\},\quad
\omega_7=\{7,16\},\quad \omega_8=\{8,17\},$$
$$\omega_9=\{9,18\},\quad \omega_{10}=\{10\},\quad
\omega_{11}=\{19\},\quad \omega_{12}=\{20,22\},$$
$$  \text{and}\quad \omega_{13}=\{21,23\},$$
are $\lu\Phi_1\pu_2$-orbits. In particular, $\cL$ is
$\lu\Phi_1\pu_2$-invariant.

Next consider the subgroup $H_1=\lu\Phi_1,\Phi_2\pu$.
As $\Phi_1$ and $\Phi_2$ commute, $\lu\Phi_1\pu_2$ is
normal in~$H_1$. As $\Phi_2(t_2)=t_8$ and $\Phi_2^2=1$,
we have $\Phi_2(t_8)=t_2$. Now it follows from
Lemma~\ref{l:orbits}, applied to $\lu\Phi_1\pu_2\normaleq H_1$,
that $\Phi_2$ bijectively maps $\omega_2=\{2,13\}$ and
$\omega_8=\{8,17\}$ each onto the other. Therefore
$$ \omega_2\cup\omega_8=\{2,13,8,17\}=\{2,8,13,17\} $$
is an $H_1$-orbit. Similarly we see that the following sets
are $H_1$-orbits:
$$\Omega_1=\{1,10\}=\omega_1\cup\omega_{10}\,, \quad
\Omega_2=\{2,8,13,17\}=\omega_2\cup\omega_8\,, \quad
\Omega_3=\{3,11\}=\omega_3\,, $$
$$\Omega_4=\{4,7,12,16\}=\omega_4\cup\omega_7\,, \quad
\Omega_5=\{5,9,15,18\}=\omega_5\cup\omega_9\,, \quad
\Omega_6=\{6,14\}=\omega_6\,, $$
$$\Omega_7=\{19\}=\omega_{11}\,, \quad
\Omega_8=\{20,21,22,23\}=\omega_{12}\cup\omega_{13}
\,. $$

Further we consider subgroup $H_2=\lu\Phi_1,\Phi_2,
\Phi_4\pu$. It follows from relations $\Phi_4\Phi_1
\Phi_4^{-1}=\Phi_1\Phi_2$ and $\Phi_4\Phi_2 \Phi_4^{-1}
= \Phi_1$ that $\Phi_4$ normalizes $H_1$, so $H_1$ is
normal in~$H_2$. Similarly, the relations $\Phi_3^2=1$,
$\Phi_3\Phi_1\Phi_3=\Phi_1\Phi_2$, $\Phi_3\Phi_2
\Phi_3=\Phi_2$, and $\Phi_3\Phi_4\Phi_3=\Phi_4^{-1}$
imply that $\Phi_3$ normalizes $H_2$, so $H_2\normaleq G$.

(One can observe (though this is not necessary) that under
the isomorphism between $G$ and $S_4$, described in
the proofs of Lemma~\ref{l:S4} and
Proposition~\ref{pr:G_is_S4},
the subgroups $H_1$ and $H_2$ correspond to normal
subgroups $V$ and $A_4$ of~$S_4$.)

By Lemma~\ref{l:Phi_act} $\Phi_4$ permutes cyclically $t_1$, $t_3$,
and $t_6$. So by Lemma~\ref{l:orbits} $\Phi_4$ cyclically permutes
the $H_1$-orbits $\Omega_1=\{1,10\}$, $\Omega_3=
\{3,11\}$, and $\Omega_6=\{6,14\}$. Therefore the set
$\Sigma_1=\Omega_1\cup\Omega_3\cup\Omega_6=\{1,3,6,
10,11,14\}$ is an $H_2$-orbit. Next, as each of the tensors
$t_i$, where $i=2,4,5,19,23$, is $\Phi_4$-invariant, we see
that each of the $H_1$-orbits $\Omega_2$, $\Omega_4$,
$\Omega_5$, $\Omega_7$, and $\Omega_8$ is
$\Phi_4$-invariant, and therefore is an $H_2$-orbit.

As $\Phi_3$ interchanges $t_2$ and $t_5$ and normalizes
$H_2$, it interchanges the $H_2$-orbits $\Omega_2$ and
$\Omega_5$. Therefore $\Sigma_2=\Omega_2\cup
\Omega_5$ is a $G$-orbit. Next, as $\Phi_3$ fixes elements
$t_6\in\Sigma_1$, $t_4\in\Omega_4$, $t_{19}\in\Omega_7$,
and $t_{23}\in\Omega_8$, we see that $\Phi_3$ leaves
invariant their $H_2$-orbits $\Sigma_1$, $\Omega_4$,
$\Omega_7$, and $\Omega_8$. Thus we obtain that the
sets $\Sigma_1=\{1,3,6,10,11,14\}$, $\Sigma_2=\Omega_2
\cup\Omega_5=\{2,5,8,9,13,15,17,18\}$, $\Omega_4=\{4,7,
12,16\}$, $\Omega_7=\{19\}$, and $\Omega_8=\{20,21,22,
23\}$ are $G$-orbits. So their union $\cL=\Sigma_1\cup
\Sigma_2\cup\Omega_4\cup\Omega_7\cup\Omega_8$
is invariant under~$G$.
\eoproof

\subsection{The full group $\Aut(\cL)$}
	\label{subs:AutL_full}
According to Proposition \ref{pr:L_inv}, the group
$G=\lu\Phi_i\mid i=1,2,3,4\pu$ is a subgroup
of $\Aut(\cL)$. In this subsection we prove that $G$
is the {\em full} automorphism group of~$\cL$.

Let $\Aut(\cL)_0\leq\Aut(\cL)$ be the subgroup of all
elements that correspond to the identity permutation of
$\{L_1,L_2,L_3\}$. Also put $Q_1=\lu\Phi_3,\Phi_4\pu$.
\begin{lemma}   \label{l:AutL0Q}
$\Aut(\cL)=\Aut(\cL)_0Q_1$.
\end{lemma}
{\em Proof.} Let $\pi:\Gamma(t)\lra S_3$ be the
homomorphism taking each $g\in\Gamma(t)$ to the
corresponding permutation of $\{L_1,L_2,L_3\}$. Then
$\pi(\Phi_4)=(123)$ and $\pi(\Phi_3)=(12)(3)$, whence
$\pi(Q_1)=\lu(123),(12)(3)\pu=S_3$. Therefore for each
element $g\in\Aut(\cL)$ there exists an element $g'\in Q_1$
such that $\pi(g)=\pi(g')$. Put $g''=g(g')^{-1}$, then
$g=g''g'$. Also $\pi(g'')=\pi(g)\pi(g')^{-1}=e$, and therefore
$g''\in\Aut(\cL)_0$. So $g\in\Aut(\cL)_0Q_1$. As $g$ was an
arbitrary element of $\Aut(\cL)$, we obtain that $\Aut(\cL)
=\Aut(\cL)_0Q_1$.
\eoproof

It follows from Proposition~\ref{pr:Gamma0_H} that the
elements of $\Aut(\cL)_0$ are
precisely the elements of $\Aut(\cL)$ of the form $g=T(a,b,c)$,
for some $a,b,c\in GL_3(K)$.

Introduce a notion which will play important role in the rest
of the paper.

\paragraph{Tensor projections.}
Let $U\ot V$ be the product of two spaces. For any  two
subspaces $X,Y\sse U$ we have $X\ot V\cap Y\ot V=(X\cap
Y)\ot V$. It follows that for any subspace $L\sse U\ot V$
there exists the least (i.e., the unique minimal) subspace
$X\sse U$ such that $L\sse X\ot V$. We call $X$ the
{\em tensor projection} of $L$ to $U$, and denote this by
$X=\tpr_UL$.

The tensor projection $\tpr_VL$ to the second factor is defined
similarly. Also, for an element $x\in U\ot V$ we write $\tpr_Ux$
for $\tpr_U\lu x\pu$.

Generally, if $\wt U=U_1\ot\ldots\ot U_m$ is the product
of several spaces, and $\ov U=U_{i_1}\ot\ldots\ot U_{i_l}$,
where $1\leq i_1<\ldots<i_l\leq m$, is a ``subproduct'',
then we can define $\tpr_{\ov U}L$, the tensor projection to
$\ov U$, for any subspace $L\sse\wt U$.

It is more or less clear that operation of taking tensor
projections has transitivity property; for example,
$$ \tpr_{U_2}(\tpr_{U_1\ot U_2}L)=\tpr_{U_2}L $$
for any $L\sse U_1\ot U_2\ot U_3$.

Let $U\ot V$ be the product of two spaces, and $X\sse
U\ot V$ be a set of nonzero elements of~$U\ot V$. Define
\begin{equation}    \label{f:tpr_set}
\tpr_UX=\{\tpr_Ux\mid x\in X\}.
\end{equation}
This is a set of nonzero subspaces of~$U$. It may happen
that $\tpr_Ux=\tpr_Uy$ for two distinct $x,y\in X$, $x\ne y$.
So we shall consider $\tpr_UX$ as a {\em multiset}, i.e., a
set with multiplicities (at least in the case if $X$ is finite).
Note that we can define $\tpr_UX$ by formula~(\ref{f:tpr_set})
for a multiset $X$ also.

The operation of taking tensor projections has certain
invariance
properties. The following lemma is evident, but its accurate
proof (which is left to the reader) may be tedious.
\begin{lemma}   \label{l:tpr_inv}
Let $\wt U=U_1\ot\ldots\ot U_m$ be a tensor product of
several spaces, $X\sse\wt U$ be a finite (multi)subset of
nonzero tensors, and let $\fe\in S(U_1,\ldots, U_m)$ be a
decomposable automorphism of $\wt U$ such that $\fe(X)
=X$. Suppose that $\fe$ takes factor $U_i$ to $U_j$, and let
$\psi:U_i\lra U_j$ be the corresponding isomorphism ($\psi$
is defined up to constant). Then $\psi$ takes (multi)set
$\tpr_{U_i}X$ to $\tpr_{U_j}X$.
\end{lemma}
\vspace{2ex}

Return to considering group $\Aut(\cL)$. We need the 
following lemma.
\begin{lemma}   \label{l:3sp_lines}
Let $V=\lu e_1,e_2,e_3\pu$ be a three-dimensional space,
and let $\pi_{23}\in GL(V)$ acts by $\pi_{23}:e_1\mapsto e_1$,
$e_2\leftrightarrow e_3$. Let $\fe\in GL(V)$ be a
transformation preserving the multiset of lines
$$ \cX=\{1\cdot\lu e_1\pu, 4\cdot\lu e_2\pu, 4\cdot\lu e_3
\pu, 2\cdot\lu e_1-e_2\pu, 2\cdot\lu e_1-e_3\pu \}.$$
Then either $\fe=\lambda\cdot\id_V$ or $\fe=\lambda\pi_{23}$,
where $\lambda\in K^\ast$.
\end{lemma}
{\em Proof.} It is clear that $\fe$ preserves each of the three
sets of lines $\{\lu e_1\pu\}$, $\{\lu e_2\pu, \lu e_3\pu\}$,
and $\{\lu e_1-e_2\pu, \lu e_1-e_3\pu \}$. So $\fe$ either leaves
each of the lines $\lu e_i\pu$ invariant, or preserves
$\lu e_1\pu$ and interchanges $\lu e_2\pu$ and $\lu e_3
\pu$.

In the first case we have $\fe(e_i)=a_ie_i$, where $a_i\in
K^\ast$. Then $\fe(e_1-e_2)=a_1e_1-a_2e_2$. The latter
vector must be proportional to $e_1-e_2$ or to $e_1-e_3$,
so it is proportional to $e_1-e_2$ and $a_1=a_2$. Similarly
$a_1=a_3$, and therefore $\fe=a_1\cdot\id_V$.

In the second case $\fe$ acts as $e_1\mapsto a_1e_1$,
$e_2\mapsto a_2e_3$, $e_3\mapsto a_3e_2$. The line $\lu
e_1-e_2\pu$ goes to $\lu a_1e_1-a_2e_3\pu$, whence $a_2
=a_1$. Similarly $a_3=a_1$, and therefore $\fe=a_1\pi_{23}$.
\eoproof

\begin{lemma}   \label{l:tpr_L1_ogr}
Consider the following set of one-dimensional subspaces
in the space $M=M_{33}(K)$:
\begin{eqnarray*} \cC=\{ && \lu e_{11}-e_{21}\pu,
\lu e_{21}+e_{22}\pu, \lu -e_{11}+e_{31}\pu,
\lu e_{31}+e_{32}\pu, \lu e_{13}-e_{33}\pu,
\lu e_{32}+e_{33}\pu, \\
&& \lu e_{13}-e_{23}\pu, \lu e_{22}+e_{23}\pu,
\lu e_{12}\pu, \lu e_{23}\pu, \lu e_{21}\pu,
\lu e_{31}\pu, \lu e_{33}\pu \}.
\end{eqnarray*}
Let $a,b\in GL_3(K)$ be elements such that the transformation
$\fe:x\mapsto axb$ preserves~$\cC$. Then
$$ a\in\{1,\pi_{23}\}\quad \text{and}\quad b
\in\{1,\pi_{13}\}$$
up to scalar factors.
\end{lemma}
{\em Proof.}
Let $V$ and $V'$ be spaces of 3-columns and 3-rows,
respectively. We can identify $M=M_3(K)$ with $V\ot V'$
by the isomorphism $\alpha:V\ot V'\lra M_3(K)$ defined by
$\alpha(v\ot v')=vv'$ (cf. Subsection~\ref{subs:mnp_prel}).

Note that a matrix $x\in M$ is of rank 1 if and only if the
corresponding tensor $\alpha^{-1}(x)\in V\ot V'$ is
decomposable, $\alpha^{-1}(x)=v\ot v'$. The set $\cC'=
\alpha^{-1}(\cC)$, corresponding to $\cC$ under $\alpha$,
consists of 13 lines in $V\ot V'$ spanned by decomposable
tensors, namely
\begin{eqnarray*} \cC'=\alpha^{-1}(\cC)=\{
&& \lu(e_1-e_2)\ot e^1\pu, \lu e_2\ot (e^1+e^2)\pu,
\lu (e_1-e_3)\ot e^1\pu, \\
&& \lu e_3\ot (e^1+e^2)\pu, \lu (e_1-e_3)\ot e^3\pu,
\lu e_3\ot(e^2+e^3)\pu, \lu (e_1-e_2)\ot e^3\pu, \\
&& \lu e_2\ot(e^2+e^3)\pu, \lu e_1\ot e^2\pu,
\lu e_2\ot e^3\pu, \lu e_2\ot e^1\pu, \lu e_3\ot e^1\pu,
\lu e_3\ot e^3\pu \}.
\end{eqnarray*}

Let $\fe':V\ot V'\lra V\ot V'$ be the automorphism
corresponding to $\fe$ under $\alpha$, that is, $\fe'=
\alpha^{-1}\fe\alpha$. It is easy to see that $\fe'$ is
decomposable, namely $\fe'=\fe_1\ot\fe_2$, where $\fe_1:
V\lra V$ and $\fe_2:V'\lra V'$ are defined by $\fe_1(x)=ax$
and $\fe_2(y)=yb$. Indeed, for any $v\in V$ and $v'\in V'$
we have
\begin{eqnarray*}
\fe'(v\ot v') &=& (\alpha^{-1}\fe\alpha)(v\ot v')
=\alpha^{-1}(\fe(\alpha(v\ot v')))
=\alpha^{-1}(\fe(vv'))=\alpha^{-1}(avv'b) \\
&=& \alpha^{-1}((av)(v'b))=av\ot v'b
=\fe_1(v)\ot\fe_2(v')=(\fe_1\ot\fe_2)(v\ot v').
\end{eqnarray*}

Since $\cC$ is invariant under $\fe$, we see that $\cC'$ must
be invariant under~$\fe'$. Therefore the tensor projections
$\tpr_V\cC'=\cC'_1$ and $\tpr_{V'}\cC'=\cC'_2$ must be
invariant (as multisets) under $\fe_1$ and $\fe_2$,
respectively.

We can immediately see that
$$ \cC'_1=\{1\cdot\lu e_1\pu, 4\cdot\lu e_2\pu, 4\cdot
\lu e_3\pu, 2\cdot\lu e_1-e_2\pu, 2\cdot\lu e_1-e_3\pu \}$$
and
$$ \cC'_2=\{4\cdot\lu e^1\pu, 1\cdot\lu e^2\pu, 4\cdot
\lu e^3\pu, 2\cdot\lu e^1+e^2\pu, 2\cdot\lu e^2+e^3
\pu \}.$$
It follows from Lemma~\ref{l:3sp_lines} that $a=1$ or $a=\pi_{23}$,
up to a scalar. Similarly one can prove that $b=1$ or
$b=\pi_{13}$ up to a scalar.
\eoproof

\begin{prop}    \label{pr:AutL0}
The equality $\Aut(\cL)_0=\lu\Phi_1,\Phi_2\pu$ holds.
\end{prop}
{\em Proof.}
The inclusion $\Aut(\cL)_0\supseteq\lu\Phi_1,\Phi_2\pu$
is obvious, because both $\Phi_1$ and $\Phi_2$ are of the
form $T(a,b,c)$. We need to prove the inverse inclusion.

Let $u=u_1\ot u_2\ot u_3$ be a decomposable tensor of
$L_1\ot L_2\ot L_3$. The triple $(\rk(u_1), \rk(u_2),\rk(u_3))$
will be called the {\em type} of~$u$. The tensors of type
$(1,1,1)$ in $\cL$ are $t_i$ with $i=2$, $5$, $8$, $9$,
$13$,$15$, $17$, $18$, $19$, $20$, $21$, $22$, $23$,
of type $(2,1,1)$ --- $t_i$ with $i=1,10$, $(1,2,1)$
--- with $i=3,11$, $(1,1,2)$ --- $i=6,14$, and $(2,2,2)$ ---
$i=4$, $7$, $12$, $16$ (and there are no tensors of
other types, say $(2,2,3)$, in~$\cL$).

By $\cB$ we denote the set of all $t_i$ of type $(1,1,1)$.

Observe that for any $x\in GL_3(K)$ and $y\in M_3(K)$ we
have $\rk(xy)=\rk(yx)=\rk(y)$. Hence for arbitrary
decomposable tensor $u=u_1\ot u_2\ot u_3\in L$ and
arbitrary transformation of the form $g=T(a,b,c)$ the types
of tensors $u$ and $g(u)$ coincide. It follows that any
element $g\in\Aut(\cL)_0$ preserves~$\cB$.

Further, let $g=T(a,b,c)$ be any element of $\Aut(\cL)_0$.
As $g$ preserves $\cB$, the set of tensor projections
$\cC=\tpr_{L_1}\cB$ must be invariant under transformation
$\fe:x\mapsto axb^{-1}$. It is easy to see that
\begin{eqnarray*} \cC=\{ && \lu e_{11}-e_{21}\pu,
\lu e_{21}+e_{22}\pu, \lu -e_{11}+e_{31}\pu,
\lu e_{31}+e_{32}\pu, \lu e_{13}-e_{33}\pu,
\lu e_{32}+e_{33}\pu, \\
&& \lu e_{13}-e_{23}\pu, \lu e_{22}+e_{23}\pu,
\lu e_{12}\pu, \lu e_{23}\pu, \lu e_{21}\pu,
\lu e_{31}\pu, \lu e_{33}\pu \}.
\end{eqnarray*}

By Lemma~\ref{l:tpr_L1_ogr}, $g$ has the form
$g=T(\pi_{23}^\eps,\pi_{13}^\eta,c)$,
for some $\eps,\eta\in\{0,1\}$. Remembering that $\Phi_1=
T(\pi_{23},\pi_{13},1)$ and $\Phi_2=T(\pi_{23},1,\pi_{23})$,
we see that there exist (uniquely defined) $\gamma,\delta
\in\{0,1\}$ and $g'$ of the form $g'=T(1,1,c')$ such that
$g=\Phi_1^\eps\Phi_2^\eta g'$. So it is sufficient to show
that any element $g'$ of the form $g'=T(1,1,c')$, leaving
$\cB$ invariant, coincide with the identity map.

Obviously, $g'$ acts on $L_1$ trivially (up to a scalar). Note
also that the tensor projections of all elements of $\cB$ to
$L_1$ are pairwise distinct. It follows that $g'$ fixes each
element of~$\cB$. So the map $x\mapsto xc^{-1}$
preserves any subspace of the form $\tpr_{L_2}v$, where
$v\in\cB$. Hence easily follows (by an argument similar to
the proof of Lemma~\ref{l:tpr_L1_ogr}) that $c$ is
a scalar, whence $g'=1$.
\eoproof
\vspace{2ex}

{\em Thus, the statement of Theorem~\ref{th:main}, 
concerning the Laderman algorithm, is established.}

\subsection{Some details of calculations}
	\label{subs:how_find}
In the arguments of Subsections
\ref{subs:lad_subgr}--\ref{subs:AutL_full} we used the
transformations $\Phi_1$, $\Phi_2$, $\Phi_3$, $\Phi_4$
``in ready form'', but the reader may ask (and the author
should explain) how these transformations were {\em
found}. (Though, it is not difficult to have an idea of this
from arguments of Subsection~\ref{subs:AutL_full}).

The key idea is to decompose each of the spaces $L_i=
M_{3,3}$ ($i=1,2,3$) as $L_i=U_i\ot V_i$, where $U_i$
(resp., $V_i$) are three copies of the space of 3-columns
(resp., 3-rows). Then $L=L_1\ot L_2\ot L_3$ decomposes as
\begin{equation}    \label{f:L_fine}
L=U_1\ot V_1\ot U_2\ot V_2\ot U_3\ot V_3\,.
\end{equation}
It follows from Theorem~\ref{th:Gamma_t_full} that any
element of $\Gamma(t)$ is a decomposable automorphism of $L$
with respect to decomposition~(\ref{f:L_fine}).
Next, a decomposable tensor $u=u_1\ot u_2\ot u_3\in L$
is decomposable with respect to~(\ref{f:L_fine}) if and
only if it is of type $(1,1,1)$. It was noted above that
$\Aut(\cL)$ must preserve the set $\cB$ of all elements of
$\cL$ of type $(1,1,1)$.

We can consider $\cB$ as a set of 13 tensors decomposable
with respect to~(\ref{f:L_fine}). It is convenient to
write $\cB$ as a table (see Table 1).
\begin{figure}[h]
\caption{Table 1}
\begin{tabular}{|c||c|c|c|c|c|c|}
\hline  N & $U_1$  & $V_1$ & $U_2$  & $V_2$  & $U_3$
& $V_3$ \\
\hline
\hline 2 & $1-2$ & $1$ & $-1+2$ & $2$ & $1+2$ & $2$ \\
\hline 5 & $2$  & $1+2$  & $1$  & $-1+2$  & $2$  & $1+2$ \\
\hline 8 & $-1+3$  & $1$  & $1-2$  & $3$  & $1+3$  & $3$ \\
\hline 9 & $3$  & $1+2$  & $1$  & $-1+3$  & $3$  & $1+3$ \\
\hline 13 &  $1-3$ & $3$  & $2-3$  & $2$  & $1+2$  & $3$ \\
\hline 15 & $3$  & $2+3$  & $3$  & $-1+2$  & $2$  & $1+3$ \\
\hline 17 & $1-2$  & $3$  & $2-3$  & $3$  & $1+3$  & $2$ \\
\hline 18 &  $2$ & $2+3$  & $3$  & $-1+3$  & $3$  & $1+2$ \\
\hline 19 & $1$  & $2$  & $2$  & $1$  & $1$  & $1$ \\
\hline 20 & $2$  & $3$  & $3$  & $2$  & $2$  & $2$ \\
\hline 21 & $2$  & $1$  & $1$  & $3$  & $3$  & $2$ \\
\hline 22 & $3$  & $1$  & $1$  & $2$  & $2$  & $3$ \\
\hline 23 & $3$  & $3$  & $3$  & $3$  & $3$  & $3$ \\
\hline
\end{tabular}
\end{figure}

For example, the 3-rd row of the table reads as follows:
$$t_8=(-e_1+e_3)\ot e^1\ot(e_1-e_2)\ot e^3\ot(e_1+e_3)\ot
e^3= (-e_{11}+e_{31})\ot(e_{13}-e_{23})\ot(e_{13}+e_{33}).$$

Assume that $g\in\Aut(\cL)$ preserves each of the factors
$L_1$, $L_2$, and $L_3$. Then $g$ has the form $g=T(a,b,c)$,
in particular $g$ preserves all factors $U_i$, $V_i$ of
decomposition~(\ref{f:L_fine}). Since $g$ preserves $\cB$,
it follows that the transformation $x\mapsto ax$ preserves
the multiset
$$\tpr_{U_1}\cB=\{1\cdot\lu e_1\pu, 4\cdot\lu e_2\pu,
4\cdot\lu e_3\pu, 2\cdot\lu e_1-e_2\pu, 2\cdot\lu e_1-e_3
\pu \},$$
whence $a=1$ or $a=\pi_{23}$, up to a scalar. Similarly one
can show that $b=\pi_{13}^\eta$ and $c=\pi_{23}^\theta$,
for some $\eta,\theta\in Z_2=\{0,1\}$. Therefore
$$g=T(\pi_{23}^\eps,\pi_{13}^\eta,\pi_{23}^\theta)$$
for some $\eps,\eta,\theta\in Z_2$. It is easy to check that
any element of the latter form with $\eps+\theta+\eta=0$
preserves $\cB$, while the element with $(\eps,\eta,\theta)
=(0,0,1)$ (and therefore any element with $\eps+\theta+
\eta=1$) does not.

Then the author has checked that elements with $\eps+
\theta+\eta=0$ preserve the set $\cL\setminus\cB$
(i.e., the set of all elements of $\cL$ whose type is different
from $(1,1,1)$) also, and so preserve the whole~$\cL$. Thus,
the group $\Aut(\cL)\cap\Gamma^0(t)$ turns out to be
completely described.

Next we should consider elements of $\Aut(\cL)$
corresponding to nontrivial permutation of the factors
$L_1$, $L_2$, and $L_3$.

The observation that $\cL$ contains 13 and 4 elements of
types $(1,1,1)$ and $(2,2,2)$, respectively, and 2 elements
of each of the types $(2,1,1)$, $(1,2,1)$, and $(1,1,2)$,
suggests that the group of permutations of factors induced
by $\Aut(\cL)$ is either $Z_3$ or~$S_3$.

Indeed, using Lemma~\ref{l:tpr_inv} and the table above, it is not
hard to find candidates for automorphisms corresponding
to nontrivial permutations of $L_1$, $L_2$, $L_3$.
A candidate for an automorphism corresponding to
$(12)(3)$ can be found from the table very easily. In fact,
this candidate is nothing else but~$\Phi_3$. As to an
automorphism corresponding to $(123)$, to find it is a more
complicated task.

Finally note that the checking that $\cL$ is invariant under
all $\Phi_i$ was made by the author directly, without using
Lemma~\ref{l:orbits} and relations of
Proposition~\ref{pr:Phi_rels}. (These
relations were found later, in order to streamline the
argument in the present text.)

\section{Automorphisms of the Hopcroft algorithm}
	\label{sec:hop}
\subsection{Hopcroft algorithm}
	\label{subs:hop_alg}

Recall the description of the Hopcroft algorithm
(in computational form), according to~\cite{Hop-Mus}. Let
$$ X=\begin{pmatrix} x_{11} & x_{12} \\
x_{21} & x_{22} \\ x_{31} & x_{32} \end{pmatrix}, \quad\text{and}\quad
Y=\begin{pmatrix} y_{11} & y_{12} & y_{13} \\
y_{21} & y_{22} & y_{23} \end{pmatrix}. $$
Then the coefficients $z_{ij}$ of the matrix
$$ Z=XY=\begin{pmatrix}
z_{11} & z_{12} & z_{13} \\ z_{21} & z_{22} & z_{23} \\
z_{31} & z_{32} & z_{33} \end{pmatrix} $$
can be computed by formulae
$$ z_{11}=p_1+p_2\,,\quad z_{22}=p_3+p_4\,,\quad
z_{33}=p_5+p_6\,, $$
$$ z_{12}=-p_2-p_3+p_7-p_8\,,\quad z_{21}=-p_1-p_4+p_8-p_9\,, $$
$$ z_{13}=-p_1-p_5-p_{13}+p_{15}\,,\quad
z_{31}=-p_2-p_6+p_{13}-p_{14}\,, $$
$$ z_{23}=-p_3-p_6+p_{11}-p_{12}\,, \quad
z_{32}=-p_4-p_5+p_{10}-p_{11}\,, $$
where
$$ p_1=(x_{11}-x_{12})y_{11}\,,\quad p_2=x_{12}(y_{11}
+y_{21}),\quad p_3=x_{21}y_{12}\,,$$
$$ p_4=x_{22}y_{22}\,,\quad p_5=x_{31}(y_{13}
+y_{23}),\quad p_6=(-x_{31}+x_{32})y_{23}\,,$$
$$ p_7=(x_{11}+x_{21})(y_{11}+y_{12}+y_{21}+y_{22}),$$
$$ p_8=(x_{11}-x_{12}+x_{21})(y_{11}+y_{21}+y_{22}),$$
$$ p_9=(x_{11}-x_{12}+x_{21}-x_{22})(y_{21}+y_{22}),$$
$$ p_{10}=(x_{22}+x_{32})(y_{12}+y_{13}+y_{22}+y_{23}),$$
$$ p_{11}=(x_{22}-x_{31}+x_{32})(y_{12}+y_{13}+y_{23}),$$
$$ p_{12}=(-x_{21}+x_{22}-x_{31}+x_{32})(y_{12}+y_{13}),$$
$$ p_{13}=(x_{12}+x_{31})(y_{11}-y_{23}),\quad
p_{14}=(-x_{12}-x_{32})(y_{21}+y_{23}),$$
$$ p_{15}=(x_{11}+x_{31})(y_{11}+y_{13}).$$

Further, present Hopcroft algorithm in tensor form.
Consider the following decomposable tensors in the
space $M_{32}\ot M_{23}\ot M_{33}$:
$$ t_1=(e_{11}-e_{12})\ot e_{11}\ot(e_{11}-e_{31}-e_{12}), $$
$$t_2=e_{12}\ot(e_{11}+e_{21})\ot(e_{11}-e_{21}-e_{13}),$$
$$t_3=e_{21}\ot e_{12}\ot(-e_{21}+e_{22}-e_{32}),$$
$$t_4=e_{22}\ot e_{22}\ot(-e_{12}+e_{22}-e_{23}),$$
$$t_5=e_{31}\ot(e_{13}+e_{23})\ot(e_{33}-e_{31}-e_{23}),$$
$$t_6=(-e_{31}+e_{32})\ot e_{23}\ot(e_{33}-e_{13}-e_{32}),$$
$$t_7=(e_{11}+e_{21})\ot(e_{11}+e_{12}+e_{21}+e_{22})
\ot e_{21}\,,$$
$$t_8=(e_{11}-e_{12}+e_{21})\ot(e_{11}+e_{21}+e_{22})
\ot(e_{12}-e_{21}),$$
$$t_9=(e_{11}-e_{12}+e_{21}-e_{22})\ot(e_{21}+e_{22})
\ot (-e_{12}),$$
$$t_{10}=(e_{22}+e_{32})\ot(e_{12}+e_{13}+e_{22}+
e_{23}) \ot e_{23}\,,$$
$$t_{11}=(e_{22}-e_{31}+e_{32})\ot(e_{12}+e_{13}+e_{23})
\ot(-e_{23}+e_{32}),$$
$$t_{12}=(-e_{21}+e_{22}-e_{31}+e_{32})\ot(e_{12}+
e_{13})\ot(-e_{32}),$$
$$t_{13}=(e_{12}+e_{31})\ot(e_{11}-e_{23})\ot(e_{13}
-e_{31}), $$
$$t_{14}=(e_{12}+e_{32})\ot(e_{21}+e_{23})\ot e_{13}\,,$$
$$t_{15}=(e_{11}+e_{31})\ot(e_{11}+e_{13})\ot e_{31}\,.$$
\begin{prop}    \label{pr:hop_tens}
The set $\cH=\{t_1\,,\ldots,t_{15}\}$ is the tensor
form of the Hopcroft algorithm.
\end{prop}
{\em Proof.} A direct computation, left to the reader.
\eoproof

\subsection{The group $\Aut(\cH)$}
	\label{subs:hop_group}
In this subsection we find the automorphism group
of the Hopcroft algorithm. Our arguments are similar
to those for Laderman's algorithm, cf. Subsections
\ref{subs:lad_subgr}--\ref{subs:AutL_full}. So we
provide only the results of
computations,usually, leaving the details to the
reader.

Let $L_1=M_{32}$, $L_2=M_{23}$, $L_3=M_{33}$, and
$L=L_1\ot L_2\ot L_3$. We can consider transformations
of $L$ of the form $T(a,b,c)$, where $a,c\in GL_3(K)$
and $b\in GL_2(K)$.

Put $d=\begin{pmatrix} 0 & -1 \\ 1 & -1 \end{pmatrix}
\in GL_2(K)$, then $d^3=1$ and $d^{-1}=\begin{pmatrix}
-1 & 1 \\ -1 & 0 \end{pmatrix}$. Also put
$$\pi_{123}=\begin{pmatrix} 0 & 0 & 1 \\ 1 & 0 & 0 \\
0 & 1 & 0 \end{pmatrix}.$$
Let $e_{ij}$ be a matrix unit in one of the spaces
$M_{32}$, $M_{23}$, or $M_{33}$.
Observe that the multiplication of the
matrix $e_{ij}$ by $\pi_{123}$ on the left, resp. by
$\pi_{123}^{-1}$ on the right, shifts the subscript $i$ (resp., $j$)
by 1: $\pi_{123} e_{ij}=e_{i+1,j}$ (if the product
$\pi_{123}e_{ij}$ makes sense, that is, if
$e_{ij}\in M_{32}$ or $M_{33}$), and similarly
$e_{ij} \pi_{123}^{-1}=e_{i,j+1}$, if
$e_{ij}\in M_{23}$ or $M_{33}$. Here subscripts $i+1$ ($j+1$) are
taken modulo 3, so that $3+1=1$.

Consider transformations
$$ \Phi_1=T(\pi_{123},d,\pi_{123}): x\ot y\ot z\mapsto
\pi_{123}x\begin{pmatrix} -1 & 1 \\ -1 & 0\end{pmatrix}
\ot \begin{pmatrix} 0 & -1 \\ 1 & -1\end{pmatrix} y
\pi_{123}^{-1}\ot \pi_{123}z\pi_{123}^{-1}\,, $$
$$ \Phi_2=T(\pi_{13},\pi_{12},\pi_{13}): x\ot y\ot z
\mapsto \pi_{13}x\pi_{12}\ot \pi_{12}y\pi_{13}\ot
\pi_{13}z\pi_{13}\,, $$
and
$$\Phi_3: x\ot y\ot z\mapsto y^t\pi_{12}\eps_1\ot
\eps_1\pi_{12}x^t\ot z^t $$
(here notation $\pi_{12}$, $\pi_{13}$, and $\eps_1$
have the same meaning as in Subsection~\ref{subs:lad_subgr}).
\begin{lemma}   \label{l:hop_Phi_rels}
The transformations $\Phi_1$, $\Phi_2$, and $\Phi_3$
satisfy the following relations:
$$ \Phi_1^3=\Phi_2^2=\Phi_3^2=1,\quad \Phi_2\Phi_1
\Phi_2=\Phi_1^{-1}, $$
$$\Phi_3\Phi_1=\Phi_1\Phi_3\,,\quad \Phi_3\Phi_2=
\Phi_2\Phi_3\,.$$
\end{lemma}
{\em Proof.} Recall that the map by the rule
$(a,b,c)\mapsto T(a,b,c)$ is a group homomorphism
(from $GL_3(K)\times GL_2(K)\times GL_3(K)$ to
$GL(L)$). So the relations $\Phi_1^3=\Phi_2^2=1$
follow from $\pi_{123}^3=\pi_{13}^2=1$ and $d^3=
\pi_{12}^2=1$. Similarly, the relation $\Phi_2\Phi_1
\Phi_2=\Phi_1^{-1}$ follows from $\pi_{13}\pi_{123}
\pi_{13}=\pi_{123}^{-1}$ and $\pi_{12}d\pi_{12}=
d^{-1}$. The remaining three relations can be proved
by direct computations (cf. the proof of
Proposition~\ref{pr:Phi_rels}).
\eoproof

\begin{lemma}   \label{l:S3Z2}
Let $G$ be a group generated by three elements $a_1$,
$a_2$, $a_3$. Suppose that $a_i$ satisfy relations
\begin{equation}    \label{f:S3Z2_rels}
a_1^3=a_2^2=a_3^2=1,\quad a_2a_1a_2=a_1^{-1}\,,
\quad a_3a_1=a_1a_3\,,\quad a_3a_2=a_2a_3\,.
\end{equation}
Suppose also that $a_1\ne1$ and $a_3\ne1$. Then
$G\cong S_3\times Z_2$.
\end{lemma}
{\em Proof.} The argument is similar to that
in the proof of Lemma~\ref{l:S4}. Note first that
system~(\ref{f:S3Z2_rels}) is equivalent to
\begin{equation}    \label{f:S3Z2_rels2}
a_1^3=a_2^2=a_3^2=1,\quad a_2a_1=a_1^2a_2\,,
\quad a_3a_1=a_1a_3\,,\quad a_3a_2=a_2a_3\,.
\end{equation}

Next, consider the group $S_3\times Z_2$. The
elements of $Z_2$ will be denoted by $0$ and~$1$. In
$S_3\times Z_2$ consider the elements $b_1=((123),0)$,
$b_2=((12),0)$, and $b_3=(e,1)$. Clearly, they satisfy
relations (\ref{f:S3Z2_rels}) and~(\ref{f:S3Z2_rels2}).

Further, observe that if $X$ is any group generated
by three elements $c_i$ satisfying relations~(\ref{f:S3Z2_rels2})
(with $a_i$ replaced by $c_i$), then any element of
$X$ can be written as $c_1^{l_1}c_2^{l_2}c_3^{l_3}$,
where $0\leq l_1\leq2$, $0\leq l_2,l_3\leq1$. Hence
$|X|\leq12=|S_3\times Z_2|$.

Finally, it is easy to see that any minimal normal subgroup of
$S_3\times Z_2$ is either $\lu b_1\pu_3$ or $\lu b_3\pu_2$.

Using the facts observed it is not hard to give a proof
similar to that of Lemma~\ref{l:S4}. The details are left
to the reader.
\eoproof
\vspace{2ex}

The following statement is an obvious corollary of
Lemmas \ref{l:hop_Phi_rels} and~\ref{l:S3Z2}.
\begin{cor} \label{cor:hop_S3Z2}
The group $G=\lu\Phi_1,\Phi_2,\Phi_3\pu$ is isomorphic to
$S_3\times Z_2$.
\end{cor}
\begin{lemma}   \label{l:hop_Phi_act}
The following relations for action of transformations
$\Phi_i$ on tensors $t_j$ hold:
\begin{eqnarray*}
\Phi_1: & & t_1\mapsto t_4\mapsto t_5\,,\quad
t_2\mapsto t_3\mapsto t_6\,,\quad t_7\mapsto t_{12}
\mapsto t_{14}\,, \\
& & t_9\mapsto t_{10}\mapsto t_{15}\,,\quad t_8\mapsto
t_{11}\mapsto t_{13}\,; \\
\Phi_2: & & t_1\mapsto t_6\,,\quad t_{14}\mapsto t_{15}
\,,\quad t_8\mapsto t_{11}\,;\\
\Phi_3: & & t_1\mapsto t_2\,,\quad t_7\mapsto t_9\,,
\quad t_8\mapsto t_8\,.
\end{eqnarray*}
\end{lemma}
{\em Proof.} A direct computation. \eoproof

\begin{prop}    \label{pr:hop_inv}
Consider the action of the group $G=\lu\Phi_1, \Phi_2, \Phi_3\pu$
on the set of nonzero decomposable tensors in~$L$. Then the sets
$$ \Omega_1=\{t_1,\ldots,t_6\} =\{1,2,3,4,5,6\},$$
$$\Omega_2=\{7,9,10,12,14,15\},$$
and
$$\Omega_3=\{8,11,13\}$$
are
$G$-orbits. In particular, the set
$$\cH=\Omega_1\cup\Omega_2\cup\Omega_3=\{t_1,
\ldots,t_{15}\}$$
is invariant under~$G$.
\end{prop}
{\em Proof.} The argument is similar to the proof of
Proposition~\ref{pr:L_inv}. First of all, it follows
from Lemma~\ref{l:hop_Phi_act}
and the relation $\Phi_1^3=1$ that the sets
$$ \omega_1=\{t_1,t_4,t_5\}=\{1,4,5\},\quad
\omega_2=\{2,3,6\},\quad \omega_3=\{7,12,14\},$$
$$\omega_4=\{8,11,13\},\quad \text{and}\quad \omega_5=
\{9,10,15\}$$
are orbits of the group $\lu\Phi_1\pu_3$.
Next, as $\Phi_2$ normalizes $\lu\Phi_1\pu_3$, it
follows from Lemma~\ref{l:orbits} and the relation
$$ \Phi_2:t_1\mapsto t_6\,,\quad t_{14}\mapsto t_{15}
\,,\quad t_8\mapsto t_{11}\,,$$
that $\Phi_2$ interchanges $\omega_1$ and $\omega_2$,
$\omega_3$ and $\omega_5$, and preserves~$\omega_4$.
So the sets $\Omega_1=\omega_1\cup\omega_2$,
$\Omega_2= \omega_3\cup\omega_5$, and $\Omega_3=\omega_4$,
are orbits under the group $\lu\Phi_1,\Phi_2\pu$.
Finally, as $\Phi_3$ normalizes (even centralizes)
$\lu\Phi_1,\Phi_2\pu$, it follows from the action of
$\Phi_3$ on $t_1$, $t_7$, and $t_8$, that each of
the $\lu\Phi_1,\Phi_2\pu$-orbits $\Omega_1$,
$\Omega_2$, and $\Omega_3$ is invariant under~$\Phi_3$.
So $\Omega_i$ is an orbit under~$G$.
\eoproof

\begin{cor} \label{cor:hop_AutH_sub}
$G=\lu\Phi_1,\Phi_2,\Phi_3\pu$ is a subgroup
of $\Aut(\cH)$.
\end{cor}

In the rest of this section we prove that, in fact,
$\Aut(\cH)=G$.

Let $\Aut(\cH)_0\leq\Aut(\cH)$ be the subgroup of
all elements preserving each of the factors $L_1$,
$L_2$, and $L_3$. The following lemma and its
proof are similar to Lemma~\ref{l:AutL0Q}.
\begin{lemma}   \label{l:hop_AutH_dec}
We have $\Aut(\cH)=\Aut(\cH)_0\lu\Phi_3\pu_2$.
\end{lemma}
{\em Proof.} Let $\pi:\Gamma(t)\lra S_3$ be the
homomorphism that assigns to each element
$g\in\Gamma(t)$ the corresponding permutation of
the factors $L_1$, $L_2$, and $L_3$. As $\dim L_1
=\dim L_2=6$ and $\dim L_3=9$, we have $\pi(\Gamma(t))
\sse\{e,(12)(3)\}$ (and, actually, $\pi(\Gamma(t))
=\{e,(12)(3)\}$ by Theorem~\ref{th:Gamma_t_full}).
On the other hand, it is clear that $\pi(\Phi_3)
=(12)(3)$. The rest of the proof is similar to that
of Lemma~\ref{l:AutL0Q}.
\eoproof

Let $V$ and $V'$ be spaces of 3-columns and 3-rows,
respectively, $(e_1,e_2,e_3)$ and $(e^1,e^2,e^3)$
be the usual bases of $V$ and $V'$, and let $GL_3(K)$
acts on $V$ and $V'$ in the usual way. Let $\sigma
\in S_3$ be a permutation. By $\wt\sigma$ we denote
the  element of $GL_3(K)$ permuting basis vectors
$(e_i)$ according to~$\sigma$.
\begin{lemma}   \label{l:3+3_lines}
Let $\fe\in GL_3(K)$ be an element such that the
sets of three lines
$$\cX=\{\lu e_1+e_2\pu, \lu e_1+e_3\pu, \lu e_2+e_3\pu
\}$$
and
$$\cY=\{\lu e^1\pu,\lu e^2\pu,\lu e^3\pu\}$$
in $V$ and $V'$ are invariant under~$\fe$. Then
$\fe=\lambda\wt\sigma$ for some $\sigma\in S_3$
and $\lambda\in K^\ast$.
\end{lemma}
{\em Proof.} Note that for any $\sigma\in S_3$ the transformation
$\wt\sigma$ preserves both $\cX$ and $\cY$. Moreover, the
permutation by which $\wt\sigma$ acts on $\cY$ coincides
with~$\sigma$. Now let $\fe$ be as in the hypothesis of the lemma,
and $\sigma$ be the permutation by which $\fe$ acts on~$\cY$. Then
the transformation $\fe'=\wt\sigma^{-1}\fe$ leaves each of the
lines $\lu e^i\pu$ invariant. Therefore
$\fe'=\diag(\lambda_1,\lambda_2, \lambda_3)$ for some
$\lambda_i\in K^\ast$. Also, $\fe'$ preserves~$\cX$. So the line
$\fe'(\lu e_1+e_2\pu)=\lu\lambda_1e_1+\lambda_2 e_2\pu$ must be in
$\cX$, and therefore it coincides with $\lu e_1+e_2\pu$, whence
$\lambda_1=\lambda_2$. Similarly $\lambda_1= \lambda_3$, whence
$\fe'=\diag(\lambda_1, \lambda_1,\lambda_1)=\lambda_1E$. So
$\fe=\wt \sigma\fe'=\lambda_1\wt\sigma$. \eoproof

\begin{prop}    \label{pr:AutH0}
The equality $\Aut(\cH)_0=\lu\Phi_1,\Phi_2\pu$
holds.
\end{prop}
{\em Proof.} The argument is mainly similar to
the proof of Proposition~\ref{pr:AutL0}.
Any element of $\Aut(\cH)_0$ is of the form
$T(a,b,c)$, for some $(a,b,c)\in GL_3(K)\times
GL_2(K)\times GL_3(K)$.

For any $m,n\in\bbN$ (not necessary $m=n$) and
any $x\in GL_m(K)$, $y\in GL_n(K)$, $z\in M_{mn}$
we have $\rk(xz)=\rk(zy)=\rk(z)$. So for any
decomposable tensor $u=u_1\ot u_2\ot u_3\in
M_{mn}\ot M_{np}\ot M_{pm}$ and any element
$g=T(a,b,c)$, where $a\in GL_m(K)$, $b\in
GL_n(K)$, $c\in GL_p(K)$, the tensors $u$ and
$g(u)$ are of the same type. Therefore
$\Aut(\cH)_0$ preserves the subset of all tensors
of type $(1,1,1)$ in~$\cH$. It is easy to see
that this subset is
$$ \Omega_2=\{t_i\mid i=7,9,10,12,14,15\}. $$
Thus, it is sufficient to prove the following
statement:
\begin{quote}
($\ast$) {\em If a transformation $g=T(a,b,c)$, where $a,c\in
GL_3(K)$ and $b\in GL_2(K)$, leaves the set $\Omega_2$ invariant,
then $g\in \lu\Phi_1,\Phi_2\pu$.}
\end{quote}

Let $D$ and $F$ (resp., $D'$ and $F'$) be two
copies of the space of 3-columns (resp., 3-rows),
and let $E$ and $E'$ be spaces of 2-columns
and 2-rows, respectively. Consider tensor
product
$$ N=D\ot E'\ot E\ot F'\ot F\ot D'.$$
We can identify $N$ with $L=M_{32}\ot M_{23}
\ot M_{33}$ by the isomorphism $\tau:N\lra L$
defined by
$$ \tau(d\ot e'\ot e\ot f'\ot f\ot d')=
de'\ot ef'\ot fd'$$
(cf. Subsection~\ref{subs:Gamma0}).

Let $\cB$ and $g'$ be the subset and the
transformation of $N$, corresponding to
$\Omega_2$ and $g$, respectively, with respect to the isomorphism~$\tau$. That is, $\cB=
\tau^{-1}(\Omega_2)$ and $g'=\tau^{-1}g\tau$.
Then $g'(\cB)=(\tau^{-1}g\tau)(\tau^{-1}
(\Omega_2))=\tau^{-1}(g(\Omega_2))=\tau^{-1}
(\Omega_2)=\cB$, that is, $g'$ preserves~$\cB$.

It is easy to write $\cB$ and $g'$ explicitly.
Namely,
\begin{eqnarray*}
\cB=\{ && (e_1+e_2)\ot e^1\ot(e_1+e_2)\ot
(e^1+e^2)\ot e_2\ot e^1, \\
&& (e_1+e_2)\ot (-e^1+e^2)\ot e_2\ot(e^1+e^2)
\ot e_1\ot e^2, \\
&& (e_2+e_3)\ot e^2\ot(e_1+e_2)\ot
(e^2+e^3)\ot e_2\ot e^3, \\
&& (e_2+e_3)\ot (e^1-e^2)\ot e_1\ot(e^2+e^3)
\ot e_3\ot e^2, \\
&& (e_1+e_3)\ot e^2\ot e_2\ot(e^1+e^3)\ot e_1
\ot e^3, \\
&& (e_1+e_3)\ot e^1\ot e_1\ot(e^1+e^3)
\ot e_3\ot e^1\}.
\end{eqnarray*}
Also it is easy to see that $g'$ acts according
to the formula
\begin{equation}    \label{f:hop_g'}
g'(d\ot e'\ot e\ot f'\ot f\ot d')= ad\ot
e'b^{-1} \ot be\ot f'c^{-1}\ot cf\ot d'a^{-1}.
\end{equation}
Indeed, we have
$$ g(\tau(d\ot e'\ot e\ot f'\ot f\ot d'))
=g(de'\ot ef'\ot fd')$$
$$=T(a,b,c)(de'\ot ef'\ot fd')=ade'b^{-1}\ot
bef'c^{-1}\ot cfd'a^{-1}.$$
But the latter expression coincides with
$$ \tau(ad\ot e'b^{-1}\ot be\ot f'c^{-1}\ot cf
\ot d'a^{-1}).$$
Therefore,
$$ g'(d\ot e'\ot e\ot f'\ot f\ot d')=(\tau^{-1}
g\tau)(d\ot e'\ot e\ot f'\ot f\ot d')$$
$$ =\tau^{-1}(ade'b^{-1}\ot bef'c^{-1}\ot cfd'a^{-1})
=ad\ot e'b^{-1}\ot be\ot f'c^{-1}\ot cf\ot d'a^{-1},$$
which proves formula~(\ref{f:hop_g'}).

Equality $g'(\cB)=\cB$, together with
formula~(\ref{f:hop_g'}), imply that the tensor
projection $\tpr_D\cB$ is invariant under the
transformation $d\mapsto ad$ ($d\in D$). It is
immediately seen that
$$\tpr_D\cB=\{\lu e_1+e_2\pu, \lu e_1+e_3\pu, \lu e_2+e_3\pu\}. $$
Therefore the transformation $d\mapsto ad$
preserves the latter set. Similarly,
$$\tpr_{D'}\cB=\{\lu e^1\pu,\lu e^2\pu,\lu e^3\pu\}$$
must be invariant under transformation
$d'\mapsto d'a^{-1}$. So the element $a\in GL_3
(K)$ satisfies hypothesys of Lemma~\ref{l:3+3_lines},
whence $a=\lambda\wt\sigma$ for some $\lambda
\in K^\ast$ and a permutation $\sigma\in S_3$.
Thus, $g=T(\lambda\wt\sigma,b,c)=T(\wt\sigma,
b,c)$.

Remembering that $\Phi_1=T(\pi_{123},
\begin{pmatrix} 0 & -1 \\ 1 & -1 \end{pmatrix},
\pi_{123})$ and $\Phi_2=T(\pi_{13},\pi_{12},
\pi_{13})$, and taking into account that $(123)$ and $(13)(2)$
generate $S_3$, we see that there exists an element
$g_1\in\lu\Phi_1,\Phi_2\pu$ of the form
$g_1=T(\wt\sigma,b_1,c_1)$. Therefore the element $g_2=g_1^{-1}g$
is of the form $T(1,b_2,c_2)$. Moreover, it is clear that $g_2$
preserves~$\Omega_2$. So it is sufficient to show that if an
element $g_2$ of the form $T(1,b_2,c_2)$ preserves $\Omega_2$,
then $g_2\in\lu\Phi_1,\Phi_2\pu$. We shall prove even more, namely
that $g_2=1$.

Let $g_2'=\tau^{-1}g_2\tau$ be the
transformation of $N$, corresponding to~$g_2$.
Clearly, $g'_2$ preserves~$\cB$. Further, 
formula~(\ref{f:hop_g'}) immediately implies that
for any element $v\in N$ the tensor projections
of $v$ and $g'_2(v)$ to $D\ot D'$ coincide:
$$ \tpr_{D\ot D'}g'_2(v)=\tpr_{D\ot D'}v,
\quad \forall\: v\in N. $$
But the tensor projections of all elements of $\cB$
to $D\ot D'$ are pairwise distinct, whence
$g'_2(v)=v$. That is, $g'_2$ fixes each element
of~$\cB$. It follows that the transformation
$f\mapsto c_2f$ preserves tensor projection to $F$
of each element of $\cB$, so preserves each
of three lines $\lu e_1\pu$, $\lu e_2\pu$, and
$\lu e_3\pu$. Similarly, the transformation
$f'\mapsto f'c_2^{-1}$ preserves
$F'$-projection of each element of $\cB$,
that is, preserves each of the three lines
$\lu e^1+e^2\pu$, $\lu e^1+e^3\pu$, and
$\lu e^2+e^3\pu$. Hence easily follows that $c_2$
is a scalar.

Finally, the transformation $e\mapsto b_2e$
preserves tensor projection to $E$ of each element
of $\cB$, and therefore preserves each of the
lines $\lu e_1\pu$, $\lu e_2\pu$, and $\lu e_1+
e_2\pu$. Hence $b_2$ is a scalar also.

As both $b_2$ and $c_2$ are scalars, we obtain
that $g_2=T(1,b_2,c_2)=1$ ($=\id_L$).
\eoproof
\vspace{2ex}

We summarize the results of the present section
in the following proposition.
\begin{prop}    \label{pr:AutH}
The group $\Aut(\cH)$ coincides with $G=\lu
\Phi_1,\Phi_2,\Phi_3\pu$. The latter group is
isomorphic to $S_3\times Z_2$.
\end{prop}

\vspace{2ex}

{\em The latter proposition proves the part of 
Theorem~\ref{th:main} concerning the Hopcroft algorithm.}

\end{document}